\documentclass[aps,pra,reprint,a4paper,showpacs,nofootinbib]{revtex4-1}
\usepackage{amsmath,amssymb,bm,bbm,psfrag,graphicx,feynmp}

\newcommand{\e}{\mathrm{e}}
\DeclareMathOperator{\erfc}{erfc}

\DeclareMathOperator{\sgn}{sgn}

\newenvironment{gap}{
  \begin{fmfgraph*}(25,10)    
    \fmfstraight    
    \fmfleft{lb,lt}
    \fmfright{rb,rt}
    \fmftop{t4}
    \fmfbottom{b4}
    \fmf{phantom}{t4,vlt}
    \fmf{phantom}{b4,vlb}  
    \fmf{phantom}{vrb,rb}
    \fmf{phantom}{vrt,rt}
    \fmfpoly{smooth,tension=1.5,label=\tiny $\Delta$}{vrt,vlt,vlb,vrb}    
    \fmf{phantom}{lt,t1,t2,t3,t4}
    \fmf{phantom}{lb,b1,b2,b3,b4}
    \fmffreeze
  }{\end{fmfgraph*}}

\begin{document}

\date{\today}
\title{Collective modes, stability and superfluid transition of a
quasi-two-dimensional dipolar Fermi gas}
\author{L.\,M.\,Sieberer$^{1,2}$}
\author{M.\,A.\,Baranov$^{1,2,3}$}
\affiliation{$^1$Institute for Theoretical Physics, University of Innsbruck, 6020
  Innsbruck, Austria} 
\affiliation{$^2$Institute for Quantum Optics and Quantum Information of the
  Austrian Academy of Sciences, 6020 Innsbruck, Austria} 
\affiliation{$^3$RRC ``Kurchatov Institute'', Kurchatov Square 1, 123182
Moscow, Russia}

\begin{abstract}
  We examine collective modes, stability, and BCS pairing in a
  quasi-two-dimensional gas of dipolar fermions aligned by an external field.
  By using the (conserving) Hartree-Fock approximation, which treats direct and
  exchange interactions on an equal footing, we obtain the spectrum of
  single-particle excitations and long wavelength collective modes (zero sound)
  in the normal phase. It appears that exchange interactions result in strong
  damping of zero sound when the tilting angle between the dipoles and the
  normal to the plane of confinement is below some critical value. In
  particular, zero sound cannot propagate if the dipoles are perpendicular to
  the plane of confinement. At intermediate coupling we find unstable modes that
  can lead either to collapse of the system or the formation of a density
  wave. The BCS transition to a superfluid phase, on the other hand, occurs at
  arbitrarily weak strengths of the dipole-dipole interaction, provided the
  tilting angle exceeds a critical value. We determine the critical temperature
  of the transition taking into account many-body effects as well as virtual
  transitions to higher excited states in the confining potential, and discuss
  prospects of experimental observations.
\end{abstract}

\pacs{67.85.-d,03.75.Ss,74.78.-w}
\maketitle

\begin{fmffile}{diagram} \setlength{\unitlength}{1mm}
\begin{fmfshrink}{.5}

\section{Introduction}

\label{sec:introduction} In recent years, experimentalists have achieved major
breakthroughs in preparing samples of diatomic molecules in the rovibrational
ground state and cooling them towards quantum degeneracy \cite%
{ni08:_high_phase_space_densit_gas_polar_molec,
  deiglmayr08:_format_ultrac_polar_molec_rovib_groun_state, danzl10}. With
heteronuclear molecules, in particular, rotational degrees of freedom can be
excited in a controlled way by applying external electric fields, and are
associated with large electric dipole moments \cite%
{baranov08:_theor,carr09:_cold,krems09:_cold_molec,lahaye09}. This possibility
of inducing strong and anisotropic dipole-dipole interactions between the
molecules opens fascinating prospects for the observation of various many-body
effects and novel quantum phases \cite%
{baranov02:_super_pairin_in_polar_dipol_fermi_gas,%
  baranov04:_bcs_pairin_in_trapp_dipol_fermi_gas,%
  wang06:_quant_fluid_self_assem_chain_polar_molec,%
  buechler07:_stron_correl_quant_phases_cold_polar_molec,%
  bruun08:_quant_phases_of_two_dimen,%
  baranov08:_theor,%
  krems09:_cold_molec,%
  lahaye09, cooper09:_stabl_topol_super_phase_ultrac,
  pikovski10:_inter_super_bilay_system_fermion_polar_molec,%
  baranov11:_bilay,%
  levinsen11:_topol}.

The above-mentioned experimental studies of heteronuclear polar molecules 
\cite{ni08:_high_phase_space_densit_gas_polar_molec,
deiglmayr08:_format_ultrac_polar_molec_rovib_groun_state} suffer from losses
due to chemical reactions such as $\mathrm{KRb} + \mathrm{KRb} \to \mathrm{K}%
_2 + \mathrm{Rb}_2$ \cite%
{ni10:_dipol,ospelkaus10:_quant_state_contr_chemic_react}, which place
severe limitations on the achievable densities in three-dimensional samples.
These reactions are significantly suppressed if one confines the molecules
to a quasi-two-dimensional (quasi-2D) geometry and orients their dipole
moments perpendicular to the plane of the 2D translational motion \cite%
{quemener10:_elect,micheli10:_univer_rates_react_ultrac_polar,quemener11:_dynam}%
, as has been verified experimentally by de Miranda \textit{et al.}~\cite%
{miranda11:_contr}. Moreover, in some of the polar molecules that consist of
alkali atoms, atom-exchanging reactions are endothermic and, therefore, do
not occur \cite{zuchowski10:_react}. Thus, it seems most promising for
future investigations of dipolar molecules to either focus on species that
do not undergo chemical reactions, or consider samples that are strongly
confined to a quasi-2D regime.

In this paper, we consider a quasi-2D gas of fermionic dipoles, aligned by an
external field (see Fig.~\ref{fig:tilted_dipoles}). The simplest case
corresponds to the dipoles being oriented perpendicular to the plane of
confinement, so that the pairwise dipole-dipole interaction is isotropic and
repulsive. At non-zero values of the tilting angle $\theta _{0}$ it becomes
anisotropic and, for $\theta _{0}>\arcsin (1/\sqrt{3})$, partially attractive. A
discussion of this physical setup within the framework of Fermi liquid theory
was given in \cite{chan10:_anisot_fermi}, where, e.\thinspace g.,
single-particle properties such as the anisotropic self-energy and the resulting
deformation of the Fermi surface from the spherical shape corresponding to the
non-interacting case were calculated to first order in perturbation
theory. While this perturbative approach provides reliable answers in the weak
coupling regime, moderate interaction strengths require more sophisticated
methods such as the Hartree-Fock approximation (HFA), which was used in
Ref.~\cite%
{yamaguchi10:_densit_wave_instab_in_two} to obtain the spectrum of
single-particle excitations at zero and finite temperature. At zero temperature,
the results were found to agree very well with the outcome of a variational
approach that was initially used to study Fermi surface deformations in the 3D
case \cite{miyakawa08:_phase_space_defor_of_trapp} and adapted to the 2D case by
the authors of Ref.~\cite%
{bruun08:_quant_phases_of_two_dimen}.

Collective modes in single-, bi- and multilayered structures of dipolar Fermi
gases were studied in \cite{li10:_collec} using the random phase approximation
(RPA), which neglects exchange interactions. As a result, in particular for the
single-layer setup, these studies predict the spectrum of long wavelength
collective excitations to be sensitive to microscopic details of the two-body
interaction potential in the form of a short-distance cut-off, which is needed
to handle the singular behavior of a dipole-dipole interaction. In a quasi-2D
setting, the characteristic length of the harmonic confinement takes the role of
the cut-off, resulting in a confinement-dependent value of the RPA speed of zero
sound \cite%
{fischer06:_stabil_of_quasi_two_dimen,kestner10:_compr}. The authors of \cite%
{sun10:_spont_fermi}, however, correctly remark that a cut-off dependent
constant term in the momentum space representation of the interparticle
potential corresponds to a short-range contact interaction in real space and,
therefore, must not have an effect in a single-component Fermi gas. Thus, the
existence of the RPA zero sound mode is questionable. In this paper, we study
zero sound collective modes on the basis of the so-called conserving HFA
developed in
Refs.~\cite{baym61:_conser_laws_correl_funct,baym62:_self_consis_approx_many_body_system}. The
advantage of this method is that it provides a way to fully include exchange
contributions in a given order of perturbation theory, such that the results are
consistent with conservation of particle number, energy, and momentum, as well
as with fermionic statistics of particles. We show that the existence of a zero
sound collective mode in a quasi-2D dipolar Fermi gas for small values of the
tilting angle $\theta _{0}$ and the dependence of the sound velocity on a
short-distance cut-off are artifacts of the RPA. In particular, we find that the
propagation of zero sound is not possible if the dipoles are aligned
perpendicular to the plane of confinement or if they are tilted only slightly --
in consistence with the homogeneous 3D setting \cite%
{ronen10:_zero_fermi,chan10:_anisot_fermi} in which there is no propagating zero
sound in the directions perpendicular (or close to perpendicular) to the
direction of dipole polarization.

The issue of stability of the normal phase of the system against collapse was
addressed by Chan \textit{et al.}~\cite{chan10:_anisot_fermi} following
Pomeranchuk's approach \cite{pomeranchuk58:_stabil_of_fermi_liquid}: The normal
phase is thermodynamically stable if an arbitrary distortion of the Fermi
surface results in an increase of the ground state energy. In Ref.~%
\cite{chan10:_anisot_fermi}, however, distortions do not refer to the deformed
Fermi surface, but rather to the circular one of the non-interacting system,
i.\thinspace e., the authors of this reference are performing the stability
analysis around a configuration that does not extremize the ground state
energy. Hence their expression for the change in the ground state energy
contains a term that is linear in the distortion [Eq.~(81) in
\cite{chan10:_anisot_fermi}], which is absent if one takes the deformed Fermi
surface as reference center [see Eq.~\eqref{eq:46} below]. An alternative
approach by Bruun and Taylor \cite%
{bruun08:_quant_phases_of_two_dimen} uses a variational ansatz for the shape of
the Fermi surface, on the basis of which the compressibility is calculated. The
collapse instability is then identified with a negative value of this
quantity. The stability of the normal phase against density fluctuations with a
finite momentum (density wave instability) was investigated in RPA in
Refs.~\cite{sun10:_spont_fermi} and -- extending the discussion to finite
temperatures and taking into account the deformation of the Fermi surface --
\cite{yamaguchi10:_densit_wave_instab_in_two}. It was found, that a density wave
transition takes place in a broad region in the parameter space (the coupling
strength and the tilting angle $\theta _{0}$), where the system is stable
against collapse. An improvement of the RPA treatment was achieved in
Ref.~\cite{parish11:_densit_fermi} by an approximate treatment of correlations
beyond the RPA in the form of a local-field correction in the density-density
response function (the same method with a different form of the local-field
correction was used in \cite{zinner11:_densit} for a specific value of the
tilting angle $\theta _{0}=\arccos 1/%
\sqrt{3}$). The result is that the density wave instability should be expected
at higher values of the interaction strength than predicted by the RPA. This
approach, however, provides results that are very sensitive to the form of the
local-field correction, and the justification of a particular choice is not
clear. We address the issue of stability on the basis of the conserving HFA that
provides a consistent (and within this approximation scheme exact) treatment of
exchange contributions. For the case $\theta _{0}=0$, the same approach was used
in Ref.~\cite{babadi11:_densit} to study the density wave instability in a
dipolar mono- and multilayer sytems. Our result (see
Sec.~\ref{sec:dens-wave-inst}) for the critical value of the coupling strength
for $\theta _{0}=0$ agrees very well with that from
Ref.~\cite{babadi11:_densit}.

The critical temperature $T_{c}$ of the transition to the superfluid phase
for the quasi-2D dipolar Fermi gas was obtained by Bruun and Taylor \cite%
{bruun08:_quant_phases_of_two_dimen} in the BCS approach with the
dipole-dipole interaction restricted to the dominant $p$-wave channel. We
extend this work by taking into account the full angular dependence of the
dipole-dipole interaction, as well as by calculating the preexponential
factor in the expression for the critical temperature. The latter requires
taking into account both the many-body contributions to the interparticle
interaction (the so-called Gor'kov--Melik-Barkhudarov (GM) corrections~\cite%
{gorkov61:_contr_to_theor_of_super}) and virtual transitions to excited
states in the trapping potential, which ultimately result in a non-trivial
dependence of the critical temperature on the trapping frequency, the gas
density, and the tilting angle.

This paper is organized as follows: In Sec.~\ref{sec:system} we present the
microscopic model for the quasi-2D dipolar Fermi gas and identify the
relevant parameters and parameter regimes. We review key quantities in the
many-body problem and the equations that relate them and discuss our
strategy to solve these equations in the intermediate coupling regime in
Secs.~\ref{sec:appr-interm-coupl} and \ref{sec:quasi-2d-dipolar}. An
investigation of the single-particle excitation spectrum in the normal phase
is given in Sec.~\ref{sec:single-part-excit}, and is a prerequisite for the
study of collective modes in Sec.~\ref{sec:coll-excit-hfa}. We address the
issue of instability towards, respectively, collapse and the formation of a
density wave in Secs.~\ref{sec:long-wavel-inst} and \ref{sec:dens-wave-inst}%
. Results for the critical temperature of the transition to the superfluid
phase are given in Sec.~\ref{sec:superfl-trans}. Finally, Sec.~\ref%
{sec:concluding-remarks} is devoted to a summary of our findings and a
discussion of the prospects of observing the described phenomena in
experiments. Details of our numerical methods and analytical expressions for
the matrix elements of the dipole-dipole interaction are given in the
appendices.

\section{System}

\label{sec:system} We consider a gas of dipolar fermions of mass $m$ with
dipole moments $\mathbf{d} = d \hat{\mathbf{d}}$, which are polarized along
the direction $\hat{\mathbf{d}} = (\sin \theta_0,0,\cos \theta _0)$, i.\,e., 
$\theta_0$ is the angle between the orientation of dipoles and the $z$-axis
(see Fig.~\ref{fig:tilted_dipoles}). The gas is strongly confined to the $xy 
$-plane by a harmonic trapping potential $V(z) = m \omega_0^2 z^2/2$. Here,
strong confinement means that the transverse extension of the gas cloud,
which is on the order of $l_0 = \sqrt{\hbar/m \omega_0}$, is small as
compared to the mean interparticle separation in the $xy$-plane. The latter
quantity is proportional to the inverse Fermi momentum $p_F^{(0)}$, which,
in turn, is determined by the area density $n_{2D}$, $p_F^{(0)} = \hbar 
\sqrt{4 \pi n_{2D}}$. Thus, the small parameter characterizing the
tight-confinement or quasi-2D limit is $\eta \equiv p_F^{(0)} l_0/\hbar \ll 1
$. This condition implies that the Fermi energy $\varepsilon_F^{(0)}$ is
much smaller than the trapping potential level spacing, $%
\varepsilon_F^{(0)}/\hbar \omega_0 \ll 1$.

We also assume ultracold temperatures, $T \ll \varepsilon_F^{(0)}$, such
that the average kinetic energy of particles is given by the Fermi energy $%
\varepsilon_F^{(0)}$. 
\begin{figure}[tbp]
\psfrag{theta0}{$\theta_0$} \psfrag{x}{$x$} \psfrag{y}{$y$} \psfrag{z}{$z$} 
\centering
\includegraphics[width=.8\linewidth]{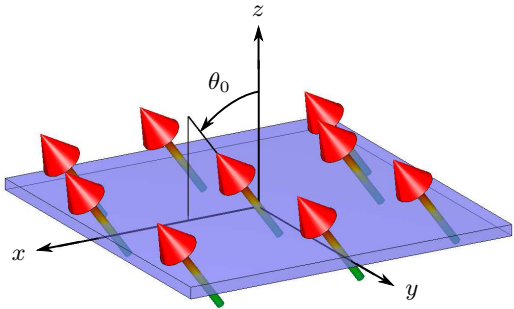}
\caption{(Color online) Fermionic dipoles confined to the $xy$-plane. The
  dipoles are aligned in the $xz$-plane and form an angle $\protect\theta_0$
  with the $z $%
  -axis.}
\label{fig:tilted_dipoles}
\end{figure}

The Hamiltonian of this system reads 
\begin{multline}  \label{eq:1}
H = \int d \mathbf{r} \, \widehat{\psi}^{\dag}(\mathbf{r}) \left[ -\frac{%
\hbar^2}{2m} \Delta + \frac{1}{2} m \omega_0^2 z^2 -\mu^{\prime}\right] 
\widehat{\psi}(\mathbf{r}) \\
+ \frac{1}{2} \int d \mathbf{r} \, d \mathbf{r}^{\prime}\, \widehat{\psi}%
^{\dag}(\mathbf{r}) \, \widehat{\psi}^{\dag}(\mathbf{r}^{\prime}) \, V_d(%
\mathbf{r}-\mathbf{r}^{\prime}) \, \widehat{\psi}(\mathbf{r}^{\prime}) \, 
\widehat{\psi}(\mathbf{r}),
\end{multline}
where $\widehat{\psi}(\mathbf{r})$ is the fermionic field operator, $\mu
^{\prime}$ denotes the chemical potential, and $V_d(\mathbf{r}) =
(d^2/r^3)[1 - 3(\hat{\mathbf{r}} \cdot \hat{\mathbf{d}})^2]$ is the
dipole-dipole interaction. Here we omit the contribution of the short-range
part of the interparticle interaction: In the considered case of a
single-component Fermi gas, it results only in $p$-wave scattering, which is
small assuming that $p_F^{(0)} r_0/\hbar \ll 1$, where $r_0$ is the radius
of the short-range part of the interparticle interaction.

A characteristic length of the dipole-dipole interaction is given by $r_d =
md^2/\hbar^2 \gg r_0$. This is the length scale below which the
dipole-dipole interaction substantially influences the relative wave
function of two particles. We assume $r_d \ll l_0$, such that interparticle
collisions are essentially three-dimensional. This gives us another small
parameter $g \equiv p_F^{(0)}r_d/\hbar \ll 1$.

Under the above conditions, the motion of particles in the $z$-direction is
limited to the ground state of the confining potential $\phi_0(z)$ -- the
lowest harmonic oscillator (HO) level, and a single-particle wave function
is $\psi (\mathbf{r}) = \varphi(\bm{\rho}) \, \phi_0(z)$, where $\varphi (%
\bm{\rho})$ describes the in-plane motion [$\bm{\rho} = (x,y)$].

As a result, first order interaction effects (see Secs.~\ref%
{sec:single-part-excit}, \ref{sec:coll-excit-hfa}, and \ref%
{sec:inst-spat-homog}) can be described by an effective interaction 
\begin{equation}
V_{0}(\bm{\rho})\equiv \int dz\,dz^{\prime }\,\phi _{0}(z)^{2}\,V_{d}(%
\bm{\rho},z-z^{\prime })\,\phi _{0}(z^{\prime })^{2}  \label{eq:2}
\end{equation}%
with the Fourier transform [$\mathbf{p}=(p_{x},p_{y})$] \cite%
{fischer06:_stabil_of_quasi_two_dimen} 
\begin{equation}
V_{0}(\mathbf{p})=\sqrt{2}\pi \tfrac{d^{2}}{l_{0}}\,w\bigl(\tfrac{pl_{0}}{%
\sqrt{2}\hbar }\bigr)\left[ u(\mathbf{p})\sin ^{2}\!\theta
_{0}-2\,P_{2}(\cos \theta _{0})\right] ,  \label{eq:3}
\end{equation}%
where $u(\mathbf{p})\equiv (p_{x}^{2}-p_{y}^{2})/p^{2}$ and $w(x)\equiv x\,%
\mathrm{e}^{x^{2}}\erfc(x)$.\footnote{\label{fn:1}This expression is unique
up to a momentum independent additive constant, which depends on the
regularization of the Fourier integral at the origin. Such a constant,
however, corresponds to a short-range interaction and has no physical effect
in a single component Fermi gas because its contributions vanish upon proper
antisymmetrization. Therefore, we set this constant to zero.} When $p\approx
p_{F}^{(0)}$, the argument of $w\bigl(\tfrac{pl_{0}}{\sqrt{2}\hbar }\bigr)$
is as small as $pl_{0}/\hbar \approx \eta \ll 1$ and we have 
\begin{equation}
V_{0}(\mathbf{p})\approx \pi \tfrac{d^{2}p}{\hbar }\left[ u(\mathbf{p})\sin
^{2}\!\theta _{0}-2\,P_{2}(\cos \theta _{0})\right] .  \label{eq:4}
\end{equation}%
We note that in this limit the effective interaction is independent of the
confinement length $l_{0}$.

Processes of second order in the interaction (see Sec.~\ref%
{sec:superfl-trans}), on the other hand, involve virtual transitions to
excited states $\phi _{n}(z)$ with $n>0$ of the harmonic confining
potential. The matrix elements of the interaction for these transitions, in
momentum representation for the in-plane motion, are 
\begin{multline}
V_{n_{1},n_{2},n_{3},n_{4}}(\mathbf{p})\equiv \int d\bm{\rho} \,
dz\,dz^{\prime }\,V_{d}(\bm{\rho},z-z^{\prime })  \label{eq:5} \\
\times \mathrm{e}^{-i\mathbf{p}\cdot \bm{\rho}/\hbar }\,\phi
_{n_{1}}(z)\,\phi _{n_{2}}(z^{\prime })\,\phi _{n_{3}}(z)\,\phi
_{n_{4}}(z^{\prime }).
\end{multline}%
For $n_{1}=n_{2}=n_{3}=n_{4}=0$ we obtain the effective 2D interaction %
\eqref{eq:3}. Note that although the relative magnitude of a single virtual
excitation is as small as $g\eta ^{2}\ll 1$ (see App.~\ref{sec:matr-elem-vdd}%
), the totality of these processes is essential for the correct description
of the contribution of short distances $\lesssim l_{0}$ (or virtual energies 
$\gtrsim \hbar \omega _{0}$) to the interparticle scattering. We mention
also, that within second order the $p$-wave contribution due to the
short-range part of the interparticle interaction can still be neglected
(see Ref.~\cite{baranov11:_bilay} for discussion).

\section{Approaching the intermediate coupling regime}

\label{sec:appr-interm-coupl} The considered problem is characterized by two
parameters: $g = p_F^{(0)} r_d/\hbar$ and $\eta = p_F^{(0)} l_0/\hbar$. The
first parameter describes the strength of the interparticle interaction with
respect to the mean kinetic energy: $g \ll 1$ characterizes the weakly
interacting regime, while we have $g \gtrsim 1$ in the regime of
intermediate and strong interactions. The second parameter $\eta$ describes
the strength of the confinement: $\eta \ll 1$ corresponds to strong
confinement. When $g$ and $\eta$ are small, $g,\eta \ll 1$, one can use
perturbation theory to calculate various quantities. Many physical effects,
however, occur at intermediate values of $g$ (we will always assume strong
confinement with $\eta \ll 1$), see Fig.~\ref{fig:phase_diagram}, for which
one cannot limit oneself to lowest order diagrams or to a specific sequence
of diagrams (ladder diagrams in a dilute system). To obtain analytic
expressions in this case, one can use analyticity arguments to extrapolate
expressions obtained in the weak coupling regime to intermediate coupling
strength. Of course, the accuracy of such expressions cannot be estimated.
However, (provided the relevant physics is present in the weak coupling
regime) they can be used to make qualitative statements on the behavior of
the system and often obtain reasonable quantitative estimates. Following
this strategy, one has to select a ``reasonable'' set of Feynman diagrams,
which allows one to write down a closed set of integral equations for
relevant physical quantities. This sequence should, of course, catch the
relevant physics and be consistent with general physical principles such as
conservation laws and particle statistics. For the purposes of this paper we
will use the (conserving) HFA (see Refs.~\cite%
{baym61:_conser_laws_correl_funct,baym62:_self_consis_approx_many_body_system}
and discussion below), which describes the motion of particles in an average
potential with exchange effects taken into account and, therefore, can be
used to describe phenomena for which interparticle collisions are not
important (zero sound in our case). Note that taking exchange contributions
into account is crucial: They guarantee the cancellation of all interaction
contributions in the case of a short-range interparticle interaction (when $%
V_0(\mathbf{p})$ is momentum-independent, $V_0(\mathbf{p}) = \mathrm{const.}$%
), as it should be in a single-component Fermi gas.

\section{Quasi-2D dipolar Fermi liquid}

\label{sec:quasi-2d-dipolar} Properties of a normal (non-superfluid) Fermi
system are conveniently described in terms of Green's functions. The
single-particle Green's function (see, for example, Ref.~\cite{LL:IX}) for a
two-dimensional system (in the following we shall be using units in which $%
\hbar = 1$), 
\begin{equation}  \label{eq:6}
G(p) = -i \int dt \, d \mathbf{r} \, \mathrm{e}^{i (\omega t - \mathbf{p}
\cdot \mathbf{r})} \langle \mathrm{T} \{ \widehat{\psi}(t, \mathbf{r}) \, 
\widehat{\psi}^{\dag}(0, \mathbf{0}) \} \rangle,
\end{equation}
where $\mathrm{T}$ stands for the time-ordering operator ($\mathrm{T}$%
-product) and $p = (\omega, \mathbf{p})$, carries information on
single-particle excitations, while collective behavior resulting from
two-particle correlations is described by the two-particle Green's function 
\begin{multline}  \label{eq:7}
G_2(p_1, p_2; p_3, p_4) = \int \prod_{j = 1}^4 dt_j \, d \mathbf{r}_j \, 
\mathrm{e}^{i (\omega_j t_j - \mathbf{p}_j \cdot \mathbf{r}_j)} \\
\times \langle \mathrm{T} \{ \, \widehat{\psi}(t_1, \mathbf{r}_1) \, 
\widehat{\psi}(t_2, \mathbf{r}_2) \, \widehat{\psi}^{\dag }(t_4, \mathbf{r}%
_4) \, \widehat{\psi}^{\dag }(t_3, \mathbf{r}_3) \} \rangle \\
= G(p_1) \, G(p_2) \\
\times \left[ \delta(p_1 - p_3) \, \delta(p_2 - p_4) - \delta(p_1 - p_4) \,
\delta(p_2 -p_3) \right] \\
+ G(p_1) \, G(p_2) \, \Gamma (p_1, p_2; p_3, p_4) \, G(p_3) \, G(p_4),
\end{multline}
or the closely related vertex function $\Gamma(p_1, p_2; p_3, p_4)$:
Single-particle excitations correspond to poles of $G(p)$, collective modes
and instabilities of the many-body system are encoded in poles of the vertex
function $\Gamma(p_1, p_2; p_3, p_4)$. In a homogeneous system, this
quantity depends only on three independent momenta, and it is convenient to
introduce $\Gamma(p_1, p_2; q) \equiv \Gamma(p_1, p_2; p_1 + q, p_2 - q)$
such that $q$ is the transferred momentum, which satisfies the
Bethe-Salpeter equation in the particle-hole channel \cite{LL:IX} 
\begin{multline}  \label{eq:8}
\Gamma(p_1, p_2; q) = \tilde{\Gamma}_{\mathrm{ph}}(p_1, p_2; q) \\
-i \int \tilde{\Gamma}_{\mathrm{ph}}(p_1, p^{\prime}+ q; q) \, G(p^{\prime}+
q) \, G(p^{\prime}) \\
\times \Gamma (p^{\prime}, p_2; q) \frac{d p^{\prime}}{(2 \pi)^3},
\end{multline}
where $\tilde{\Gamma}_{\mathrm{ph}}$ denotes the particle-hole irreducible
vertex, i.\,e., the sum of connected vertex diagrams with two incoming and
two outgoing fermionic lines which cannot be divided into two parts by
cutting two fermion lines of opposite direction.

The single-particle Green's functions can be expressed in terms of the
self-energy function $\Sigma(p)$ through the Dyson equation \cite{LL:IX} 
\begin{equation}  \label{eq:9}
G(p)^{-1} = G^{(0)}(p)^{-1} - \Sigma(p),
\end{equation}
where the non-interacting Green's function is 
\begin{equation}  \label{eq:10}
G^{(0)}(p)^{-1} = \omega - \xi(\mathbf{p}),
\end{equation}
with $\xi(\mathbf{p}) = p^2/2m - \mu $ and the shifted chemical potential $%
\mu = \mu^{\prime}- \omega_0/2$. $\Sigma$, in turn, is connected with the
vertex function by the equation of motion (Schwinger-Dyson equation) \cite%
{LL:IX} 
\begin{multline}  \label{eq:11}
\Sigma(p) = i \int \left[ V_0(\mathbf{p} - \mathbf{p}_1) - V_0(\mathbf{0}) %
\right] G(p_1) \frac{dp_1}{(2 \pi)^3} \\
+ \int \Gamma (p_1 + p_2 -p, p; p_1, p_2) \, G(p_1) \, G(p_2) \\
\times G(p_1 +p_2 -p) \, V_0(\mathbf{p} - \mathbf{p}_2) \frac{dp_1}{(2 \pi)^3%
} \frac{dp_2}{(2 \pi)^3}.
\end{multline}

The solution of the coupled system of equations \eqref{eq:8}, \eqref{eq:9}
and \eqref{eq:11} is specified by the irreducible vertex $\tilde{\Gamma}_{%
\mathrm{ph}}$, which is the sum of an infinite set of Feynman diagrams. As a
result, one cannot write the irreducible vertex $\tilde{\Gamma}_{\mathrm{ph}%
} $ in a closed form in terms of the Green's function $G$ and the vertex $%
\Gamma$, and some approximation procedure of choosing a subset of
contributions is needed. This procedure should be consistent with
conservation laws and statistics of the system. A prescription for
generating such a conserving approximation is to replace the right-hand side
(RHS) of Eq.~\eqref{eq:11} for $\Sigma$ by a functional of $G$ and $V_0$~%
\cite%
{baym61:_conser_laws_correl_funct,baym62:_self_consis_approx_many_body_system}%
. The one-particle propagator is then to be obtained self-consistently from
the approximate equation for $\Sigma$ and \eqref{eq:9}, and the approximate
irreducible vertex $\tilde{\Gamma}_{\mathrm{ph}}$ which determines $\Gamma$
via \eqref{eq:8}, can be found by suitable functional differentiation. In
other words, by fixing an appropriate expression for the self-energy $\Sigma 
$, one uniquely determines the expression for $\tilde{\Gamma}_{\mathrm{ph}}$
in order to make the approximation conserving.

As discussed in Refs.~\cite%
{baym61:_conser_laws_correl_funct,baym62:_self_consis_approx_many_body_system}%
, the simplest example of a conserving approximation taking exchange effects
into account is the HFA, which we will use in this paper. In this
approximation, the self-energy is given diagrammatically as \\[-4mm]
\begin{equation}  \label{eq:12}
\parbox{8mm}{
        \begin{fmfgraph*}(8,12)
          \fmfleft{l} \fmfright{r}
          \fmf{plain,width=thick}{l,vl}
          \fmf{plain,width=thick}{vr,r}
          \fmfpoly{smooth,tension=.2,label=\tiny $\Sigma$}{vl,vr}
        \end{fmfgraph*}}=%
\parbox{12mm}{
        \begin{fmfgraph}(12,12)
          \fmfleft{l} \fmfright{r}
          \fmf{plain,width=thick}{r,v1}
          \fmf{plain,width=thick}{v2,l}
          \fmf{phantom,tension=.25}{v1,v2}
          \fmffreeze
          \fmf{wiggly}{v1,v2}
          \fmffreeze
          \fmf{fermion,left,width=thick}{v2,v1}
        \end{fmfgraph}}+ 
\parbox{6mm}{
        \begin{fmfgraph}(6,20)
          \fmfleft{l} \fmfright{r}
          \fmftop{t}
          \fmf{plain,width=thick}{r,v1}
          \fmf{plain,width=thick}{v1,l}
          \fmffreeze
          \fmf{wiggly}{v1,v2} \fmf{phantom}{v2,t} \fmffreeze
          \fmf{fermion,left,width=thick}{t,v2} \fmf{plain,left,width=thick}{v2,t}
        \end{fmfgraph}}
\end{equation}
\\[-10mm]
or analytically 
\begin{equation}  \label{eq:13}
\Sigma(\mathbf{p}) = i \int \left[ V_0(\mathbf{p} - \mathbf{p}^{\prime}) -
V_0(\mathbf{0}) \right] \, G(p^{\prime}) \frac{dp^{\prime}}{(2 \pi)^3},
\end{equation}
and is frequency independent. The corresponding particle-hole irreducible
vertex is 
\begin{equation}  \label{eq:14}
\parbox{12mm}{
        \begin{fmfgraph*}(12,10)
          \fmfleft{lb,lt}
          \fmfright{rb,rt}
          \fmf{fermion,width=thick}{lt,vlt}
          \fmf{fermion,width=thick}{vlb,lb}  
          \fmf{fermion,width=thick}{rb,vrb}
          \fmf{fermion,width=thick}{vrt,rt}
          \fmfpoly{smooth,tension=.8,label=\tiny $\tilde{\Gamma}_{\mathrm{ph}}$}{vrt,vlt,vlb,vrb}
        \end{fmfgraph*}}=%
\parbox{20mm}{\fmfframe(4,4)(4,4){\begin{fmfgraph*}(12,10)
            \fmfv{label=\small $p_1$}{lt}
            \fmfv{label=\small $p_1 + q$,label.angle=-90}{lb}
            \fmfv{label=\small $p_2 - q$,label.angle=90}{rt}
            \fmfv{label=\small $p_2$}{rb}
            \fmfleft{lb,lt}
            \fmfright{rb,rt}            
            \fmf{phantom}{lt,vl}
            \fmf{phantom}{vl,lb}  
            \fmf{phantom}{rb,vr}
            \fmf{phantom}{vr,rt}
            \fmf{wiggly}{vl,vr}            
            \fmffreeze  
            \fmfi{fermion,width=thick}{vloc(__lt){(1,-1)} .. {(0,-1)}vloc(__vl)}
            \fmfi{fermion,width=thick}{vloc(__vl){(0,-1)} .. {(-1,-1)}vloc(__lb)}
            \fmfi{fermion,width=thick}{vloc(__rb){(-1,1)} .. {(0,1)}vloc(__vr)}
            \fmfi{fermion,width=thick}{vloc(__vr){(0,1)} .. {(1,1)}vloc(__rt)}
          \end{fmfgraph*}}} - 
\parbox{20mm}{\fmfframe(4,4)(4,4){\begin{fmfgraph*}(12,10)
            \fmfv{label=\small $p_1$}{lt}
            \fmfv{label=\small $p_1 + q$,label.angle=-90}{lb}
            \fmfv{label=\small $p_2 - q$,label.angle=90}{rt}
            \fmfv{label=\small $p_2$}{rb}
            \fmfleft{lb,lt}
            \fmfright{rb,rt}
            \fmf{phantom}{lt,vt}
            \fmf{phantom}{vb,lb}  
            \fmf{phantom}{rb,vb}
            \fmf{phantom}{vt,rt}
            \fmf{wiggly}{vt,vb}
            \fmffreeze  
            \fmfi{fermion,width=thick}{vloc(__lt){(1,-1)} .. {(1,0)}vloc(__vt)}
            \fmfi{fermion,width=thick}{vloc(__vt){(1,0)} .. {(1,1)}vloc(__rt)}
            \fmfi{fermion,width=thick}{vloc(__rb){(-1,1)} .. {(-1,0)}vloc(__vb)}
            \fmfi{fermion,width=thick}{vloc(__vb){(-1,0)} .. {(-1,-1)}vloc(__lb)}
          \end{fmfgraph*}}},
\end{equation}
or 
\begin{equation}  \label{eq:15}
\begin{split}
\tilde{\Gamma}_{\mathrm{ph}}(p_1, p_2; q) & = V_0(\mathbf{q}) - V_0(\mathbf{p%
}_1 - \mathbf{p}_2 + \mathbf{q}) \\
& \equiv \tilde{\Gamma}_{\mathrm{ph}}(\mathbf{p}_1 - \mathbf{p}_2 + \mathbf{%
q }, \mathbf{q}).
\end{split}%
\end{equation}
and is also frequency independent. Eq.~\eqref{eq:8} then shows that $%
\Gamma(p_1, p_2; q)$ does not depend on the frequencies $\omega_1$ and $%
\omega_2$, i.\,e., $\Gamma(p_1, p_2; q) = \Gamma(\mathbf{p}_1, \mathbf{p}_2;
q)$, and we have 
\begin{multline}  \label{eq:16}
\Gamma(\mathbf{p}_1, \mathbf{p}_2; q) = \tilde{\Gamma}_{\mathrm{ph}}(\mathbf{%
p}_1 - \mathbf{p}_2 + \mathbf{q}, \mathbf{q}) \\
-i \int \tilde{\Gamma}_{\mathrm{ph}}(\mathbf{p}_1 - \mathbf{p}^{\prime}, 
\mathbf{q}) \, G(p^{\prime}+ q) \, G(p^{\prime}) \\
\times \Gamma(\mathbf{p}^{\prime}, \mathbf{p}_2; q) \frac{dp^{\prime}}{(2
\pi)^3}.
\end{multline}

\section{Single-particle excitations in HFA}

\label{sec:single-part-excit} From Eqs.~\eqref{eq:9}, \eqref{eq:10}, and %
\eqref{eq:13} we obtain the single-particle Green's function in momentum
space representation 
\begin{equation}  \label{eq:17}
G(\omega, \mathbf{p}) = \frac{1}{\omega - \varepsilon(\mathbf{p}) + i 0 %
\sgn[\varepsilon(\mathbf{p})]},
\end{equation}
where the quasi-particle dispersion relation is 
\begin{equation}  \label{eq:18}
\varepsilon(\mathbf{p}) = \xi(\mathbf{p}) + \Sigma(\mathbf{p}),
\end{equation}
and the self-energy $\Sigma$, after performing the integration over $%
\omega^{\prime}$ in Eq.~\eqref{eq:13}, can be written as [note that the
direct term does not contribute, as $V_0(\mathbf{0}) = 0$ for the effective
dipole-dipole interaction \eqref{eq:3}] 
\begin{equation}  \label{eq:19}
\Sigma(\mathbf{p}) = -\int V_0(\mathbf{p} - \mathbf{p}^{\prime}) \, n(%
\mathbf{p}^{\prime}) \frac{d \mathbf{p}^{\prime}}{(2 \pi)^2}.
\end{equation}
In this expression $n(\mathbf{p}) = \theta[-\varepsilon (\mathbf{p})] $ is
the Fermi-Dirac distribution at zero temperature. Equation \eqref{eq:19} has
to be solved self-consistently together with the particle number equation 
\begin{equation}  \label{eq:20}
n_{2D} = \int \frac{d \mathbf{\mathbf{p}}}{(2 \pi)^2} \, n(\mathbf{p}).
\end{equation}

The problem of finding the solution to Eqs.~\eqref{eq:19} and \eqref{eq:20}
is simplified considerably by noting that the evaluation of the RHS of both
equations does not require full knowledge of the quasi-particle dispersion
relation $\varepsilon(\mathbf{p})$ but only of the Fermi momentum $p_F$,
which is determined by the requirement that the quasi-particle energy %
\eqref{eq:18} is equal to zero for $\mathbf{p} = \mathbf{p}_F \equiv p_F 
\hat{\mathbf{p}}$, 
\begin{equation}  \label{eq:21}
\varepsilon(\mathbf{p}_F) = p_F^2/2m - \mu + \Sigma(\mathbf{p}_F) = 0.
\end{equation}
The solution $p_F$ to this equation actually depends on the direction $\hat{%
\mathbf{p}} = (\cos \phi, \sin \phi)$. However, to shorten the notation, in
the following we will often write $p_F$ and $\mathbf{p}_F$ instead of $%
p_F(\phi)$ and $\mathbf{p}_F(\phi)$ respectively.

We proceed by specializing \eqref{eq:19} to the Fermi surface $\mathbf{p} = 
\mathbf{p}_F$ and inserting the resulting expression for $\Sigma(\mathbf{p}%
_F)$ in \eqref{eq:21}, 
\begin{equation}  \label{eq:22}
\frac{p_F^2}{2m} = \mu + \frac{1}{m} \int \frac{d \phi^{\prime}}{2 \pi}
\int_0^{p_F^{\prime}} dp^{\prime}\, p^{\prime}\, \nu \, V_0(\mathbf{p}_F - 
\mathbf{p}^{\prime}),
\end{equation}
where $p_F^{\prime}\equiv p_F(\phi^{\prime})$ and $\nu = m/2 \pi$ is the
density of states. An expression for the chemical potential $\mu$ can be
obtained by taking the integral of this formula over the angle $\phi$ and
making use of the fact that the particle number equation \eqref{eq:20} is
equivalent to the condition 
\begin{equation}  \label{eq:23}
\int \frac{d \phi}{2 \pi} \, p_F^2 = p_F^{(0)2},
\end{equation}
if we express the density as $n_{2D} = p_F^{(0)2}/4 \pi$. We find 
\begin{equation}  \label{eq:24}
\mu = \varepsilon_F^{(0)} - \frac{1}{m} \int \frac{d \phi}{2 \pi} \frac{d
\phi^{\prime}}{2 \pi} \int_0^{p_F^{\prime}} dp^{\prime}\, p^{\prime}\, \nu
\, V_0(\mathbf{p}_F - \mathbf{p}^{\prime}).
\end{equation}
Eqs.~\eqref{eq:22} and \eqref{eq:24} form a closed system for the deformed
Fermi surface $p_F$ and we obtain the joint solution to these equations
numerically by means of an iterative scheme which is described in detail in
App.~\ref{sec:calculation-hf-self}. The resulting quasi-particle dispersion
relation \eqref{eq:18} is shown in Fig.~\ref{fig:quasi-particle-dispersion},
along with the linear approximation at the Fermi surface 
\begin{equation}  \label{eq:25}
\varepsilon(\mathbf{p}) \approx v_F(\phi ) [p - p_F(\phi)].
\end{equation}
Here $v_F(\phi)$ is the radial component of the Fermi velocity: 
\begin{equation}  \label{eq:26}
\begin{split}
\mathbf{v}_F(\phi) & \equiv \nabla \varepsilon[\mathbf{p}_F(\phi)] \\
& = \hat{\mathbf{p}} \, v_F(\phi) + \hat{\mathbf{e}}_{\phi} \, v_F^{\perp
}(\phi).
\end{split}%
\end{equation}

For future reference we summarize the corresponding results obtained in
perturbation theory \cite{chan10:_anisot_fermi}. To first order in the
dipole-dipole interaction, on the RHS of Eq.~\eqref{eq:19} we insert the
distribution function of a non-interacting Fermi gas, $n(\mathbf{p}) =
\theta(p_F^{(0)} - p)$. The self-energy function can then be expressed in
terms of complete elliptic integrals. We omit the cumbersome analytical
expression and content ourselves with stating the result for $\mathbf{p} = 
\mathbf{p}_F^{(0)} = p_F^{(0)} \hat{\mathbf{p}}$, 
\begin{equation}  \label{eq:27}
\Sigma^{(1)}(\mathbf{p}_F^{(0)}) = - \tfrac{16}{9\pi} g \varepsilon_F^{(0)} %
\left[ \tfrac{3}{5} \sin^2 \! \theta_0 \cos(2 \phi) - 2 \, P_2(\cos
\theta_0) \right].
\end{equation}
We write the chemical potential as $\mu = \varepsilon_F^{(0)} + \delta \mu$.
The first order correction $\delta \mu$ is given by 
\eqref{eq:24} [with $p_F'$
on the RHS replaced by $p_F^{(0)}$], 
\begin{equation}  \label{eq:28}
\delta \mu = \tfrac{32}{9 \pi} g \varepsilon_F^{(0)} \, P_2(\cos \theta_0).
\end{equation}
Combining these results with \eqref{eq:21} we find the equilibrium
deformation of the Fermi surface $\delta p_F(\phi) = p_F(\phi) - p_F^{(0)}$, 
\begin{equation}  \label{eq:29}
\begin{split}
\delta p_F(\phi) & = \frac{m}{p_F^{(0)}} \left[ \delta \mu -\Sigma^{(1)}(%
\mathbf{p}_F^{(0)}) \right] \\
& = \tfrac{8}{15\pi}gp_F^{(0)} \sin^2 \! \theta_0 \cos(2 \phi).
\end{split}%
\end{equation}
From its definition in \eqref{eq:26}, the radial component of the Fermi
velocity is then 
\begin{equation}  \label{eq:30}
v_F(\phi) = v_F^{(0)} \left[ 1 + \tfrac{4}{3 \pi} g \, P_2(\cos \theta_0) - 
\tfrac{2}{5 \pi} g \sin^2 \! \theta_0 \cos(2 \phi) \right],
\end{equation}
Finally, the (radial) effective mass is defined as $m^{*}(\phi) \equiv
p_F(\phi)/v_F(\phi)$. Then, for the deviation $\delta m(\phi) = m^{*}(\phi)
- m$ of the effective mass from the bare mass we have 
\begin{equation}  \label{eq:31}
\begin{split}
\frac{\delta m(\phi)}{m} & = -\frac{m}{p_F^{(0)}} \frac{\partial
\Sigma^{(1)}(\mathbf{p}_F^{(0)})}{\partial p} \\
& = -\tfrac{4}{3 \pi} g \, P_2(\cos \theta_0)+ \tfrac{14}{15 \pi} g \sin^2
\! \theta_0 \cos(2 \phi).
\end{split}%
\end{equation}
\begin{figure}[tbp]
\psfrag{x1}[t][]{\scriptsize $0$} \psfrag{x2}[t][]{\scriptsize $0.5$}  %
\psfrag{x3}[t][]{\scriptsize $1$} \psfrag{x4}[t][]{\scriptsize $1.5$}  %
\psfrag{x5}[t][]{\scriptsize $2$} \psfrag{y1}[r][]{\scriptsize $-2$}  %
\psfrag{y2}[r][]{\scriptsize $0$} \psfrag{y3}[r][]{\scriptsize $2$}  %
\psfrag{y4}[r][]{\scriptsize $4$} \psfrag{xlabel}[t][b]{\small $p/p_F^{(0)}$}
\psfrag{ylabel}[b][t]{\small $\varepsilon(\mathbf{p})/\varepsilon_F^{(0)}$}  %
\psfrag{g1theta0pi4}[][t]{} \psfrag{g2theta0pi2}[][t]{} \centering
\includegraphics[width=.8\linewidth]{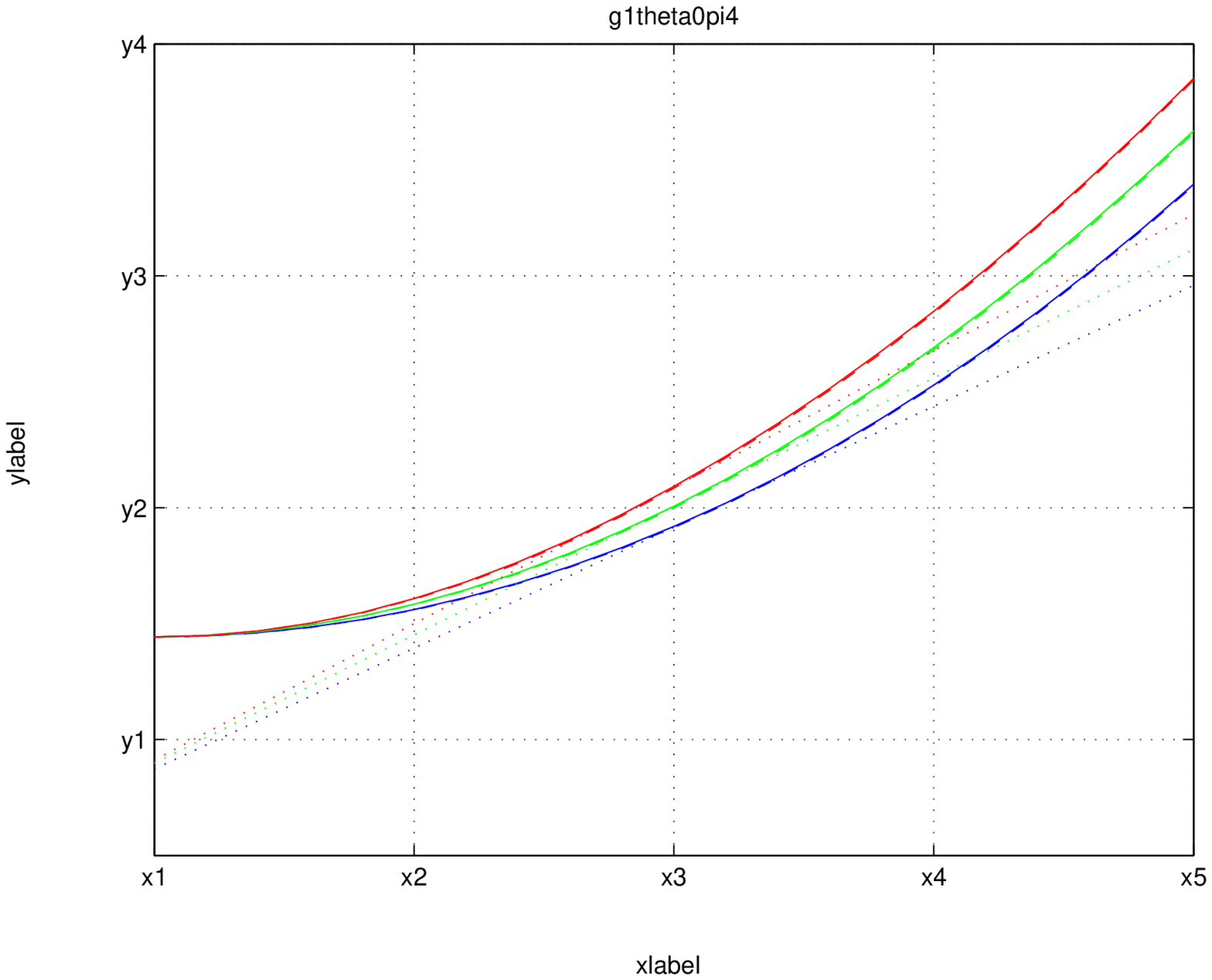} \\%
[.4cm]
\includegraphics[width=.8\linewidth]{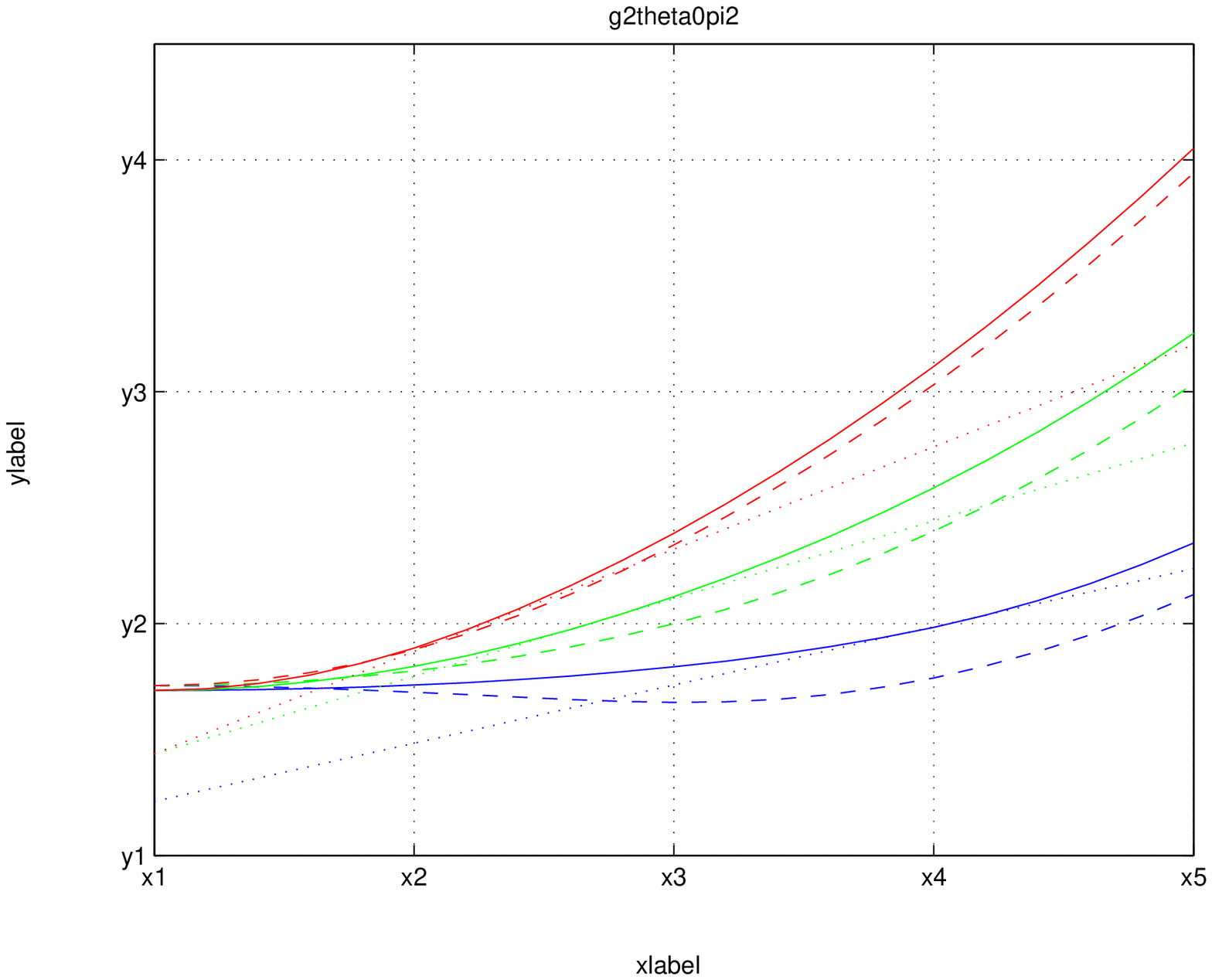}
\caption{(Color online) Quasi-particle dispersion relation for $g = 1$,
  $\protect\theta_0 = \protect\pi/4$ (top) and $g = 2$, $\protect\theta_0 =
  \protect\pi/2$ (bottom) in HFA (solid lines) and first order perturbation
  theory (dashed lines) for some values of $\protect\phi_{\mathbf{p}}$
  [$\mathbf{p} = p \, (\cos \protect\phi_{\mathbf{p}}, \sin
  \protect\phi_{\mathbf{p}})$] (lowermost blue curves: $%
  \protect\phi_{\mathbf{p}} = 0$, middle green curves:
  $\protect\phi_{\mathbf{p}} = \protect%
  \pi/4$, topmost red curves: $\protect\phi_{\mathbf{p}} = \protect\pi/2$). The
  linear approximation \eqref{eq:25} at the Fermi surface is shown as dotted
  lines. Note that the results obtained in HFA and perturbation theory agree
  very well for $g = 1$, $\protect\theta_0 = \protect\pi/4$, whereas deviations
  are clearly visible for $g = 2$, $\protect\theta_0 = \protect\pi/2 $.}
\label{fig:quasi-particle-dispersion}
\end{figure}

\section{Collective excitations in HFA: zero sound}

\label{sec:coll-excit-hfa} Collective modes $\omega =\omega (\mathbf{q})$ of
the system correspond to poles of $\Gamma (\mathbf{p}_{1},\mathbf{p}_{2};q)$
with respect to the variable $\omega $, which is the frequency component of $%
q$. In the vicinity of a pole we have $\Gamma \gg \tilde{\Gamma}_{\mathrm{ph}%
}$ and, therefore, we may neglect the first term on the RHS of Eq.~%
\eqref{eq:16}. In the resulting homogeneous equation the second argument $%
\mathbf{p}_{2}$ of $\Gamma $ acts as a parameter. Hence, near its pole, the
function $\Gamma $ can be represented as a product $\chi (\mathbf{p}_{1};%
\mathbf{q})\,\chi ^{\prime }(\mathbf{p}_{2};\mathbf{q})$ of two functions.
After cancelling $\chi ^{\prime }(\mathbf{p}_{2};\mathbf{q})$ on both sides
of Eq.~\eqref{eq:16} and integrating over $\omega ^{\prime }$, we obtain 
\begin{multline}
\chi (\mathbf{p};\mathbf{q})=\int \frac{d\mathbf{p^{\prime }}}{(2\pi )^{2}}\,%
\tilde{\Gamma}_{\mathrm{ph}}(\mathbf{p}-\mathbf{p}^{\prime },\mathbf{q})
\label{eq:32} \\
\times \frac{n(\mathbf{p}^{\prime })-n(\mathbf{p}^{\prime }+\mathbf{q})}{%
\omega +\varepsilon (\mathbf{p}^{\prime })-\varepsilon (\mathbf{p}^{\prime }+%
\mathbf{q})+i0\sgn(\omega )}\,\chi (\mathbf{p}^{\prime };\mathbf{q}).
\end{multline}

The quantity $\varepsilon(\mathbf{p}^{\prime}+ \mathbf{q}) - \varepsilon(%
\mathbf{p}^{\prime})$ on the RHS of the last equation is just the energy
cost of creating a particle-hole pair by exciting a particle from a state $%
\mathbf{p}^{\prime}$ within the Fermi surface to a state $\mathbf{p}%
^{\prime}+ \mathbf{q}$ outside the Fermi surface. Therefore, a stable
collective mode is possible only when the energy (frequency) $\omega$ of the
mode lies outside the particle-hole continuum (PHC) (so we can omit the
imaginary term in the denominator). Otherwise the integrand has a pole at $%
\omega = \varepsilon(\mathbf{p}^{\prime}+ \mathbf{q}) - \varepsilon(\mathbf{p%
}^{\prime})$, which ultimately leads to strong Landau damping of the
collective mode \cite{LL:IX}.

In the long wavelength limit $\lvert \mathbf{q} \rvert \rightarrow 0$, the
main contribution to the integral \eqref{eq:32} comes from states in the
vicinity of the Fermi surface, and we can rewrite this equation as 
\begin{equation}
\chi (\mathbf{p})=\int \frac{d\phi ^{\prime }}{2\pi }\,\nu \,\tilde{\Gamma}_{%
\mathrm{ph}}(\mathbf{p}-\mathbf{p}_{F}^{\prime},\mathbf{0})\frac{%
p_{F}^{\prime }/m}{\hat{\mathbf{p}}^{\prime }\cdot \mathbf{v}_{F}^{\prime }}%
\frac{\mathbf{q}\cdot \mathbf{v}_{F}^{\prime }}{\omega -\mathbf{q}\cdot 
\mathbf{v}_{F}^{\prime }}\,\chi (\mathbf{p}_{F}^{\prime }),  \label{eq:33}
\end{equation}%
where $\chi (\mathbf{p})\equiv \chi (\mathbf{p};\mathbf{0})$, $p_{F}^{\prime
}\equiv p_{F}(\phi ^{\prime })$, and $\mathbf{v}_{F}^{\prime }\equiv \mathbf{%
v}_{F}(\phi ^{\prime })$. The function $\chi (\mathbf{p})$ is thus
completely determined by its values on the Fermi surface and by setting $%
\mathbf{p}=\mathbf{p}_{F}$ we can obtain an equation for the restriction of $%
\chi (\mathbf{p})$ to the Fermi surface, 
\begin{equation}
\chi (\mathbf{p}_{F})=\int \frac{d\phi ^{\prime }}{2\pi }\,F(\phi ,\phi
^{\prime })\frac{p_{F}^{\prime }/m}{\hat{\mathbf{p}}^{\prime }\cdot \mathbf{v%
}_{F}^{\prime }}\frac{\mathbf{q}\cdot \mathbf{v}_{F}^{\prime }}{\omega -%
\mathbf{q}\cdot \mathbf{v}_{F}^{\prime }}\,\chi (\mathbf{p}_{F}^{\prime }),
\label{eq:34}
\end{equation}%
where we define the dimensionless quasi-particle interaction function ($f$%
-function) as 
\begin{equation}
\begin{split}
F(\phi ,\phi ^{\prime })& \equiv \nu \,\tilde{\Gamma}_{\mathrm{ph}}(\mathbf{p%
}_{F}-\mathbf{p}_{F}^{\prime },\mathbf{0}) \\
& =\nu \left[ V_{0}(\mathbf{0})- \,V_{0}(\mathbf{p}_{F}-\mathbf{p}%
_{F}^{\prime }) \right] \\
& =g\frac{\left| \mathbf{p}_{F}-\mathbf{p}_{F}^{\prime }\right| }{p_{F}^{(0)}%
} \\
& \mathrel{\phantom{=}}\times \left[ P_{2}(\cos \theta _{0})-\frac{1}{2}u(%
\mathbf{p}_{F}-\mathbf{p}_{F}^{\prime })\sin ^{2}\theta _{0}\right] .
\end{split}
\label{eq:35}
\end{equation}%
Note that only the exchange interaction contributes to the $f$-function. For 
$\theta _{0}=0$, i.\thinspace e., when the dipoles are perpendicular to the $%
xy$-plane and the system is symmetric with respect to rotations around the $%
z $-axis, the Fermi momentum is isotropic and equal to $p_{F}^{(0)}$ [see
Eq.~\eqref{eq:23}], and \eqref{eq:35} simplifies to 
\begin{equation}
F(\phi ,\phi ^{\prime })=2g\sin (\lvert \phi -\phi ^{\prime }\rvert /2),
\quad \theta_0 = 0.  \label{eq:36}
\end{equation}

Equation \eqref{eq:34} shows that $\omega$ depends linearly on $q = \left|  
\mathbf{q} \right|$, 
\begin{equation}  \label{eq:37}
\omega = v_F^{(0)} s \, q.
\end{equation}
Due to the anisotropy of the dipole-dipole interaction, $s$ will in general
be a function of the propagation direction $\phi_{\mathbf{q}}$ [$\mathbf{q}
= q \, (\cos \phi_{\mathbf{q}}, \sin \phi_{\mathbf{q}})$]. The symmetry of
the problem, however, requires the dependence of $s$ on $\phi_{\mathbf{q}}$
to be $\pi$-periodic and even, hence, it is sufficient to restrict ourselves
to the range $0 \leq \phi_{\mathbf{q}} \leq \pi/2$.

As it has already been pointed out, the excitation energy $\omega$ of the
collective mode has to be separated from the PHC. Equation \eqref{eq:34}
shows that in the long wavelength limit $q \rightarrow 0$ this requirement
reduces to 
\begin{equation}  \label{eq:38}
v_F^{(0)} s > v_{\mathrm{ph}},
\end{equation}
where the quantity $v_{\mathrm{ph}}$ is the slope of the upper boundary of
the PHC in the direction $\hat{\mathbf{q}}$ at $q = 0$, and can be computed
by taking the maximum over all values of the angle $\phi^{\prime}$, 
\begin{equation}  \label{eq:39}
v_{\mathrm{ph}} \equiv \max_{\phi^{\prime}} \{ \hat{\mathbf{q}} \cdot 
\mathbf{v}_F^{\prime}\}.
\end{equation}

For the purpose of solving Eq.~\eqref{eq:34} numerically it is convenient to
replace $\chi$ by another function 
\begin{equation}  \label{eq:40}
\nu(\phi ) \equiv \frac{1}{\hat{\mathbf{p}} \cdot \mathbf{v}_F} \frac{\hat{%
\mathbf{q}} \cdot \mathbf{v}_F}{v_F^{(0)} s - \hat{\mathbf{q}} \cdot \mathbf{%
v}_F} \, \chi (\mathbf{p}_F).
\end{equation}
Equation \eqref{eq:34} then becomes 
\begin{equation}  \label{eq:41}
(v_F^{(0)} s - \hat{\mathbf{q}} \cdot \mathbf{v}_F) \, \nu(\phi ) = \frac{1}{%
m} \frac{\hat{\mathbf{q}} \cdot \mathbf{v}_F}{\hat{\mathbf{p}} \cdot \mathbf{%
v}_F} \int \frac{d \phi^{\prime}}{2 \pi} \, F(\phi,\phi^{\prime}) \,
p_F^{\prime}\, \nu(\phi^{\prime}).
\end{equation}
One should note that Eq.~\eqref{eq:41} -- in contrast to Eq.~\eqref{eq:34}
-- allows for non-trivial solutions even for $g = 0$ [or $%
F(\phi,\phi^{\prime}) = 0$]. These solutions with $s \in [-1,1]$ take the
form of $\delta$-function in $\phi$ and correspond to single particle-hole
pairs from the PHC.

We solve Eq.~\eqref{eq:41} numerically by discretizing it on an evenly
spaced grid of 2000 points in the variable $\phi$ and rewriting the integral
on the RHS as a sum according to the trapezoidal integration rule.
Convergence tests show that increasing the number of grid points further by
a factor of 2 results in a change in $s$ that is less than $10^{-5}$ (for
all values of the parameters in the range $0.1 \leq g \leq 2$, and $0\leq
\theta_0,\phi_{\mathbf{q}} \leq \pi/2$; going to $g \ll 1$ is not possible
in this approach, see discussion below).

Our results for $g=1$ are shown in Figs.~\ref{fig:zs_region}, \ref%
{fig:s_of_theta0} and \ref{fig:s_of_phi_q}. 
\begin{figure}[tbp]
\psfrag{a}[lt][b]{\small $\theta_0$} 
\psfrag{b}[b][t][1][180]{\small
    $\phi_{\mathbf{q}}$} \psfrag{c}[t][]{\small $0$} 
\psfrag{d}[t][]{\small
    $\pi/8$} \psfrag{e}[t][]{\small $\pi/4$} \psfrag{f}[t][]{\small $3
\pi/8$}  \psfrag{g}[t][]{\small $\pi/2$} \psfrag{h}[r][]{\small $0$}  %
\psfrag{i}[r][]{\small $\pi/8$} \psfrag{j}[r][]{\small $\pi/4$}  %
\psfrag{k}[r][]{\small $3 \pi/8$} \psfrag{l}[r][]{\small $\pi/2$}  %
\psfrag{m}[l][]{\small $0$} \psfrag{n}[l][]{\small $0.005$}  %
\psfrag{o}[l][]{\small $0.01$} \psfrag{p}[l][]{\small $0.015$}  %
\psfrag{q}[l][]{\small $0.02$} \psfrag{r}[l][]{\small $0.025$}  %
\psfrag{s}[][]{I}  \psfrag{t}[][]{II} %
\includegraphics[width=.9\linewidth]{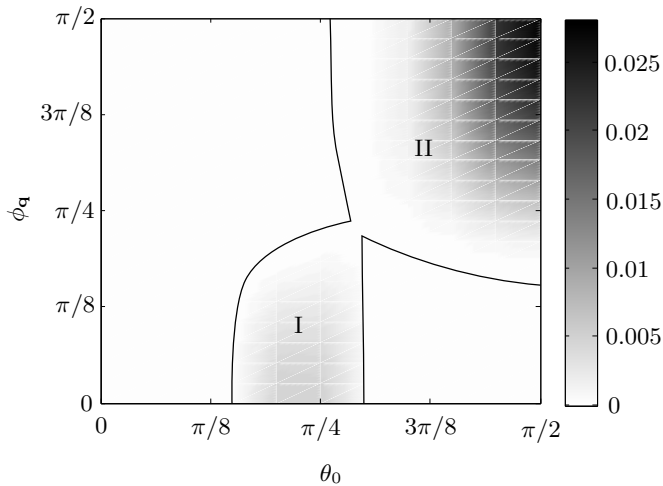} 
\caption{The condition \eqref{eq:38} for the propagation of zero sound is met in
  regions I and II, where the gray levels encode the difference $%
  v_{F}^{(0)}s-v_{\mathrm{ph}}$ for the largest eigenvalue $s$ of Eq.~\eqref{eq:41}.}
\label{fig:zs_region}
\end{figure}
We see that the existence of zero sound and the value of the sound velocity
strongly depend on the propagation direction, on the tilting angle, and on
the strength of the interaction. There is no dissipationless zero sound mode
if the tilting angle is smaller than some critical value, see Fig.~\ref%
{fig:zs_region}, as then all numerically calculated eigenvalues $s$ are
below or equal to the limiting velocity $v_{\mathrm{ph}}$ of the PHC, thus
violating the propagation criterion \eqref{eq:38}. This is to some extend
similar to the propagation of zero sound in the homogeneous three
dimensional case, which was studied in Refs.~\cite%
{ronen10:_zero_fermi,chan10:_anisot_fermi}: The authors
of these references found that there is no undamped propagation of zero
sound for a wide range of angles perpendicular to the direction of the
polarization of dipoles. The reason for this at first glance
counterintuitive statement is that the contribution of the direct
interaction (which is repulsive in coordinate space) to $\tilde{\Gamma}_{%
\mathrm{ph}}$ [or to $F(\phi ,\phi ^{\prime })$] vanishes, see Eqs.~%
\eqref{eq:15} and \eqref{eq:4}, such that the long wavelength collective
behavior of the dipolar gas is governed by the exchange interaction (the $f$%
-function contains only the exchange contribution). (The collective modes
without the exchange contribution were considered in Ref.~\cite%
{kestner10:_compr}, and, as a consequence, their result is determined by the
momentum-independent term in the Fourier transform $V_{0}(\mathbf{q})$ of
the dipole-dipole interaction which is omitted in our paper because it is
cancelled by the corresponding exchange contribution.)

\begin{psfrags}
      \psfrag{x1}[t][]{\scriptsize $0$}
      \psfrag{x2}[t][]{\scriptsize $\pi/8$}
      \psfrag{x3}[t][]{\scriptsize $\pi/4$}
      \psfrag{x4}[t][]{\scriptsize $3 \pi/8$}
      \psfrag{x5}[t][]{\scriptsize $\pi/2$}
      \begin{figure}
        \psfrag{y1}[r][]{\scriptsize $0.9$}
        \psfrag{y2}[r][]{\scriptsize $1.1$}
        \psfrag{y3}[r][]{\scriptsize $1.3$}
        \psfrag{y4}[r][]{\scriptsize $1.5$}
        \psfrag{xi1}[t][]{\scriptsize $0.68$}    
        \psfrag{xi2}[t][]{\scriptsize $0.72$}
        \psfrag{yi1}[r][]{\scriptsize $1.08$}
        \psfrag{yi2}[r][]{\scriptsize $1.12$}    
        \psfrag{xlabel}[t][b]{\small $\theta_0$}
        \psfrag{ylabel}[b][t]{\small $s$}
        \centering
        \includegraphics[width=.8\linewidth]{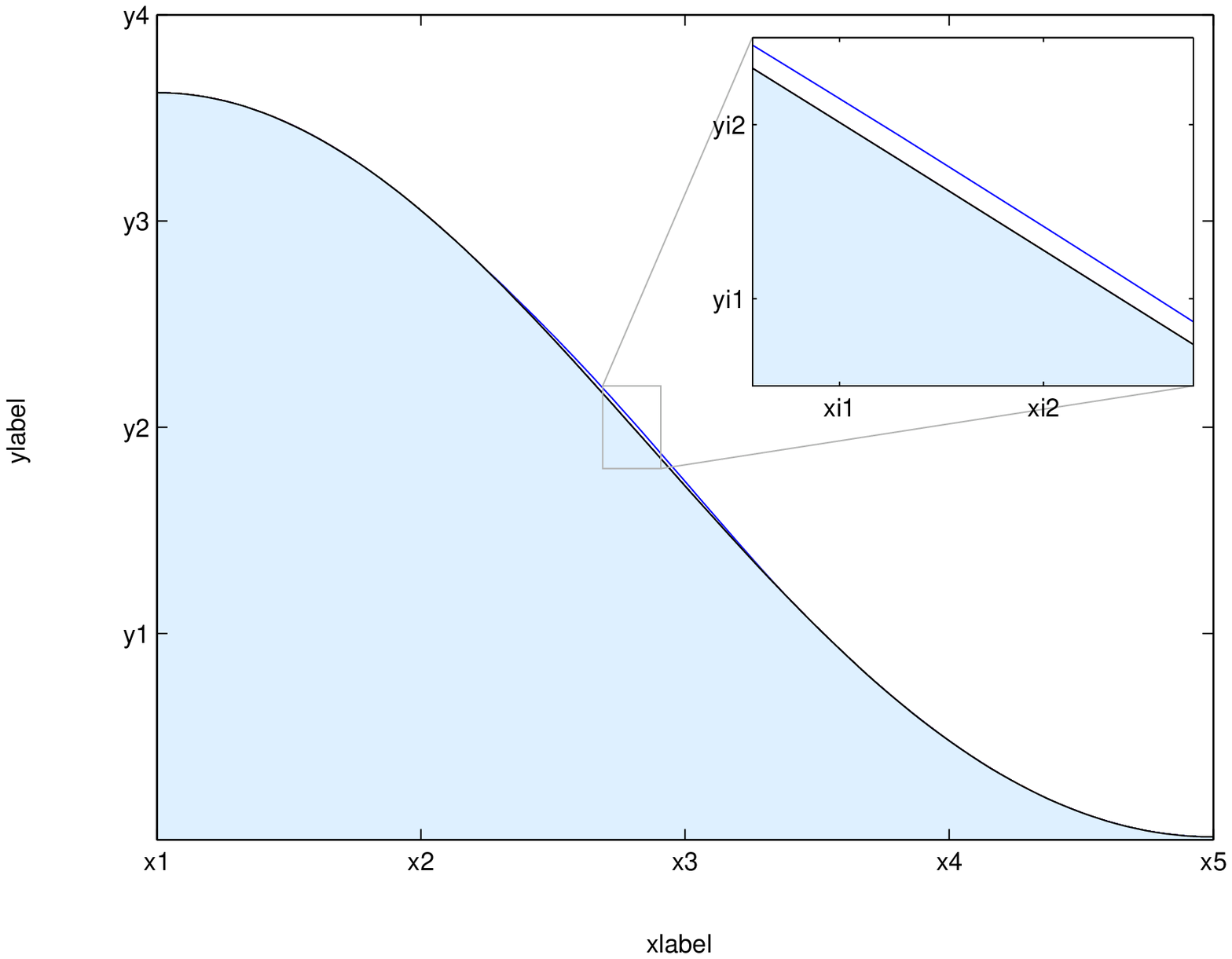}    
        \\[.3cm]
        \includegraphics[width=.8\linewidth]{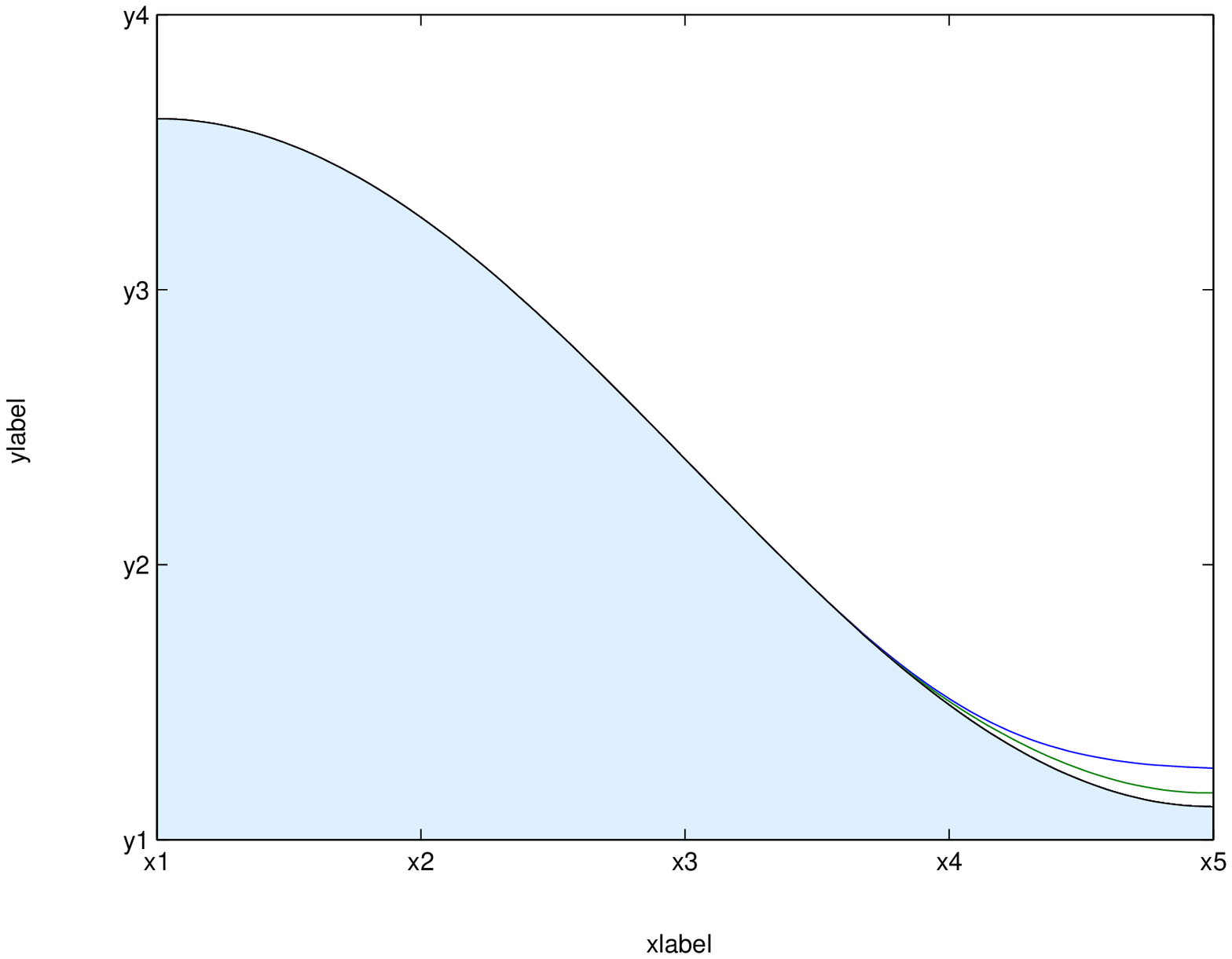}
        \caption{(Color online) The speed of zero sound as a function of the
          tilting angle $\theta_0$ for $g = 1$ and $\phi_{\mathbf{q}} = 0,
          \pi/2$ (top, bottom). Note that for $\phi_{\mathbf{q}} = \pi/2$ and
          $\theta_0$ close to $\pi/2$ we have two zero sound modes. The region
          that is forbidden by the condition \eqref{eq:38} is shaded.}
        \label{fig:s_of_theta0}
      \end{figure}
      \begin{figure}
        \psfrag{116}{}
        \psfrag{121}{}
        \psfrag{y1}[r][]{\scriptsize $0.5$}
        \psfrag{y2}[r][]{\scriptsize $0.6$}
        \psfrag{y3}[r][]{\scriptsize $0.7$}
        \psfrag{y4}[r][]{\scriptsize $0.8$}
        \psfrag{y5}[r][]{\scriptsize $0.9$}
        \psfrag{y6}[r][]{\scriptsize $1$}
        \psfrag{y7}[r][]{\scriptsize $1.1$}
        \psfrag{y8}[r][]{\scriptsize $1.2$}
        \psfrag{xlabel}[t][b]{\small $\phi_{\mathbf{q}}$}
        \psfrag{ylabel}[b][t]{\small $s$}
        \centering
        \includegraphics[width=.8\linewidth]{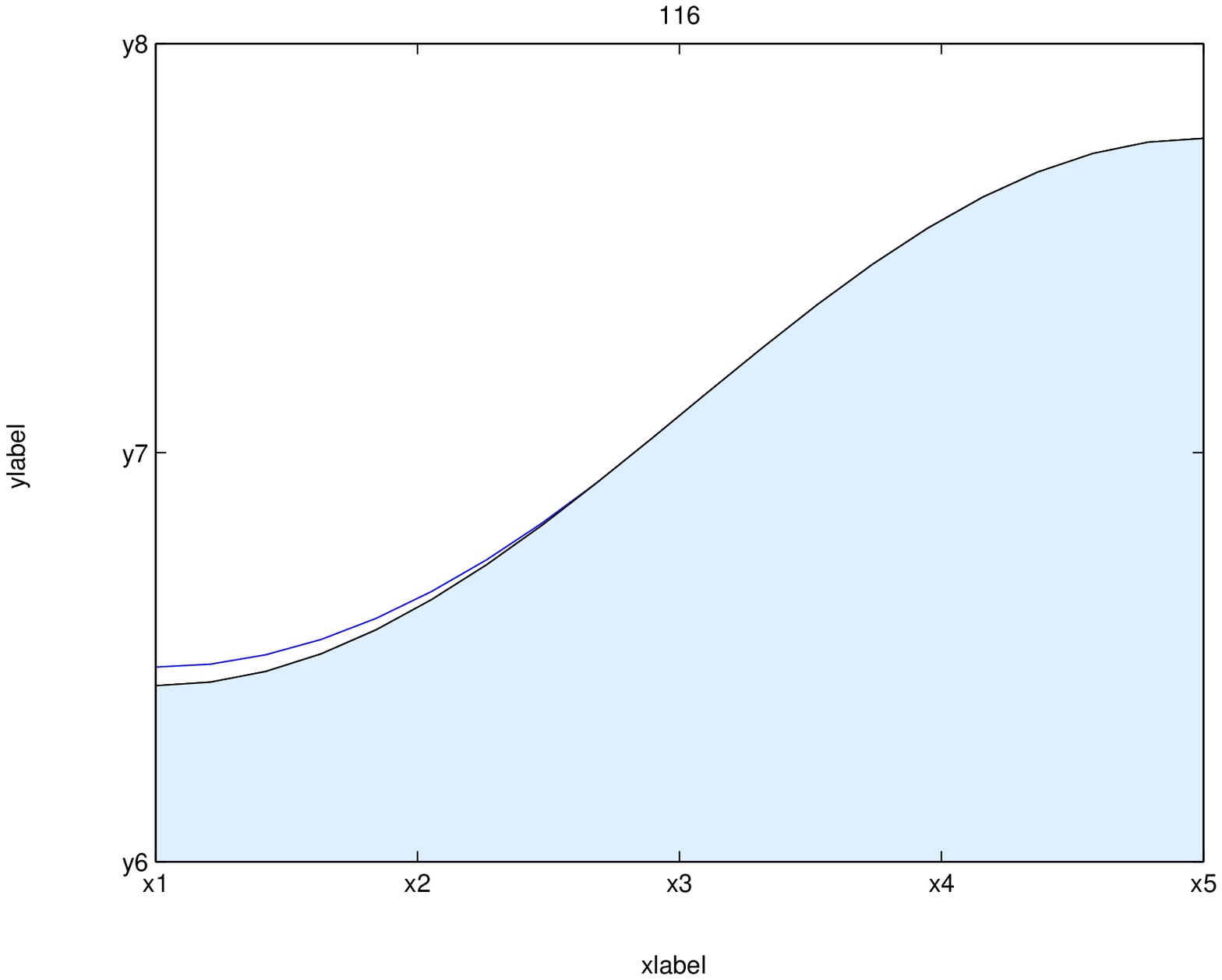}
        \\[.3cm]
        \includegraphics[width=.8\linewidth]{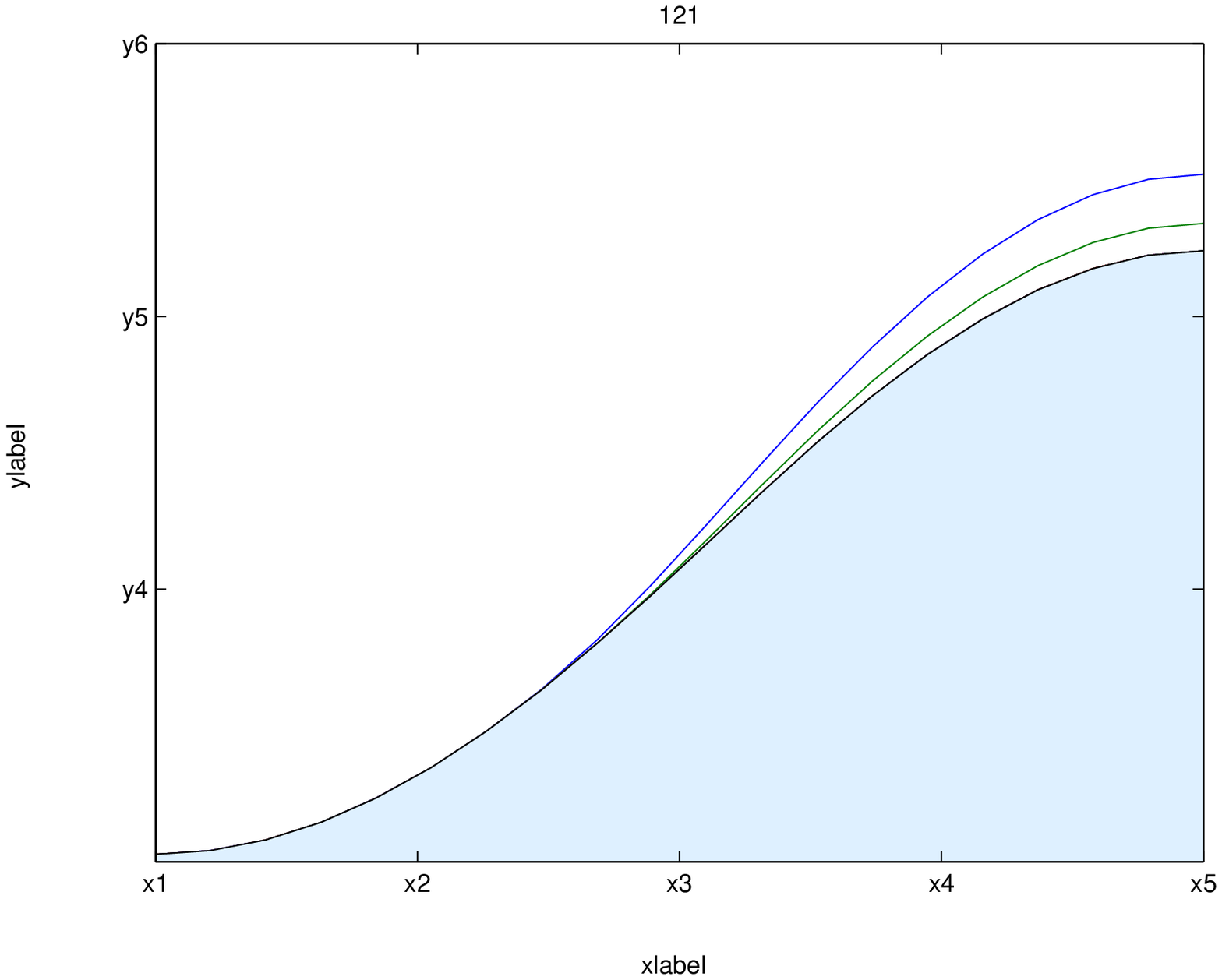}
        \caption{(Color online) The speed of zero sound as a function of the
          propagation direction $\phi_{\mathbf{q}}$ for $g = 1$ and $\theta_0 =
          \pi/4, \pi/2$ (top, bottom). Note that for $\theta_0 = \pi/2$ and
          $\phi_{\mathbf{q}} \gtrsim \pi/4$ we have two zero sound modes. As in
          Fig.~\ref{fig:s_of_theta0}, the region that is forbidden by the
          condition \eqref{eq:38} is shaded.}
        \label{fig:s_of_phi_q}
      \end{figure}  
    \end{psfrags}

We first find a zero sound mode that satisfies the propagation criterion  %
\eqref{eq:38} (see Figs.~\ref{fig:zs_region} and \ref{fig:s_of_theta0}, top)
for $\theta _{0}\approx 0.5$ and $\phi _{\mathbf{q}}=0$ (i.\thinspace e., in
the direction of the projection of the dipoles $\mathbf{d}$ on the  $xy$%
-plane). It is a longitudinal mode that is concentrated symmetrically 
around its propagation direction. As $\theta _{0}$ increases, this mode can 
propagate in a broader range of angles up to $\phi _{\mathbf{q}}\approx \pi 
/4$ (region I in Fig.~\ref{fig:zs_region}; see also  Fig.~\ref%
{fig:s_of_phi_q}, top); However, the corresponding sound velocity  drops
below the propagation boundary $v_{\mathrm{ph}}$ at $\theta  _{0}\approx 0.9$%
.

A different mode emerges from the continuum at $\theta_0 \approx \pi/4$ and  
$\phi_{\mathbf{q}} = \pi/2$ (region II in Fig.~\ref{fig:zs_region}). This 
mode is antisymmetrically peaked around the direction of propagation. At 
even higher values of $\theta_0$ we find more than one modes that satisfy  %
\eqref{eq:38} (see Figs.~\ref{fig:s_of_theta0}, bottom, and  \ref%
{fig:s_of_phi_q}, bottom).

This behavior of collective modes remains the same when we increase the 
interaction strength to $g \approx 2$. In particular, the angles $\theta_0 
\approx 0.5$ and $\theta_0 \approx \pi/4$ at which the symmetric and 
antisymmetric modes appear, respectively, are practically left  unchanged.
However, we find that the zero sound modes become more  ``distinct'',
i.\,e., the curves in Figs.\ \ref{fig:s_of_theta0} and  \ref{fig:s_of_phi_q}
are separated further from the PHC.

For $g < 1$ the above-mentioned peaking of modes around the forward 
direction is even more pronounced, and a high number of grid points is 
required to properly resolve these modes, making it impossible to go to very
small values $g \ll 1$ with our numerical method. At $0.1 \leq g < 1$ it 
becomes increasingly difficult to make quantitative statements as with 
regard to the region where the condition \eqref{eq:38} met, since the 
quantity of interest $v_F^{(0)} s - v_{\mathrm{ph}}$ is of the same order of
magnitude as its estimated error. Qualitatively, however, we find the same 
behavior as described above for $1 \leq g$.

We also solve Eq.~\eqref{eq:32} with finite values of $\mathbf{q}$. Some 
results are shown in Fig.~\ref{fig:cmdr} and demonstrate that the dispersion
$\omega = \omega (\mathbf{q})$ of the collective mode obtained in this 
manner agrees well with the linear approximation from Eq.~\eqref{eq:37}. For
details of the numerical procedure see App.~\ref{sec:numer-solut-bse}. 
\begin{figure}[tbp]
\psfrag{x1}[t][]{\scriptsize $0$} \psfrag{x2}[t][]{\scriptsize $0.2$}  %
\psfrag{x3}[t][]{\scriptsize $0.4$} \psfrag{x4}[t][]{\scriptsize $0.6$}  %
\psfrag{y1}[r][]{\scriptsize $0$} \psfrag{y2}[r][]{\scriptsize $0.5$}  %
\psfrag{y3}[r][]{\scriptsize $1$} \psfrag{y4}[r][]{\scriptsize $1.5$}  %
\psfrag{xi1}[t][]{\scriptsize $0$} \psfrag{xi2}[t][]{\scriptsize $0.05$}  %
\psfrag{xi3}[t][]{\scriptsize $0.$} \psfrag{yi1}[r][]{\scriptsize $0$}  %
\psfrag{yi2}[r][]{\scriptsize $0.1$} \psfrag{yi3}[r][]{\scriptsize $0.2$}  %
\psfrag{xlabel}[t][b]{\small $q/p_F^{(0)}$} 
\psfrag{ylabel}[b][t]{\small
    $\omega/\varepsilon_F^{(0)}$} \centering
\includegraphics[width=.8\linewidth]{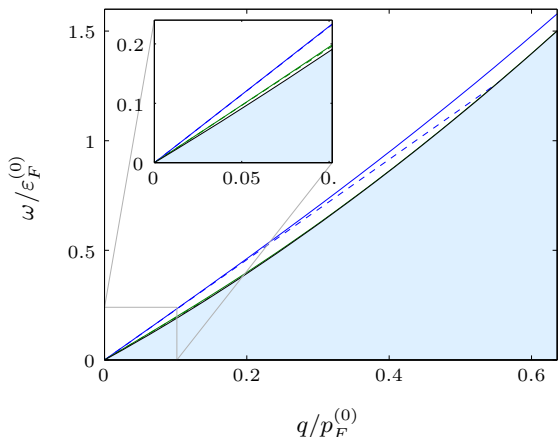}
\caption{(Color online) Collective mode dispersion relation for
  $\protect\phi_{\mathbf{q}} = \protect\pi/2$, $g = 2$ and $\protect\theta_0 =
  \protect\pi/2$ (solid lines). For comparison we also plot the linear
  approximation \eqref{eq:37} as dashed lines. The inset shows that at long
  wavelengths we have two zero sound modes.}
\label{fig:cmdr}
\end{figure}

\section{Instability of a spatially homogeneous system}

\label{sec:inst-spat-homog}

Apart from poles on the real axis corresponding to excitation frequencies of
collective modes, Eq.~\eqref{eq:32} can also have purely imaginary
eigenvalues that indicate the existence of unstable modes growing
exponentially with time. At the onset of an instability we have an
eigenvalue $\omega =0$, and, depending on the corresponding value of the
wave vector $q = \lvert \mathbf{q} \rvert$, one has a long wavelength
instability ($q=0$) resulting in local collapse of the system, or finite
wavelength instability ($q\neq 0$) leading to breaking of translational
invariance and formation of a periodic spatial structure (density waves).

\subsection{Long wavelength instability}

\label{sec:long-wavel-inst}

Equation \eqref{eq:41} shows that, if the instability occurs at $%
q\rightarrow 0$, the boundary of the instability region is determined by 
\begin{equation}
-\nu (\phi )=\frac{1}{m\,v_{F}(\phi )}\int \frac{d\phi ^{\prime }}{2\pi }%
\,F(\phi ,\phi ^{\prime })\,p_{F}(\phi ^{\prime })\,\nu (\phi ^{\prime }),
\label{eq:42}
\end{equation}%
and, hence, the instabilities could occur only in the regime of intermediate
coupling, $g\approx 1$.\footnote{%
In Eq.~\eqref{eq:41}, on the other hand, the prefactor on the LHS can become
arbitrarily small when $v_{F}^{(0)}s$ gets close to the limiting velocity of
the PHC $v_{\mathrm{ph}}$, and there is no generic restriction on the size
of $g$ for this equation to have a solution, i.\thinspace e., for zero sound
to occur.}

Equation \eqref{eq:42} is equivalent to the Pomeranchuk criterion \cite%
{pomeranchuk58:_stabil_of_fermi_liquid} on the Landau $f$-function to ensure
stability of a three-dimensional isotropic Fermi liquid \cite{LL:IX}. We
briefly review this method and its generalization to the two-dimensional
anisotropic case.

In the framework of Fermi liquid theory \cite{AGD}, the change of the
quasi-particle momentum distribution function $\delta n(\mathbf{p})$ results
in the change in the energy density 
\begin{multline}
\delta \mathcal{E}=\int \frac{d\mathbf{p}}{(2\pi )^{2}}\,[\varepsilon (%
\mathbf{p})-\mu ]\,\delta n(\mathbf{p})  \label{eq:43} \\
+\frac{1}{2}\int \frac{d\mathbf{p}}{(2\pi )^{2}}\frac{d\mathbf{p}^{\prime }}{%
(2\pi )^{2}}\,f(\mathbf{p},\mathbf{p}^{\prime })\,\delta n(\mathbf{p}%
)\,\delta n(\mathbf{p}^{\prime }),
\end{multline}%
where $f(\mathbf{p},\mathbf{p}^{\prime })$ is the quasi-particle interaction
function ($f$-function). A slight distortion $\delta p_{F}(\phi )$ [note
that in this section, $\delta p_{F}$ is used another way than in Secs.~\ref%
{sec:single-part-excit} and \ref{sec:superfl-trans}] of the anisotropic
Fermi surface $p_{F}(\phi )$ corresponds to a distribution function of the
form 
\begin{equation}
\delta n(\mathbf{p})=\theta[ p_{F}(\phi )+\delta p_{F}(\phi )-p]-\theta[
p_{F}(\phi )-p].  \label{eq:44}
\end{equation}

Following Pomeranchuk \cite{pomeranchuk58:_stabil_of_fermi_liquid} we expand 
$\delta \mathcal{E}$ in $\delta p_F$: The first order term vanishes due to
the fact that the energy of the Fermi liquid, considered as a functional of $%
\delta p_F$, is stationary at $\delta p_F \equiv 0$. To second order we have 
\begin{multline}  \label{eq:45}
\delta \mathcal{E} = \frac{1}{4 \pi m^2} \biggl[ \int \frac{d \phi}{2 \pi}
\, p_F(\phi) \, m \, v_F(\phi) \, \delta p_F(\phi)^2 \\
+ \int \frac{d \phi}{2 \pi} \frac{d \phi^{\prime}}{2 \pi} \, p_F(\phi) \,
p_F(\phi^{\prime}) \, F(\phi,\phi^{\prime}) \, \delta p_F(\phi) \, \delta
p_F(\phi^{\prime}) \biggr],
\end{multline}
where $F(\phi,\phi^{\prime}) \equiv \nu \, f(\mathbf{p}_F, \mathbf{p}%
_F^{\prime})$ is the dimensionless quasi-particle interaction function
already introduced in \eqref{eq:35}.

Thermodynamic stability requires that the variation of the energy density be
positive for any choice of $\delta p_F$. This condition can conveniently be
formulated by expanding the $\phi$-dependent functions in Fourier (double)
series $f(\phi) = \sum_m f_m \, \mathrm{e}^{i m \phi}$, where $f(\phi)$ is
one of $p_F(\phi)$, $\delta p_F(\phi)$, and $v_F(\phi)$, and $%
F(\phi,\phi^{\prime}) = \sum_{m, m^{\prime}} F_{m, m^{\prime}} \, \mathrm{e}%
^{i (m \phi - m^{\prime}\phi^{\prime})}$, where the coefficients $F_{m,
m^{\prime}}$ take the role of generalized Landau parameters.

Replacing the functions in the expression for $\delta \mathcal{E}$ by their
respective Fourier series expansions we obtain 
\begin{equation}  \label{eq:46}
\delta \mathcal{E} = \frac{1}{4 \pi m} \sum_{m,m^{\prime}} \delta p_m^{*} \,
M_{m,m^{\prime}} \, \delta p_{m^{\prime}},
\end{equation}
where the entries of the (self-adjoint) matrix $M$ are 
\begin{equation}  \label{eq:47}
M_{m,m^{\prime}} \equiv m \sum_k p_{m-k} \, v_{m^{\prime}-k}^{*} +
\sum_{k,l} p_{m-k} \, F_{k,l} \, p_{m^{\prime}-l}^{*}.
\end{equation}

A 2D anisotropic Fermi liquid, therefore, is thermodynamically stable if and
only if all eigenvalues of $M$ are positive. In the case of isotropic
interactions, we have $M_{m,m^{\prime }}= \delta _{m,m^{\prime }}M_{m}$, and
the requirement of stability reduces to the usual Pomeranchuk criteria,
i.\thinspace e., $M_{m}>0$ for all $m\in \mathbb{Z}$.

At the onset of an instability there exists an eigenvalue that is equal to
zero, i.\,e., there is a non-trivial solution to the equation 
\begin{equation}  \label{eq:48}
\sum_{m^{\prime}} M_{m, m^{\prime}} \, \delta p_{m^{\prime}} = 0.
\end{equation}
In terms of the original Fermi surface deformation $\delta p_F$ this
equation reads 
\begin{equation}  \label{eq:49}
\delta p_F(\phi) + \frac{1}{m \, v_F(\phi)} \int \frac{d \phi^{\prime}}{2 \pi%
} \, F(\phi,\phi^{\prime}) \, p_F(\phi^{\prime}) \, \delta
p_F(\phi^{\prime}) = 0,
\end{equation}
and if we identify $\delta p_F$ with $\nu$ we are lead back to Eq.~%
\eqref{eq:42}. The line in the $g \theta_0$-plane on which this equation has
a non-trivial solution determines the boundary to the red shaded region that
is labeled LWI in Fig.~\ref{fig:phase_diagram}. In the interior of this
region there exists an unstable long wavelength mode.

\subsection{Finite wavelength (density wave) instability}

\label{sec:dens-wave-inst} An instability at finite momentum $q=\lvert 
\mathbf{q}\rvert $ drives the system towards a state with stationary
fluctuations in the density. We address this problem numerically (see
details in Appendix B). After solving Eq.~%
\eqref{eq:141} [which is equivalent to
Eq.~\eqref{eq:32}] numerically for $q\neq 0$, we find such an instability
with $\phi _{\mathbf{q}}=\pi /2$ (i.\,e., $\mathbf{q}$ is perpendicular to
the $x$-axis -- the projection of the dipoles $\mathbf{d}$ on the $xy$%
-plane) and $q\approx 2\,p_{F}(\phi _{\mathbf{q}})$ in the blue shaded
region in Fig.~\ref{fig:phase_diagram}. As can be seen from Eq.~%
\eqref{eq:141}, the eigenvectors $\nu _{\mathbf{q}}(\mathbf{p}) \propto
\langle a_{\mathbf{p}+\mathbf{q}/2}^{\dagger } \, a_{\mathbf{p}-\mathbf{q}%
/2} \rangle $ which correspond to an instability, signal the formation of a
density modulation with momentum $\mathbf{q}$ (density wave). On the other
hand, the dependence of $\nu _{\mathbf{q}}(\mathbf{p})$ on $\mathbf{p}$
excludes a description of the transition in terms of a simple local order
parameter $\langle \widehat{\psi }^{\dagger }(\mathbf{r}) \, \widehat{\psi }(%
\mathbf{r}) \rangle =\rho (\mathbf{q})\cos (\mathbf{q} \cdot \mathbf{r})$.
In the isotropic case with $\theta _{0}=0$, where there is no preferred
direction and the system is invariant with respect to rotations around the $z
$-axis, the instability occurs at $g\approx 1.45$ and is independent of the
angle $\phi _{\mathbf{q}}$. This result is in agreement with the value $%
g\approx 1.42$ which was found by the authors of \cite{babadi11:_densit}.
Previous studies within the RPA \cite%
{sun10:_spont_fermi,yamaguchi10:_densit_wave_instab_in_two} predicted a
considerably smaller value of $g=0.5$, which is due to the fact that the RPA
overestimates the effects of the interparicle interaction as it neglects the
exchange contribution.

Note that for $\tilde{\Gamma}_{\mathrm{ph}}(\mathbf{p}-\mathbf{p}^{\prime },%
\mathbf{q})=V_{0}(\mathbf{q})$ (i.\thinspace e., when only the direct
interaction is taken into account and the exchange one is neglected), $\chi (%
\mathbf{p};\mathbf{q})$ is $\mathbf{p}$-independent and Eq.~\eqref{eq:32}
reduces to 
\begin{equation}
1-V_{0}(\mathbf{q})\,\Pi ^{(0)}(\omega ,\mathbf{q})=0,  \label{eq:50}
\end{equation}%
where 
\begin{equation}
\begin{split}
\Pi ^{(0)}(\omega ,\mathbf{q})& =%
\parbox{10mm}{
  \begin{fmfgraph*}(10,5)    
          \fmfleft{l}
          \fmfright{r}
          \fmftop{t}
          \fmfbottom{b}        
          \fmfi{fermion,label=$p' + q$,width=thick}{vloc(__l){(1,1)} .. {right}vloc(__t){right}
            .. {(1,-1)}vloc(__r)}
          \fmfi{fermion,label=$p'$,width=thick}{vloc(__r){(-1,-1)} .. {left}vloc(__b){left} .. {(-1,1)}vloc(__l)}
        \end{fmfgraph*}} \\[0.5cm]
& =\int \frac{d\mathbf{p}^{\prime }}{(2\pi )^{2}}\frac{n(\mathbf{p}^{\prime
})-n(\mathbf{p}^{\prime }+\mathbf{q})}{\omega +\varepsilon (\mathbf{p}%
^{\prime })-\varepsilon (\mathbf{p}^{\prime }+\mathbf{q})+i0\sgn(\omega )}
\end{split}
\label{eq:51}
\end{equation}%
is the 2D polarization operator. Equation~\eqref{eq:50} is used to study
long wavelength ($q\rightarrow 0$) plasmon oscillations in electrically
charged systems (see, for example, \cite{Nozieres/Pines}). Although keeping
only the direct interaction in the long wave-length limit is legitimate for
Coulomb systems (because of the divergence of the Coulomb interaction for a
small transferred momentum $q\rightarrow 0$, while the exchange one is
finite due to a non-zero momentum transfer, $\left|  \mathbf{p}-\mathbf{p}%
^{\prime }\right| \sim p_{F}$), this approximation gives physically
incorrect results in a Fermi system with a finite Fourier transform of the
interparticle interaction for small momentum transfer (like in the
considered case of a dipolar monolayer). In this case, the direct and the
exchange contributions are of the same order and keeping only the former
yields unphysical results. For a short range interparticle interaction (with
a momentum-independent Fourier transform), the two contributions have to
cancel each other, resulting in no interaction effects in a single-component
Fermi gas with a short-range interaction. Similar considerations also apply
to the analysis of instabilities in a dipolar system on the basis of Eq.~%
\eqref{eq:32}: Keeping the exchange contribution in this equation is
essential in order to obtain correct results consistent with fermionic
statistics of particles.

In order to take the exchange contribution into account in Eq.~\eqref{eq:50}%
, one can modify the polarization propagator by including an entire
interaction ladder in the polarization bubble (this is equivalent to
including the exchange interaction in $\tilde{\Gamma}_{\mathrm{ph}}$), see
also Ref. \cite{babadi11:_densit}: 
\begin{multline}
\Pi (\omega ,\mathbf{q})=\Pi ^{(0)}(\omega ,\mathbf{q})+%
\parbox{15mm}{
        \begin{fmfgraph}(15,5)        
          \fmfleft{l}
          \fmfright{r}
          \fmftopn{t}{3}
          \fmfbottomn{b}{3}
          \fmf{wiggly}{t2,b2}
          \fmfi{fermion,width=thick}{vloc(__l){(1,1)} .. {right}vloc(__t2)}
          \fmfi{fermion,width=thick}{vloc(__t2){right} .. {(1,-1)}vloc(__r)}          
          \fmfi{fermion,width=thick}{vloc(__r){(-1,-1)} .. {left}vloc(__b2)}
          \fmfi{fermion,width=thick}{vloc(__b2){left} .. {(-1,1)}vloc(__l)}
        \end{fmfgraph}}  \label{eq:52} \\
+%
\parbox{20mm}{
        \begin{fmfgraph}(20,5)        
          \fmfleft{l}
          \fmfright{r}
          \fmftopn{t}{4}
          \fmfbottomn{b}{4}
          \fmf{wiggly}{t2,b2}
          \fmf{wiggly}{t3,b3}
          \fmfi{fermion,width=thick}{vloc(__l){(1,1)} .. {right}vloc(__t2)}
          \fmfi{fermion,width=thick}{vloc(__t2){right} .. {right}vloc(__t3)}
          \fmfi{fermion,width=thick}{vloc(__t3){right} .. {(1,-1)}vloc(__r)}
          \fmfi{fermion,width=thick}{vloc(__r){(-1,-1)} .. {left}vloc(__b3)}
          \fmfi{fermion,width=thick}{vloc(__b3){left} .. {left}vloc(__b2)}
          \fmfi{fermion,width=thick}{vloc(__b2){left} .. {(-1,1)}vloc(__l)}
        \end{fmfgraph}}+\dotsb .
\end{multline}%
(Another possibility is to include a local field correction  \cite%
{parish11:_densit_fermi,zinner11:_densit}%
,  similar to the consideration of the density-density response in Coulomb 
systems. This approach, however, is very sensitive to a particular choice of
the local field correction.) With the modified polarization operator $\Pi 
(\omega ,\mathbf{q})$, the density-density correlation function (in 
frequency-momentum space) reads  
\begin{equation}
\chi (\omega ,\mathbf{q})=\frac{\Pi (\omega ,\mathbf{q})}{1-V_{0}(\mathbf{q}%
)\,\Pi (\omega ,\mathbf{q})},  \label{eq:53}
\end{equation}%
and Eq.~\eqref{eq:50} with $\Pi ^{(0)}(\omega ,\mathbf{q})$  replaced with $%
\Pi (\omega ,\mathbf{q})$ corresponds to the instability in the 
density-density correlation function. Note that when $V_{0}(\mathbf{q})$ is 
replaced by a momentum independent constant $V_{0}$ (that corresponds to a 
short-range interaction), the modified polarization operator is $\Pi (\omega
,\mathbf{q})=\Pi ^{(0)}(\omega ,\mathbf{q})\left[ 1+V_{0}\,\Pi ^{(0)}(\omega
,\mathbf{q})\right] ^{-1}$, and the density-density correlation function 
reduces to the polarization operator of a non-interacting gas, $\chi (\omega
,\mathbf{q})=\Pi ^{(0)}(\omega ,\mathbf{q})$, as it should be in a 
single-component Fermi gas.

\section{Superfluid transition}

\label{sec:superfl-trans} We now discuss the superfluid instability in a
dipolar monolayer. As we will show, this instability is sensitive to the
details of two particle scattering and, as it was already pointed out at the
end of Sec.~\ref{sec:system}, this requires us to take into account
contributions of short distances $\lesssim l_{0}$ and high energies $\gtrsim
\omega _{0}$. Therefore, in the following we will not limit ourselves 
\textit{a priori} to configurations with all particles residing in the
ground state of the trapping potential but rather allow for virtual
transitions to arbitrarily highly excited states.

\subsection{Gap equation}

\label{sec:gap-equation} The superfluid transition is characterized by the
order parameter (gap) $\Delta (\mathbf{r},\mathbf{r}^{\prime })=V_{d}(%
\mathbf{r}-\mathbf{r}^{\prime })\langle \widehat{\psi }(\mathbf{r})\,%
\widehat{\psi }(\mathbf{r}^{\prime })\rangle $, which attains non-zero
values for temperatures below the critical temperature $T_{c}$. Note that in
two dimensions at finite temperatures long-range order is actually destroyed
by phase fluctuations and the mean-field order parameter is zero. The
superfluid density, however, remains finite and the transition to the
superfluid phase follows the Berezinskii-Kosterlitz-Thouless scenario \cite%
{berezinski72:_destr_long_order_one_two,kosterlitz73:_order,kosterlitz74}
and occurs at a temperature $T_{c}^{({\mathrm{BKT}})}$. Nevertheless, we may
consider the mean-field critical temperature $T_{c}$ as a reliable estimate
of the value of $T_{c}^{({\mathrm{BKT}})}$, as the difference between the
two of them is small in the weak coupling regime \cite%
{miyake83:_fermi_liquid_theor_of_dilut}.

For $0<T_{c}-T\ll T_{c}$, the order parameter is a solution to the
homogeneous Bethe-Salpeter equation in the Cooper channel or linearized gap
equation \cite{LL:IX}, 
\begin{equation}
\parbox{12mm}{\begin{fmfgraph*}(12,10)
          \fmfleft{lb,lt}
          \fmfright{rb,rt}
          \fmf{fermion,width=thick}{lt,vlt}
          \fmf{fermion,width=thick}{lb,vlb}  
          \fmf{phantom}{vrb,rb}
          \fmf{phantom}{vrt,rt}
          \fmfpoly{smooth,tension=1.5,label=\tiny $\Delta$}{vrt,vlt,vlb,vrb}
        \end{fmfgraph*}}\hspace{-3mm}=\hspace{1mm}%
\parbox{25mm}{
        \begin{fmfgraph*}(25,10)                
          \fmfleft{lb,lt}
          \fmfright{rb,rt}
          \fmftop{ct} \fmfbottom{cb}
          \fmf{fermion,width=thick}{lt,v1lt}
          \fmf{fermion,width=thick}{lb,v1lb}  
          \fmf{phantom}{cb,v1rb}
          \fmf{phantom}{v1rt,ct}
          \fmf{phantom}{ct,v2lt}
          \fmf{phantom}{v2lb,cb}
          \fmf{phantom}{rb,v2rb}
          \fmf{phantom}{v2rt,rt}
          \fmfpoly{smooth,tension=.8,label=\tiny $\tilde{\Gamma}_{\mathrm{pp}}$}{v1rt,v1lt,v1lb,v1rb}
          \fmfpoly{smooth,tension=1.5,label=\tiny $\Delta$}{v2rt,v2lt,v2lb,v2rb}
          \fmffreeze  
          \fmfi{fermion,width=thick}{vloc(__v1rb){(1,-1)} .. {(1,1)}vloc(__v2lb)}
          \fmfi{fermion,width=thick}{vloc(__v1rt){(1,1)} .. {(1,-1)}vloc(__v2lt)}
        \end{fmfgraph*}
      }\hspace{-4mm},  \label{eq:54}
\end{equation}%
where $\tilde{\Gamma}_{\mathrm{pp}}$ denotes the particle-particle
irreducible vertex, which is the sum of connected diagrams with two incoming
and two outgoing lines which cannot be divided into two parts by cutting two
fermion lines of the same direction.

In these diagrams, thick lines correspond to interacting Matsubara Green's
functions in the presence of the transverse trapping potential. Following
Gor'kov and Melik-Barkhudarov \cite{gorkov61:_contr_to_theor_of_super} we
expand the RHS of Eq.~\eqref{eq:54} to second order in the dipole-dipole
interaction $V_d$, 
\begin{equation}  \label{eq:55}
\parbox{12mm}{\begin{fmfgraph*}(12,10)                 
          \fmfleft{lb,lt}
          \fmfright{rb,rt}
          \fmf{fermion}{lt,vlt}
          \fmf{fermion}{lb,vlb}  
          \fmf{phantom}{vrb,rb}
          \fmf{phantom}{vrt,rt} 
          \fmfpoly{smooth,tension=1.5,width=thick,label=\tiny $\Delta$}{vrt,vlt,vlb,vrb}
        \end{fmfgraph*}} \hspace{-3mm}=\hspace{-6mm} 
\parbox{25mm}{\begin{gap}
          \fmf{plain}{t2,t3}
          \fmf{plain}{b2,b3}
          \fmf{wiggly}{t3,b3}
          \fmffreeze
          \fmfi{fermion}{vloc(__t3){right} .. {(1,-1)}vloc(__vlt)}
          \fmfi{fermion}{vloc(__b3){right} .. {(1,1)}vloc(__vlb)}
        \end{gap}} \hspace{-2mm}+ \hspace{-6mm} 
\parbox{25mm}{\begin{gap}
          \fmf{plain}{t2,t3}
          \fmf{plain}{b2,b3}
          \fmf{wiggly}{t3,b3}
          \fmffreeze
          \fmfi{fermion}{vloc(__t3){right} .. {(1,-1)}vloc(__vlt)}
          \fmfipath{p}
          \fmfiset{p}{vloc(__b3){right} .. {(1,1)}vloc(__vlb)}
          \fmfi{fermion}{p}
          \fmfi{wiggly}{point length(p)/4 of p{(0,1)} .. {(.8,-1)}point 3 length(p)/4 of p}
        \end{gap}} \hspace{-2mm}+ \hspace{1mm} 
\parbox{25mm}{\begin{fmfgraph*}(25,10)
          \fmfleft{lb,lt}
          \fmfright{rb,rt}
          \fmftop{ct} \fmfbottom{cb}
          \fmf{fermion}{lt,v1lt}
          \fmf{fermion}{lb,v1lb}  
          \fmf{phantom}{cb,v1rb}
          \fmf{phantom}{v1rt,ct}
          \fmf{phantom}{ct,v2lt}
          \fmf{phantom}{v2lb,cb}
          \fmf{phantom}{rb,v2rb}
          \fmf{phantom}{v2rt,rt}
          \fmfpoly{smooth,tension=.8,label=\tiny $\delta V_d$}{v1rt,v1lt,v1lb,v1rb}
          \fmfpoly{smooth,tension=1.5,label=\tiny $\Delta$}{v2rt,v2lt,v2lb,v2rb}
          \fmffreeze  
          \fmfi{fermion}{vloc(__v1rb){(1,-1)} .. {(1,1)}vloc(__v2lb)}
          \fmfi{fermion}{vloc(__v1rt){(1,1)} .. {(1,-1)}vloc(__v2lt)}
        \end{fmfgraph*}} \hspace{-3mm}.
\end{equation}
Here, with each thin line is associated a non-interacting Matsubara Green's
function which can be written in the form \cite{AGD} 
\begin{equation}  \label{eq:56}
\begin{split}
\mathcal{G}(\omega_s, \mathbf{p}, z, z^{\prime}) & = \sum_{n = 0}^{\infty}
\, \phi_n(z) \, \phi_n(z^{\prime}) \, \mathcal{G}_n(\omega_s, \mathbf{p}), \\
\mathcal{G}_n(\omega_s, \mathbf{p}) & = \frac{1}{i \omega_s- \xi(\mathbf{p})
- \omega_0 n},
\end{split}%
\end{equation}
where $\xi(\mathbf{p}) = p^2/2m - \mu$, and fermionic Matsubara frequencies
are $\omega_s = (2 s + 1) \pi T$ for integer $s$. The chemical potential is $%
\mu = \varepsilon_F^{(0)} + \delta \mu$, where the first order correction $%
\delta \mu$ is given by Eq.~%
\eqref{eq:28} [due to the exponential smallness
    of $T_c$ (see discussion below) we may use the zero temperature value of
    $\mu$]. Thus, strictly speaking \eqref{eq:55} contains terms of infinite
order. It is, however, convenient to perform the expansion in $\delta \mu$
at a later stage.

Of the diagrams on the RHS of \eqref{eq:55} the first one gives the leading
(first order) contribution, the others correspond to (second order)
corrections: The second diagram is obtained from the first one by inserting
an exchange self-energy part (note that the direct term is absent for the
dipole-dipole interaction, see Sec.~\ref{sec:single-part-excit}). This
diagram comes with a factor of 2, as we could have equally well inserted the
self-energy part in the upper particle line. The quantity $\delta V_d$
represents second order corrections to the bare dipole-dipole interaction
and is given by the set of diagrams 
\begin{gather}  \label{eq:57}
\delta V_d = \delta V^{(a)}_d + \delta V^{(b)}_d + \delta V^{(c)}_d + \delta
V^{(d)}_d, \\[2mm]
\begin{aligned} \delta V^{(a)}_d & = \parbox{20mm}{
          \begin{fmfgraph}(15,10) \fmfstraight \fmftopn{t}{3}
            \fmfbottomn{b}{3}
            \fmf{fermion}{t2,t3}
            \fmf{fermion}{t1,t2}
            \fmf{wiggly}{t2,vn} \fmf{phantom,tension=.6}{vn,vs}
            \fmf{wiggly}{vs,b2}
            \fmf{fermion}{b2,b3}
            \fmf{fermion}{b1,b2}
            \fmffreeze \fmfi{fermion}{vloc(__vn){left} .. {right}vloc(__vs)}
            \fmfi{fermion}{vloc(__vs){right} .. {left}vloc(__vn)}
          \end{fmfgraph}
        } \hspace{-5mm}, \hspace{2mm} & \delta V^{(b)}_d & = \parbox{20mm}{
          \begin{fmfgraph}(15,10) \fmfstraight
            \fmftopn{t}{3} \fmfbottomn{b}{4} \fmf{fermion}{t2,t3}
            \fmf{fermion}{t1,t2}
            \fmf{fermion}{b3,b4}
            \fmf{fermion}{b1,b2} \fmf{wiggly}{b3,b2}
            \fmf{wiggly}{t2,v} \fmf{phantom,tension=.8}{b3,v,b2}
            \fmffreeze
            \fmfi{fermion}{vloc(__b2){up} .. {right}vloc(__v)}
            \fmfi{fermion}{vloc(__v){right} .. {down}vloc(__b3)}        
          \end{fmfgraph}
        } \hspace{-5mm}, \\[5mm] \delta V^{(c)}_d & = \parbox{20mm}{
          \begin{fmfgraph}(15,10) \fmfstraight \fmftopn{t}{4}
            \fmfbottomn{b}{3}
            \fmf{fermion}{t3,t4}
            \fmf{wiggly}{t2,t3}
            \fmf{fermion}{t1,t2}
            \fmf{phantom,tension=.8}{t3,v,t2} \fmf{wiggly}{v,b2}
            \fmf{fermion}{b2,b3}
            \fmf{fermion}{b1,b2}
            \fmffreeze        
            \fmfi{fermion}{vloc(__t2){down} .. {right}vloc(__v)}      
            \fmfi{fermion}{vloc(__v){right} .. {up}vloc(__t3)}
          \end{fmfgraph}
        } \hspace{-5mm}, & \delta V^{(d)}_d & = \parbox{20mm}{
          \begin{fmfgraph}(15,10) \fmfstraight \fmftopn{t}{4}
            \fmfbottomn{b}{4}
            \fmf{fermion}{t3,t4} 
            \fmf{fermion}{t2,t3}
            \fmf{fermion}{t1,t2}
            \fmf{wiggly,rubout}{t3,b2}
            \fmf{wiggly}{t2,b3}
            \fmf{fermion}{b3,b4}
            \fmf{fermion}{b2,b3} 
            \fmf{fermion}{b1,b2}      
          \end{fmfgraph}
        } \hspace{-5mm} . \end{aligned}  \notag
\end{gather}
These diagrams describe processes in which one of the incoming particles
polarizes the medium by exciting a virtual particle-hole pair. In diagram $%
(a)$ the particle and hole annihilate each other while interacting with the
other incoming particle, whereas in $(b,c,d)$ the hole is annihilated by one
of the incoming particles: by the second incoming particle in $(b,c)$, and
by the very same particle that created the particle-hole pair in the first
place in $(d)$.

In writing down the analytical expressions that correspond to the diagrams
in \eqref{eq:55} it is convenient to label particle lines by two-dimensional
momentum vectors $\mathbf{p} = (p_x, p_y)$ and HO quantum numbers $n$, since
the Matsubara Green's function \eqref{eq:56} is diagonal in this basis.
Accordingly, we expand the order parameter as 
\begin{equation}  \label{eq:58}
\Delta(\mathbf{p}, z, z^{\prime}) = \sum_{n, n^{\prime}= 0}^{\infty}
\phi_n(z) \, \phi_{n^{\prime}}(z^{\prime}) \, \Delta_{n, n^{\prime}}(\mathbf{%
p}).
\end{equation}
The matrix elements of the dipole-dipole interaction are defined in Eq.~%
\eqref{eq:5}, and we employ an analogous definition for matrix elements $%
\delta V_{n_1, n_2, n_3, n_4}$ of $\delta V_d$. This allows us to write the
diagrams on the RHS of \eqref{eq:55} as 
\begin{widetext}
      \begin{gather}
        \label{eq:59}
        \parbox{25mm}{\begin{gap}
            \fmf{plain}{t2,t3}
            \fmf{plain}{b2,b3}
            \fmf{wiggly}{t3,b3}
            \fmffreeze
            \fmfi{fermion}{vloc(__t3){right} .. {(1,-1)}vloc(__vlt)}
            \fmfi{fermion}{vloc(__b3){right} .. {(1,1)}vloc(__vlb)}
          \end{gap}} \hspace{-4mm} = - \int \sum_{n_3, n_4} V_{n_1,
          n_2, n_3, n_4} (\mathbf{p} - \mathbf{p}') \, T \sum_s
        \mathcal{G}_{n_3} (\omega_s, \mathbf{p}') \,
        \mathcal{G}_{n_4} (-\omega_s, -\mathbf{p}') \, \Delta_{n_3,
          n_4} (\mathbf{p}') \frac{d \mathbf{p}'}{(2 \pi)^2}, \\ \nonumber        
        \parbox{25mm}{\begin{gap}
            \fmf{plain}{t2,t3}
            \fmf{plain}{b2,b3}
            \fmf{wiggly}{t3,b3}
            \fmffreeze
            \fmfi{fermion}{vloc(__t3){right} .. {(1,-1)}vloc(__vlt)}
            \fmfipath{p}
            \fmfiset{p}{vloc(__b3){right} .. {(1,1)}vloc(__vlb)}
            \fmfi{fermion}{p}
            \fmfi{wiggly}{point length(p)/4 of p{(0,1)} .. {(.8,-1)}point 3 length(p)/4 of p}
          \end{gap}} \hspace{-4mm} = - 2 \int \sum_{n_3, n_4, n_5}
        V_{n_1, n_2, n_3, n_4} (\mathbf{p} - \mathbf{p}') \, T
        \sum_s \mathcal{G}_{n_3} (\omega_s, \mathbf{p}') \,
        \mathcal{G}_{n_4} (-\omega_s, -\mathbf{p}')
        \\ \label{eq:60} \times \Sigma^{(1)}_{n_4, n_5}
        (\mathbf{p}') \, \mathcal{G}_{n_5} (-\omega_s,
        -\mathbf{p}') \, \Delta_{n_3, n_5}  (\mathbf{p}') \frac{d
          \mathbf{p}'}{(2 \pi)^2}, \\ \label{eq:61} \Sigma^{(1)}_{n_1,
          n_2} (\mathbf{p}) = \sum_{n_3} \int T \sum_s V_{n_1, n_2,
          n_3, n_3} (\mathbf{p} - \mathbf{p}') \, \mathcal{G}_{n_3}
        (\omega_s, \mathbf{p}') \frac{d \mathbf{p}'}{(2
          \pi)^2}, \\
        \label{eq:62}
        \parbox{25mm}{\begin{fmfgraph*}(25,10)
            \fmfleft{lb,lt}
            \fmfright{rb,rt}
            \fmftop{ct} \fmfbottom{cb}
            \fmf{fermion}{lt,v1lt}
            \fmf{fermion}{lb,v1lb}  
            \fmf{phantom}{cb,v1rb}
            \fmf{phantom}{v1rt,ct}
            \fmf{phantom}{ct,v2lt}
            \fmf{phantom}{v2lb,cb}
            \fmf{phantom}{rb,v2rb}
            \fmf{phantom}{v2rt,rt}
            \fmfpoly{smooth,tension=.8,label=\tiny $\delta V_d$}{v1rt,v1lt,v1lb,v1rb}
            \fmfpoly{smooth,tension=1.5,label=\tiny $\Delta$}{v2rt,v2lt,v2lb,v2rb}
            \fmffreeze  
            \fmfi{fermion}{vloc(__v1rb){(1,-1)} .. {(1,1)}vloc(__v2lb)}
            \fmfi{fermion}{vloc(__v1rt){(1,1)} .. {(1,-1)}vloc(__v2lt)}
          \end{fmfgraph*}} \hspace{-4mm} =
        - \int \sum_{n_3, n_4} \delta V_{n_1, n_2, n_3, n_4} (\mathbf{p} - \mathbf{p}')  \, 
        T \sum_s \mathcal{G}_{n_3} (\omega_s, \mathbf{p}')  \, 
        \mathcal{G}_{n_4} (-\omega_s, -\mathbf{p}') \, \Delta_{n_3,
          n_4} (\mathbf{p}') \frac{d \mathbf{p}'}{(2 \pi)^2}.
      \end{gather}
    \end{widetext}

The sum over Matsubara frequencies in \eqref{eq:59} and \eqref{eq:62} can be
evaluated by rewriting it as a contour integral \cite{Rickayzen} and gives 
\begin{align}  \label{eq:63}
\mathcal{K}_{n,n^{\prime }}(\mathbf{p}) & \equiv T\sum_{s}\mathcal{G}%
_{n}(\omega _{s},\mathbf{p})\,\mathcal{G}_{n^{\prime }}(-\omega _{s},-%
\mathbf{p})  \notag \\
& =\frac{\tanh \Bigl[\frac{\xi (\mathbf{p})+\omega _{0}n}{2T}\Bigr]+\tanh %
\Bigl[\frac{\xi (\mathbf{p})+\omega _{0}n^{\prime }}{2T}\Bigr]}{2[2\,\xi (%
\mathbf{p})+\omega _{0}(n+n^{\prime })]}.
\end{align}

For $n,n^{\prime }\neq 0$ and in the quasi-2D limit $\omega _{0}\gg \mu $
the denominator in \eqref{eq:63} is positive for all values of $\mathbf{p}$
and the arguments of the hyperbolic tangent functions are as large as $%
\omega _{0}/T$. Then, to within exponential accuracy in this ratio, we may
approximate 
\begin{equation}
\mathcal{K}_{n,n^{\prime }}(\mathbf{p})\approx \frac{1}{2\,\xi (\mathbf{p}%
)+\omega _{0}(n+n^{\prime })}.  \label{eq:64}
\end{equation}

If only one of $n, n^{\prime}$ is different from zero, the denominator in %
\eqref{eq:63} is still positive for all $\mathbf{p}$. For concreteness, let
us assume that $n \neq 0$ and $n^{\prime}= 0$. Then we set 
\begin{equation}  \label{eq:65}
\mathcal{K}_{n, 0} \mskip-.5\thinmuskip (\mathbf{p}) \approx \frac{1 + %
\sgn[\xi(\mathbf{p})]}{2 [2 \, \xi( \mathbf{p}) + \omega_0 n]},
\end{equation}
thereby omitting a subdominant contribution to the integral \eqref{eq:59}
from a narrow shell in momentum space $\lvert \xi(\mathbf{p}^{\prime})
\rvert \lesssim T$ where $\tanh[\xi(\mathbf{p})/2 T]$ deviates appreciably
from $\sgn[\xi(\mathbf{p})]$.

Finally, in the case $n = n^{\prime}= 0$, \eqref{eq:63} reduces to 
\begin{equation}  \label{eq:66}
\mathcal{K}[\xi(\mathbf{p})] \equiv \mathcal{K}_{0, 0}(\mathbf{p}) = \tanh[%
\xi(\mathbf{p})/2 T]/2 \, \xi(\mathbf{p}).
\end{equation}
In this expression, the denominator vanishes on the Fermi surface, i.\,e.,
for $\xi(\mathbf{p}) = 0$. This leads in the limit $T/\mu \ll 1$ to a
logarithmically divergent integral $\propto \ln(T/\mu)$ on the RHS of %
\eqref{eq:59} when $n_3 = n_4 = 0$.

We have to keep contributions of second order in the gap equation only if
they are multiplied by such a large logarithm \cite%
{gorkov61:_contr_to_theor_of_super}: In \eqref{eq:60} this is the term with $%
n_{3}=n_{4}=n_{5}=0$ which contains the factor 
\begin{equation}
T\sum_{s}\mathcal{G}_{0}(\omega _{s},\mathbf{p})\,\mathcal{G}_{0}(-\omega
_{s},-\mathbf{p})^{2}=\frac{1}{2}\frac{\partial \mathcal{K}}{\partial \xi }%
[\xi (\mathbf{p})],  \label{eq:67}
\end{equation}%
and in \eqref{eq:62} it is the summand with $n_{3}=n_{4}=0$. These
simplifications allow us to rewrite Eqs.~\eqref{eq:59}, \eqref{eq:60}, and %
\eqref{eq:62} as [$\Delta (\mathbf{p})\equiv \Delta _{0,0}(\mathbf{p})$, $%
\Sigma ^{(1)}(\mathbf{p})\equiv \Sigma _{0,0}^{(1)}(\mathbf{p})$] 
\begin{widetext}
      \begin{gather}
        \label{eq:68}
        \parbox{25mm}{\begin{gap}
            \fmf{plain}{t2,t3}
            \fmf{plain}{b2,b3}
            \fmf{wiggly}{t3,b3}
            \fmffreeze
            \fmfi{fermion}{vloc(__t3){right} .. {(1,-1)}vloc(__vlt)}
            \fmfi{fermion}{vloc(__b3){right} .. {(1,1)}vloc(__vlb)}
          \end{gap}} \hspace{-4mm} = - \int \sum_{n_3, n_4} V_{n_1, n_2, n_3,
          n_4} (\mathbf{p} - \mathbf{p}') \, \mathcal{K}_{n_3, n_4}
        (\mathbf{p}') \, \Delta_{n_3, n_4} (\mathbf{p}') \frac{d \mathbf{p}'}{(2
          \pi)^2}, \\ \label{eq:69}
        \parbox{25mm}{\begin{gap}
            \fmf{plain}{t2,t3}
            \fmf{plain}{b2,b3}
            \fmf{wiggly}{t3,b3}
            \fmffreeze
            \fmfi{fermion}{vloc(__t3){right} .. {(1,-1)}vloc(__vlt)}
            \fmfipath{p}
            \fmfiset{p}{vloc(__b3){right} .. {(1,1)}vloc(__vlb)}
            \fmfi{fermion}{p}
            \fmfi{wiggly}{point length(p)/4 of p{(0,1)} .. {(.8,-1)}point 3 length(p)/4 of p}
          \end{gap}} \hspace{-4mm} = - \int V_{n_1, n_2, 0, 0} (\mathbf{p} -
        \mathbf{p}') \, \frac{\partial \mathcal{K}}{\partial
          \xi}[\xi(\mathbf{p}')] \, \Sigma^{(1)}(\mathbf{p}') \,
        \Delta(\mathbf{p}') \frac{d
          \mathbf{p}'}{(2 \pi)^2}, \\
        \label{eq:70}
        \parbox{25mm}{\begin{fmfgraph*}(25,10)
            \fmfleft{lb,lt}
            \fmfright{rb,rt}
            \fmftop{ct} \fmfbottom{cb}
            \fmf{fermion}{lt,v1lt}
            \fmf{fermion}{lb,v1lb}  
            \fmf{phantom}{cb,v1rb}
            \fmf{phantom}{v1rt,ct}
            \fmf{phantom}{ct,v2lt}
            \fmf{phantom}{v2lb,cb}
            \fmf{phantom}{rb,v2rb}
            \fmf{phantom}{v2rt,rt}
            \fmfpoly{smooth,tension=.8,label=\tiny $\delta V_d$}{v1rt,v1lt,v1lb,v1rb}
            \fmfpoly{smooth,tension=1.5,label=\tiny $\Delta$}{v2rt,v2lt,v2lb,v2rb}
            \fmffreeze  
            \fmfi{fermion}{vloc(__v1rb){(1,-1)} .. {(1,1)}vloc(__v2lb)}
            \fmfi{fermion}{vloc(__v1rt){(1,1)} .. {(1,-1)}vloc(__v2lt)}
          \end{fmfgraph*}} \hspace{-4mm} = - \int \delta V_{n_1, n_2, 0, 0}
        (\mathbf{p} - \mathbf{p}') \, \mathcal{K}[\xi(\mathbf{p}')] \,
        \Delta(\mathbf{p}') \frac{d \mathbf{p}'}{(2 \pi)^2},
      \end{gather}
    \end{widetext}where $\mathcal{K}_{n,n^{\prime }}(\mathbf{p})$ and $%
\mathcal{K}[\xi (\mathbf{p})]$ are given by Eqs.~\eqref{eq:64} and %
\eqref{eq:65}, and \eqref{eq:66} respectively.

\subsection{Renormalization}

\label{sec:renormalization} Apart from the region $p^{\prime }\approx
p_{F}^{(0)}$, major contributions to the integral in \eqref{eq:68} that are
actually divergent and need to be cut off come also from high momenta. This
region, however, is related to short interparticle distances, at which the
presence of other particles becomes irrelevant and the dynamics corresponds
to the scattering of just two particles in vacuum. In the gap equation this
two-body physics can be taken into account by expressing the bare
dipole-dipole interaction in terms of the vertex function $\gamma
_{n_{1},n_{2},n_{3},n_{4}}(E,\mathbf{p},\mathbf{p}^{\prime })$ for two
particles in vacuum with a total energy $E$ in their center-of-mass
reference frame. The Lippmann-Schwinger equation for the vertex function 
\cite{Taylor} can be represented diagrammatically as 
\begin{equation}
\parbox{12mm}{\begin{fmfgraph*}(10,10)
          \fmfleft{lb,lt}
          \fmfright{rb,rt}
          \fmf{fermion}{lt,vlt}
          \fmf{fermion}{lb,vlb}  
          \fmf{fermion}{vrb,rb}
          \fmf{fermion}{vrt,rt}
          \fmfpoly{smooth,tension=.8,label=\small $\gamma$}{vrt,vlt,vlb,vrb}
        \end{fmfgraph*}}=%
\parbox{10mm}{\begin{fmfgraph}(10,10)
          \fmfleft{lb,lt}
          \fmfright{rb,rt}
          \fmf{fermion}{lt,vt}
          \fmf{fermion}{lb,vb}  
          \fmf{fermion}{vb,rb}
          \fmf{fermion}{vt,rt}
          \fmf{wiggly}{vt,vb}  
        \end{fmfgraph}}+%
\parbox{20mm}{\begin{fmfgraph*}(20,10)
          \fmfleft{lb,lt}
          \fmfright{rb,rt}
          \fmftop{ct} \fmfbottom{cb}
          \fmf{fermion}{lt,v1t}
          \fmf{fermion}{lb,v1b}  
          \fmf{phantom}{v1b,cb}
          \fmf{phantom}{v1t,ct}
          \fmf{phantom}{ct,v2lt}
          \fmf{phantom}{cb,v2lb}
          \fmf{fermion}{v2rb,rb}
          \fmf{fermion}{v2rt,rt}
          \fmf{wiggly}{v1t,v1b}  
          \fmfpoly{smooth,tension=.8,label=$\gamma$}{v2rt,v2lt,v2lb,v2rb}
          \fmffreeze
          \fmfi{fermion}{vloc(__v1b){(1,-1)} .. {(1,1)}vloc(__v2lb)}
          \fmfi{fermion}{vloc(__v1t){(1,1)} .. {(1,-1)}vloc(__v2lt)}    
        \end{fmfgraph*}},  \label{eq:71}
\end{equation}%
where particle lines correspond to non-interacting zero temperature Green's
functions, 
\begin{equation}
\begin{split}
G^{(0)}(\omega ,\mathbf{p},z,z^{\prime })& =\sum_{n=0}^{\infty }\phi
_{n}(z)\,\phi _{n}(z^{\prime })\,G_{n}^{(0)}(\omega ,\mathbf{p}), \\
G_{n}^{(0)}(\omega ,\mathbf{p})& =\frac{1}{\omega -p^{2}/2m-\omega _{0}n+i0}.
\end{split}
\label{eq:72}
\end{equation}%
For our purposes it is convenient to rearrange the order of terms on the RHS
of \eqref{eq:71} as 
\begin{multline}
\gamma _{n_{1},n_{2},n_{3},n_{4}}(E,\mathbf{p},\mathbf{p}^{\prime
})=V_{n_{1},n_{2},n_{3},n_{4}}(\mathbf{p}-\mathbf{p}^{\prime })
\label{eq:73} \\
-\int \sum_{n_{5},n_{6}}\gamma _{n_{1},n_{2},n_{5},n_{6}}(E,\mathbf{p},%
\mathbf{q})\,K_{n_{5},n_{6}}(E,\mathbf{q}) \\
\times V_{n_{5},n_{6},n_{3},n_{4}}(\mathbf{q}-\mathbf{p}^{\prime })\frac{d%
\mathbf{q}}{(2\pi )^{2}},
\end{multline}%
where the kernel $K_{n,n^{\prime }}(E,\mathbf{p})$ is given by the integral
over frequencies, 
\begin{equation}
\begin{split}
K_{n,n^{\prime }}(E,\mathbf{p})& =-i\int G_{n}^{(0)}(E+\omega ,\mathbf{p}%
)\,G_{n^{\prime }}^{(0)}(-\omega ,-\mathbf{p})\frac{d\omega }{2\pi } \\
& =\frac{1}{p^{2}/m+\omega _{0}(n+n^{\prime })-E-i0}.
\end{split}
\label{eq:74}
\end{equation}%
Below we shall choose $E=2\mu $, in which case \eqref{eq:74} becomes 
\begin{equation}
K_{n,n^{\prime }}(2\mu ,\mathbf{p})=\frac{1}{2\,\xi (\mathbf{p})+\omega
_{0}(n+n^{\prime })-i0}.  \label{eq:75}
\end{equation}

In order to carry out the renormalization in the most transparent way we
rewrite the gap equation \eqref{eq:55} and the Lippmann-Schwinger equation %
\eqref{eq:73} schematically as 
\begin{align} 
\Delta & =-V\mathcal{K}\Delta -V\partial \mathcal{K}\Sigma ^{(1)}\Delta
-\delta V\mathcal{K}\Delta, \label{eq:76} \\ 
\gamma & =(1-\gamma K)V. \label{eq:77}
\end{align}%
We \textquotedblleft multiply\textquotedblright\ \eqref{eq:76} from the left
by $1-\gamma K$. Then, using \eqref{eq:77} and neglecting terms that contain 
$\gamma V\Sigma ^{(1)}$ and $\gamma \delta V$ (those are contributions of
third order), we obtain the renormalized gap equation 
\begin{equation}
\Delta =-\gamma (\mathcal{K}-K)\Delta -V\partial \mathcal{K}\Sigma
^{(1)}\Delta -\delta V\mathcal{K}\Delta .  \label{eq:78}
\end{equation}

In the renormalized gap equation, the kernel $\mathcal{K}_{n,n^{\prime }}(%
\mathbf{p})$ has been replaced by the difference $\mathcal{K}_{n,n^{\prime
}}(\mathbf{p})-K_{n,n^{\prime }}(2\mu ,\mathbf{p})$ which converges rapidly
at high energies, i.\thinspace e., high $p$ and large HO quantum numbers $n$
and $n^{\prime }$. To wit, for $n=n^{\prime }=0$, from Eq.~\eqref{eq:66} we
have 
\begin{equation}
\mathcal{K}_{0,0}(\mathbf{p})-K_{0,0}(2\mu ,\mathbf{p})=\frac{\tanh [\xi (%
\mathbf{p})/2T]}{2\,\xi (\mathbf{p})}-\frac{1}{2\,\xi (\mathbf{p})-i0},
\label{eq:79}
\end{equation}%
which decays as $1/p^{4}$ for $p\rightarrow \infty $, ensuring the
convergence of the integral over the momentum in the first term on the RHS
of \eqref{eq:78} without the need to introduce an additional cut-off. If
either $n$ or $n^{\prime }$ is nonzero, with \eqref{eq:65} we have 
\begin{equation}
\mathcal{K}_{n,0}(\mathbf{p})-K_{n,0}(2\mu ,\mathbf{p})\approx -\frac{\theta
[ -\xi (\mathbf{p})]}{2\,\xi (\mathbf{p})+\omega _{0}n}.  \label{eq:80}
\end{equation}%
In \eqref{eq:78} this results in a term that is $O(\mu /\omega _{0})$ and
may safely be neglected. Finally, for both $n$ and $n^{\prime }$ not equal
to zero, with the aid of \eqref{eq:64} we obtain 
\begin{equation}
\mathcal{K}_{n,n^{\prime }}(\mathbf{p})-K_{n,n^{\prime }}(2\mu ,\mathbf{p}%
)\approx 0.  \label{eq:81}
\end{equation}%
Therefore, in the propagators in \eqref{eq:78}, we may restrict ourselves to
the HO quantum numbers being equal to zero, and we obtain a closed equation
for $\Delta (\mathbf{p})\equiv \Delta _{0,0}(\mathbf{p})$: With $\gamma (E,%
\mathbf{p},\mathbf{p}^{\prime })\equiv \gamma _{0,0,0,0}(E,\mathbf{p},%
\mathbf{p}^{\prime })$ and $\delta V_{0}(\mathbf{p},\mathbf{p}^{\prime
})\equiv \delta V_{0,0,0,0}(\mathbf{p},\mathbf{p}^{\prime })$ we have 
\begin{multline}
\Delta (\mathbf{p})=-\int \biggl \{\gamma (2\mu ,\mathbf{p},\mathbf{p}%
^{\prime })\biggl[\mathcal{K}(\xi ^{\prime })-\frac{1}{2\xi ^{\prime }-i0}%
\biggr]  \label{eq:82} \\
+V_{0}(\mathbf{p}-\mathbf{p}^{\prime })\,\Sigma ^{(1)}(\mathbf{p}^{\prime
})\,\frac{\partial \mathcal{K}}{\partial \xi }(\xi ^{\prime }) \\
+\delta V_{0}(\mathbf{p},\mathbf{p}^{\prime })\,\mathcal{K}(\xi ^{\prime })%
\biggr \}\,\Delta (\mathbf{p}^{\prime })\frac{d\mathbf{p}^{\prime }}{(2\pi
)^{2}}.
\end{multline}%
To proceed we need to find an expression for the vertex function $\gamma
(2\mu ,\mathbf{p},\mathbf{p}^{\prime })$.

\subsection{Two-body vertex function}

\label{sec:two-body-vertex} Iteration of Eq.~\eqref{eq:73} yields the
familiar Born series. Terminating this series at second order we obtain for $%
n_{1}=n_{2}=n_{3}=n_{4}=0$ 
\begin{equation}
\gamma (2\mu ,\mathbf{p},\mathbf{p}^{\prime })=\gamma ^{(1)}(2\mu ,\mathbf{p}%
,\mathbf{p}^{\prime })+\gamma ^{(2)}(2\mu ,\mathbf{p},\mathbf{p}^{\prime }),
\label{eq:83}
\end{equation}%
where the first order contribution is just the Fourier transform of the
effective 2D dipole-dipole interaction 
\begin{equation}
\gamma ^{(1)}(2\mu ,\mathbf{p},\mathbf{p}^{\prime })=V_{0}(\mathbf{p}-%
\mathbf{p}^{\prime }),  \label{eq:84}
\end{equation}%
and to second order we have to include virtual excitations to higher HO
levels at the intermediate stage of the interaction, i.\thinspace e., we
have to take the sum over $n,n^{\prime }\in \mathbb{N}_{0}$, 
\begin{multline}
\gamma ^{(2)}(2\mu ,\mathbf{p},\mathbf{p}^{\prime })=  \label{eq:85} \\
-\sum_{n,n^{\prime }}\int \frac{V_{n,n^{\prime }}(\mathbf{p}-\mathbf{q}%
)\,V_{n,n^{\prime }}(\mathbf{q}-\mathbf{p}^{\prime })}{2\,\xi (\mathbf{q}%
)+\omega _{0}(n+n^{\prime })-i0}\frac{d\mathbf{q}}{(2\pi )^{2}},
\end{multline}%
where $V_{n,n^{\prime }}(\mathbf{p})\equiv V_{0,0,n,n^{\prime }}(\mathbf{p}%
)=V_{n,n^{\prime },0,0}(\mathbf{p})$. Explicit expressions for these matrix
elements -- calculated exactly as well as in WKB approximation -- are given
in App.~\ref{sec:matr-elem-vdd}.

\subsection{Asymptotic gap equation}

\label{sec:asympt-gap-equat} In the integral on the RHS of Eq.~\eqref{eq:82}
we perform the change of variables 
\begin{equation}  \label{eq:86}
\mathbf{p} = \mathbf{p}(\xi ,\phi) = \sqrt{2 m (\xi + \varepsilon_F^{(0)})}
\, \hat{\mathbf{p}},
\end{equation}
where $\hat{\mathbf{p}} = (\cos \phi ,\sin \phi)$. We shall simplify the
notation by introducing the functions (avoiding to explicitly state the
dependence on $\xi$ and $\phi$) 
\begin{equation}  \label{eq:87}
\begin{split}
f(\xi^{\prime}) & \equiv \int \frac{d \phi^{\prime}}{2 \pi} \, \nu \left[
\gamma(2 \mu, \mathbf{p}, \mathbf{p}^{\prime})+ \delta V_0(\mathbf{p}, 
\mathbf{p}^{\prime}) \right] \, \Delta (\xi^{\prime},\phi^{\prime}) \\
g(\xi^{\prime}) & \equiv \int \frac{d \phi^{\prime}}{2 \pi} \, \nu \,
\gamma(2 \mu, \mathbf{p}, \mathbf{p}^{\prime}) \, \Delta
(\xi^{\prime},\phi^{\prime}), \\
h(\xi^{\prime}) & \equiv \int \frac{d \phi^{\prime}}{2 \pi} \, \nu \, V_0(%
\mathbf{p} - \mathbf{p}^{\prime}) \, \Sigma^{(1)}( \mathbf{p}^{\prime}) \,
\Delta (\xi^{\prime},\phi^{\prime}).
\end{split}%
\end{equation}
With these definitions we can write the gap equation \eqref{eq:82} in a very
compact form, 
\begin{equation}  \label{eq:88}
\Delta = -\int_{-\mu}^{\infty} d \xi^{\prime}\left[ \mathcal{K}%
(\xi^{\prime}) \, f(\xi^{\prime}) + \mathcal{K}^{\prime}(\xi^{\prime}) \,
h(\xi^{\prime}) - \frac{g(\xi^{\prime})}{2 \xi - i 0} \right].
\end{equation}
In the second term on the RHS we integrate by parts. Neglecting boundary
terms that are of second order in the dipole-dipole interaction and are not
multiplied by the large logarithm $\ln(T/\mu)$, we obtain 
\begin{equation}  \label{eq:89}
\Delta = -\int_{-\mu}^{\infty} d \xi^{\prime}\left\{ \mathcal{K}%
(\xi^{\prime}) \left[ f(\xi^{\prime}) - h^{\prime}(\xi^{\prime}) \right] - 
\frac{g(\xi^{\prime})}{2\xi -i 0} \right\}.
\end{equation}

In the limit $T/\mu \ll 1$ the function $\mathcal{K}(\xi^{\prime})$ behaves
as $\mathcal{K}(\xi^{\prime}) \sim 1/2 \lvert \xi^{\prime}\rvert$. Hence,
the main contribution to the integral over the first term on the RHS of Eq.~%
\eqref{eq:89}, as has already been discussed above, is logarithmic in $T/\mu 
$ and comes from states near the Fermi surface where $\lvert \xi^{\prime}
\rvert \ll \mu$. In order to single out this contribution we divide the
integral over $\xi^{\prime}$ into two parts: (a) the integration of $%
\mathcal{K}(\xi^{\prime}) \left[ f(0) - h^{\prime}(0) \right]$ from $-\mu$
to $\mu$, and (b) the sum of the integrals of $\mathcal{K}(\xi^{\prime}) %
\left[ f(\xi^{\prime}) - h^{\prime}(\xi^{\prime}) - f(0) + h^{\prime}(0) %
\right]$ over $[-\mu ,\mu)$ and of $\mathcal{K} (\xi^{\prime}) \left[
f(\xi^{\prime}) - h^{\prime}(\xi^{\prime}) \right]$ over $(\mu, \infty)$. In
part (a) we use the asymptotic formula 
\begin{equation}  \label{eq:90}
\int_{-\mu}^{\mu} d \xi^{\prime}\, \frac{\tanh(\xi^{\prime}/2T)}{2
\xi^{\prime}} \sim \ln \biggl( \frac{2 \, \mathrm{e}^{\gamma} \mu}{\pi T} %
\biggr),
\end{equation}
where $\gamma \approx 0.5772$ is the Euler constant, and in part (b) we once
more [cf.~Eq.~\eqref{eq:65}] replace $\tanh(\xi^{\prime}/2T)$ by $\sgn%
(\xi^{\prime})$ and integrate by parts. Consequently, keeping terms that are 
$O(g^2)$ only when they are multiplied by the large logarithm $\ln(T/\mu)$,
Eq.~\eqref{eq:89} takes the form, 
\begin{multline}  \label{eq:91}
\Delta = -\ln \biggl( \frac{2 \mathrm{e}^{\gamma} \mu}{\pi T} \biggr) \left[
f(0) - h^{\prime}(0) \right] \\
+ \frac{1}{2} \int_{-\mu}^{\infty} d \xi^{\prime}\ln \! \left| \frac{%
\xi^{\prime}}{\mu} \right| \partial_{\xi^{\prime}} \left[ f(\xi^{\prime}) -
h^{\prime}(\xi^{\prime}) - g(\xi^{\prime}) \right] \\
-\int_{-\mu}^0 d \xi^{\prime}\ln \! \left| \frac{\xi^{\prime}}{\mu} \right|
\partial_{\xi^{\prime}} \left[ f(\xi^{\prime}) - h^{\prime}(\xi^{\prime}) %
\right] + i \frac{\pi}{2} \, g(0).
\end{multline}
The value $\xi = 0$ corresponds to the momentum 
\begin{equation}  \label{eq:92}
p = \sqrt{2 m \mu} = \sqrt{2 m (\varepsilon_F^{(0)}+ \delta \mu)} \approx
p_F^{(0)} + \frac{m}{p_F^{(0)}} \delta \mu,
\end{equation}
where $\delta \mu$ is the first order correction to the chemical potential %
\eqref{eq:28}. Inserting this expansion as well as the explicit expressions %
\eqref{eq:87} for $f$, $g$ and $h$ in the gap equation \eqref{eq:91} we
obtain 
\begin{widetext}  
      \begin{multline}
        \label{eq:93}
        \Delta(\xi,\phi) = - \ln \biggl( \frac{2 \, \e^{\gamma}
          \varepsilon_F^{(0)}}{\pi T} \biggr) \int \frac{d \phi'}{2 \pi} \, \nu
        \, \biggl\{ \biggl[ V_0(\mathbf{p} - {\mathbf{p}_F^{(0)}}') +
        \frac{m}{p_F^{(0)}} \frac{\partial V_0(\mathbf{p} -
          {\mathbf{p}_F^{(0)}}')}{\partial p'} \, \delta \mu + \gamma^{(2)}(2
        \varepsilon_F^{(0)}, \mathbf{p}, {\mathbf{p}_F^{(0)}}') \\ + \delta
        V_0(\mathbf{p}, {\mathbf{p}_F^{(0)}}') \biggr] \, \Delta(0,\phi') -
        \frac{\partial}{\partial \xi'} \Bigl[ V_0(\mathbf{p} - \mathbf{p}') \,
        \Sigma^{(1)}(\mathbf{p}') \, \Delta(\xi',\phi') \Bigr]_{\xi' = 0}
        \biggr\} \\ - \int_{-\varepsilon_F^{(0)}}^0 d \xi' \ln \!  \left\lvert
          \frac{\xi'}{\varepsilon_F^{(0)}} \right\rvert \frac{\partial}{\partial
          \xi'} \biggl[ \int \frac{d \phi'}{2 \pi} \, \nu \, V_0(\mathbf{p} -
        \mathbf{p}') \, \Delta(\xi',\phi') \biggr] + i \frac{\pi}{2} \int
        \frac{d \phi'}{2 \pi} \, \nu \, V_0(\mathbf{p} - {\mathbf{p}_F^{(0)}}')
        \, \Delta(0,\phi').
      \end{multline}
    \end{widetext}

Our calculation of the critical temperature is carried out in two steps:
First we omit all terms on the RHS of Eq.~\eqref{eq:93} but the first one.
This is equivalent to the commonly used BCS approach and allows us to obtain
the controlling factor for the dependence of $T_c$ on $g$ as well as the
leading behavior of $\Delta(\xi ,\phi)$ in the limit $g \ll 1$. We will find
that $T_c \propto \varepsilon_F^{(0)} \exp(1/\lambda_0)$, where $\lambda_0
\propto g$. The calculation of the prefactor of $\exp(1/\lambda_0) $ in the
second step requires us to take account of all terms on the RHS of Eq.~%
\eqref{eq:93}, which is referred to as the GM approach.

\subsection{Critical temperature in the BCS approach}

\label{sec:crit-temp-bcs} Keeping only the dominant contribution on the RHS
of Eq.\,\eqref{eq:93}, we have 
\begin{equation}  \label{eq:94}
\Delta(\xi,\phi) = \ln \biggl( \frac{T}{\varepsilon_F^{(0)}} \biggr) \int 
\frac{d \phi^{\prime}}{2 \pi} \, \nu \, V_0(\mathbf{p} - {\mathbf{p}_F^{(0)}}%
^{\prime}) \, \Delta(0,\phi^{\prime}).
\end{equation}
We expand $\Delta(0,\phi)$ in terms of a complete set of eigenfunctions $%
\Phi_s(\phi)$, $s \in \mathbb{N}_0$ of the integral operator with the kernel 
$V_0(\mathbf{p}_F^{(0)} - {\mathbf{p}_F^{(0)}}^{\prime})$, which for $%
p_F^{(0)} l_0 \ll 1$ can be written as [cf.~Eq.~\eqref{eq:4}] 
\begin{multline}
V_0(\mathbf{p}_F^{(0)} - {\mathbf{p}_F^{(0)}}^{\prime}) \approx - (g/\nu)
\sin(\lvert \phi - \phi^{\prime}\rvert/2)  \label{eq:95} \\
\times \left[ \cos (\phi + \phi^{\prime}) \sin^2 \! \theta_0 +2 \, P_2(\cos
\theta_0) \right].
\end{multline}
The expansion of $\Delta(0,\phi)$ reads 
\begin{equation}  \label{eq:96}
\Delta(0,\phi) = \sum_{s = 0}^{\infty} \Delta_s \, \Phi_s(\phi),
\end{equation}
where the functions $\Phi_s(\phi)$ satisfy the eigenvalue equation 
\begin{equation}  \label{eq:97}
\int \frac{d \phi^{\prime}}{2 \pi} \, \nu \, V_0( \mathbf{p}_F^{(0)} - {%
\mathbf{p}_F^{(0)}}^{\prime}) \, \Phi_s(\phi ^{\prime}) = \lambda_s \,
\Phi_s(\phi),
\end{equation}
and are normalized to unity according to $\int (d \phi /2 \pi) \,
\Phi_s(\phi)^2 = 1$. We label the eigenvalues $\lambda_s$ such that $%
\lambda_s < \lambda_{s^{\prime}}$ for $s < s^{\prime}$. Inserting the
expansion \eqref{eq:96} in Eq.~\eqref{eq:94} and specifying the resulting
equation to $\xi = 0$ we obtain the set of equations, for $s \in \mathbb{N}%
_0 $, 
\begin{equation}  \label{eq:98}
\left[ 1 - \lambda_s \ln(T/\varepsilon_F^{(0)}) \right] \Delta_s = 0.
\end{equation}
Thus the existence of a non-trivial solution for $\Delta(0,\phi)$ requires
at least the smallest eigenvalue $\lambda_0$ to be negative [note that $%
\ln(T/\varepsilon_F^{(0)}) < 0$ since $T/\varepsilon_F^{(0)} \ll 1$]. Then
the controlling factor of the critical temperature follows immediately from $%
1 - \lambda_0 \ln(T_c/\varepsilon_F^{(0)}) = 0$, and we have $\Delta_0 \neq
0 $ whereas $\Delta_s = 0$ for $s \neq 0$.

\begin{figure}[tbp]
\psfrag{x1}[t][]{\scriptsize $\pi/4$} 
\psfrag{x2}[t][]{\scriptsize $3
        \pi/8$} \psfrag{x3}[t][]{\scriptsize $\pi/2$} 
\psfrag{y1}[r][]{\scriptsize
        $0$} \psfrag{y2}[r][]{\scriptsize $0.2$} 
\psfrag{y3}[r][]{\scriptsize
        $0.4$} \psfrag{y4}[r][]{\scriptsize $0.6$} 
\psfrag{y5}[r][]{\scriptsize
        $0.8$} \psfrag{y6}[r][]{\scriptsize $1$} 
\psfrag{xlabel}[t][b]{\small
        $\theta_0$} \psfrag{ylabel}[b][t]{\small $c_n$} \centering      
\includegraphics[width=\linewidth]{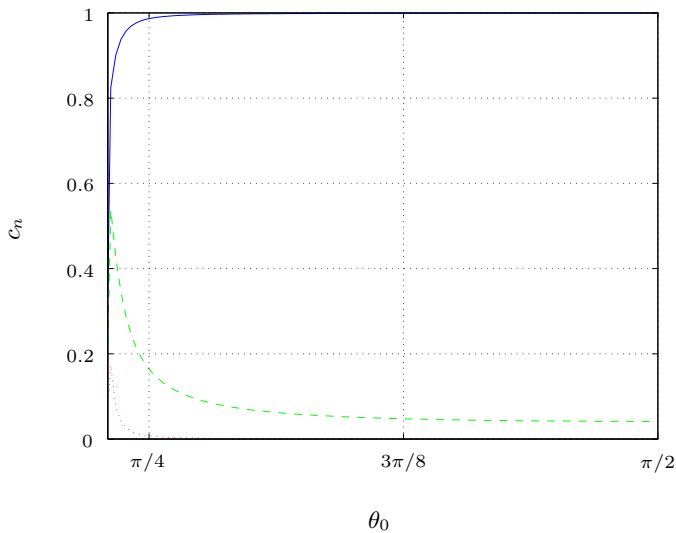}
\caption{(Color online) In order to solve \eqref{eq:97} numerically we expand
  $\Phi_0(%
  \protect\phi)$ in a Fourier cosine series [the integral operator in
  \eqref{eq:97} does not couple cosine and sine series and it turns out that $%
  \Phi_0(\protect\phi)$ can be represented as a cosine series; note that in this
  series expansions we retain only those terms that ensure the correct parity of
  the order parameter, $\Delta(\protect\xi, \protect\phi + \protect%
  \pi) = - \Delta(\protect\xi, \protect\phi)$], $\Phi_0(\protect\phi) =
  \protect\sqrt{2} \sum_{n = 1}^N c_n \cos[(2 n - 1) \protect\phi]$ with
  sufficiently large $N$, and solve the resulting linear system for the
  coefficients $c_n$. The above plot shows the coefficients $c_1$ (solid blue
  line), $c_2$ (dashed green line), and $c_3$ (dotted red line) as functions of
  $\protect\theta_0$ for $\protect%
  \theta_0 \geq \protect\theta_c$. }
\label{fig:gap_components}
\end{figure}
Solving the eigenvalue equation \eqref{eq:97} numerically reveals that, as
the tilting angle $\theta_0$ of the dipoles is increased, $\lambda_0$
becomes negative at $\theta_c \approx 0.72$, and for $\theta_0$ not too
close to $\theta_c$ the corresponding eigenfunction is well-approximated by 
\begin{equation}  \label{eq:99}
\Phi_0(\phi) \approx \sqrt{2} \cos(\phi),
\end{equation}
i.\,e., the order parameter has $p$-wave symmetry. Contributions from higher
partial waves are important only in the immediate vicinity of the critical
angle and may safely be neglected otherwise (see Fig.~\ref%
{fig:gap_components}). Within the approximation \eqref{eq:99}, the
eigenvalue $\lambda_0$ equals the diagonal matrix element 
\begin{equation}  \label{eq:100}
\begin{split}
\lambda_0 & \approx 2 \int \frac{d \phi}{2 \pi} \cos(\phi) \int \frac{d \phi
^{\prime}}{2 \pi} \, \nu \, V_0(\mathbf{p}_F^{(0)}-{\mathbf{p}_F^{(0)}}%
^{\prime}) \cos(\phi^{\prime}) \\
& = (g/\pi )(4/3 -3\sin^2 \! \theta_0).
\end{split}%
\end{equation}
In the BCS approach, therefore, for $\theta_0 \gtrsim \theta_c$ the critical
temperature is given by 
\begin{equation}  \label{eq:101}
T_c^{(\mathrm{BCS})} = \varepsilon_F^{(0)} \, \mathrm{e}^{-\pi/g \left| 4/3
-3\sin^2 \! \theta_0\right|},
\end{equation}
and Eq.~\eqref{eq:94} determines the order parameter as 
\begin{equation}  \label{eq:102}
\Delta(\xi ,\phi) = \frac{1}{\lambda_0} \int \frac{d \phi^{\prime}}{2 \pi}
\, \nu \, V_0(\mathbf{p} - {\mathbf{p}_F^{(0)}}^{\prime}) \,
\Phi_0(\phi^{\prime}).
\end{equation}

\subsection{Critical temperature in the GM approach}

\label{sec:crit-temp-gm} As we have pointed out earlier, within the BCS
approach it is possible to obtain only the controlling factor [the
exponential in Eq.~\eqref{eq:101}] of the dependence of $T_c$ on $g$. In
order to find the correct prefactor we proceed by substituting the ansatz 
\begin{equation}  \label{eq:103}
T_c = (\varpi/\varepsilon_F^{(0)}) \, T_c^{(\mathrm{BCS})}
\end{equation}
in Eq.~\eqref{eq:93}. We set $\xi = 0$ in the resulting equation, multiply
it by $\Phi_0(\phi)$ and take the integral over the angle $\phi$. Using the
normalization condition of the eigenfunction $\Phi_0$ we obtain to leading
order in $g$ 
\begin{multline}  \label{eq:104}
\ln \biggl( \frac{\varpi}{\varepsilon_F^{(0)}} \biggr) = - \frac{\delta
\lambda_0}{\lambda_0^2} - \ln \biggl( \frac{i \pi}{2 \, \mathrm{e}^{\gamma}} %
\biggr) \\
+ \frac{1}{\lambda_0} \int \frac{d \phi}{2 \pi} \, \Phi_0(\phi)
\int_{-\varepsilon_F^{(0)}}^0 d \xi^{\prime}\ln \! \left| \frac{\xi^{\prime}%
}{\varepsilon_F^{(0)}} \right| \\
\times \frac{\partial}{\partial \xi^{\prime}} \biggl[ \int \frac{d
\phi^{\prime}}{2 \pi} \, \nu \, V_0(\mathbf{p}_F^{(0)} - \mathbf{p}%
^{\prime}) \, \Delta(\xi^{\prime},\phi^{\prime}) \biggr],
\end{multline}
where the correction $\delta \lambda_0$ to the eigenvalue $\lambda_0$ is
composed of four contributions, 
\begin{equation}  \label{eq:105}
\delta \lambda_0 = \delta \lambda_0^{(\delta p_F)} + \delta
\lambda_0^{(\delta m)} + \delta \lambda_0^{(\mathrm{mb})} + \delta
\lambda_0^{(B)}.
\end{equation}
Physically, $\delta \lambda_0^{(\delta p_F)}$ and $\delta \lambda_0^{(\delta
m)}$ are due to the fact that the pairing occurs between quasi-particles and
not bare particles [these corrections contain the Fermi surface deformation %
\eqref{eq:29} and the effective mass \eqref{eq:31} respectively, see
discussion below], $\delta \lambda_0^{(\mathrm{mb})}$ has its origin in the
many-body corrections $\delta V_d$ to the bare interaction, and $\delta
\lambda_0^{(B)}$ incorporates the second order Born correction to the
two-body vertex function.

\paragraph{Self-energy corrections.}

The terms in \eqref{eq:105} that involve the self-energy $\Sigma^{(1)}$ and
the correction to the chemical potential $\delta \mu$ are 
\begin{multline}  \label{eq:106}
\delta \lambda_0^{(\delta p_F)}+ \delta \lambda_0^{(\delta m)} = \int \frac{%
d \phi}{2 \pi} \, \Phi_0(\phi) \int \frac{d \phi^{\prime}}{2 \pi} \, \nu \\
\times \biggl \{\frac{m}{p_F^{(0)}} \biggl[ \frac{\partial V_0(\mathbf{p}%
_F^{(0)}-{\mathbf{p}_F^{(0)}}^{\prime})}{\partial p}+ \frac{\partial V_0(%
\mathbf{p}_F^{(0)}-{\mathbf{p}_F^{(0)}}^{\prime})}{\partial p^{\prime}} %
\biggr] \delta \mu \, \Phi_0(\phi^{\prime}) \\
-\frac{\partial}{\partial \xi^{\prime}} \left[ V_0( \mathbf{p}_F^{(0)} - 
\mathbf{p}^{\prime}) \, \Sigma^{(1)}(\mathbf{p}^{\prime}) \,
\Delta(\xi^{\prime},\phi^{\prime}) \right]_{\xi^{\prime}= 0} \biggr \}.
\end{multline}
With the aid of Eqs.~\eqref{eq:97} and \eqref{eq:102}, the expression for $%
\delta \lambda_0^{(\delta p_F)}$can be put in the form 
\begin{multline}  \label{eq:107}
\delta \lambda_0^{(\delta p_F)} = 2\int \frac{d \phi}{2 \pi} \, \Phi_0(\phi)
\int \frac{d \phi^{\prime}}{2 \pi} \, \nu \, \delta p_F(\phi) \\
\times \hat{\mathbf{p}} \cdot \nabla V_0(\mathbf{p}_F^{(0)} - {\mathbf{p}%
_F^{(0)}}^{\prime}) \, \Phi_0(\phi^{\prime}),
\end{multline}
where the Fermi surface deformation $\delta p_F$ is given by Eq.~%
\eqref{eq:29} [the exponential smallness of $T_c$ allows us to use the
    zero temperature expression \eqref{eq:27} for the self-energy].

For $\delta \lambda_0^{(\delta m)}$ we find 
\begin{multline}  \label{eq:108}
\delta \lambda_0^{(\delta m)} = \int \frac{d \phi}{2 \pi} \, \Phi_0(\phi)
\int \frac{d \phi^{\prime}}{2 \pi} \, \nu \, \frac{\delta m(\phi^{\prime})}{m%
} \\
\times \, V_0(\mathbf{p}_F^{(0)} - {\mathbf{p}_F^{(0)}}^{\prime}) \,
\Phi_0(\phi^{\prime}),
\end{multline}
with $\delta m(\phi)$ given by Eq.~\eqref{eq:31}. Performing the angular
integrals in Eqs.~\eqref{eq:106} and \eqref{eq:108}, we find 
\begin{equation}  \label{eq:109}
\begin{split}
\delta \lambda_0^{(\delta p_F)} & = \tfrac{16}{5 \pi^2} g^2 \sin^2 \!
\theta_0 \left( \tfrac{1}{9} - \tfrac{1}{5} \sin^2 \! \theta_0 \right), \\
\delta \lambda_0^{(\delta m)} & = -\tfrac{2}{225 \pi^2} g^2 (200 -820 \sin^2
\! \theta_0 +829 \sin^4 \! \theta_0).
\end{split}%
\end{equation}

\paragraph{Many-body corrections.}

The leading many-body corrections are given by 
\begin{equation}  \label{eq:110}
\delta \lambda_0^{(\mathrm{mb})} = \nu \int \frac{d \phi}{2 \pi} \,
\Phi_0(\phi) \int \frac{d \phi^{\prime}}{2 \pi} \, \delta V_0(\mathbf{p}%
_F^{(0)}, {\mathbf{p}_F^{(0)}}^{\prime}) \, \Phi_0(\phi^{\prime}),
\end{equation}
where the analytical expressions corresponding to the diagrams \eqref{eq:57}
that make up $\delta V_0$ read 
\begin{multline}  \label{eq:111}
\delta V_0^{(a)}(\mathbf{p}, \mathbf{p}^{\prime}) = \left[ V_0(\mathbf{p}%
_{-}) \right]^2 \\
\times \int \frac{n(\mathbf{q} + \mathbf{p}_{-}/2) - n(\mathbf{q} - \mathbf{p%
}_{-}/2)}{\xi(\mathbf{q} + \mathbf{p}_{-}/2) - \xi(\mathbf{q} - \mathbf{p}%
_{-}/2) - i 0} \frac{d \mathbf{q}}{(2 \pi)^2},
\end{multline}
\vspace{-0.5cm} 
\begin{multline}  \label{eq:112}
\delta V_0^{(b)}(\mathbf{p}, \mathbf{p}^{\prime}) = -V_0(\mathbf{p}_{-})
\int V_0(\mathbf{q} - \mathbf{p}_{+}/2) \\
\times \frac{n(\mathbf{q} + \mathbf{p}_{-}/2) - n(\mathbf{q} - \mathbf{p}%
_{-}/2)}{\xi(\mathbf{q} + \mathbf{p}_{-}/2) - \xi(\mathbf{q} - \mathbf{p}%
_{-}/2) - i 0} \frac{d \mathbf{q}}{(2 \pi)^2},
\end{multline}
\vspace{-0.5cm} 
\begin{multline}  \label{eq:113}
\delta V_0^{(c)}(\mathbf{p}, \mathbf{p}^{\prime}) = -V_0(\mathbf{p}_{-})
\int V_0(\mathbf{q} + \mathbf{p}_{+}/2) \\
\times \frac{n(\mathbf{q} + \mathbf{p}_{-}/2) - n(\mathbf{q} - \mathbf{p}%
_{-}/2)}{\xi(\mathbf{q} + \mathbf{p}_{-}/2) - \xi(\mathbf{q} - \mathbf{p}%
_{-}/2) - i 0} \frac{d \mathbf{q}}{(2 \pi)^2},
\end{multline}
\vspace{-0.5cm} 
\begin{multline}  \label{eq:114}
\delta V_0^{(d)}(\mathbf{p}, \mathbf{p}^{\prime}) = -\int V_0(\mathbf{q} - 
\mathbf{p}_{-}/2) \, V_0(\mathbf{q} + \mathbf{p}_{-}/2) \\
\times \frac{n(\mathbf{q} + \mathbf{p}_{+}/2) - n(\mathbf{q} - \mathbf{p}%
_{+}/2)}{\xi(\mathbf{q} + \mathbf{p}_{+}/2) - \xi (\mathbf{q} - \mathbf{p}%
_{+}/2) - i 0} \frac{d \mathbf{q}}{(2 \pi)^2},
\end{multline}
(we are neglecting contributions that involve excited states of the
transverse trapping potential as they contain a additional factor of $%
\varepsilon_F^{(0)}/\omega_0\ll 1$). Here $\mathbf{p}_{\pm} = \mathbf{p} \pm 
\mathbf{p}^{\prime}$, and $n(\mathbf{p}) = \theta(p_F^{(0)} - p)$ is the
Fermi-Dirac distribution at zero temperature [the usage of $n(\mathbf{p})$
at zero temperature is justified by the exponential smallness of $T_c$].
Performing a numerical integration we find 
\begin{equation}  \label{eq:115}
\delta \lambda_0^{(\mathrm{mb})} = g^2 (0.37 - 1.67 \sin^2 \! \theta_0 +
1.82 \sin^4 \! \theta_0).
\end{equation}

\paragraph{Second order Born correction.}

The second order Born correction \eqref{eq:85} to the vertex function
results in the contribution 
\begin{equation}  \label{eq:116}
\delta \lambda_0^{(B)} = \int \frac{d \phi}{2 \pi} \, \Phi_0(\phi) \int 
\frac{d \phi^{\prime}}{2 \pi} \, \nu \, \gamma^{(2)}(2 \varepsilon_F^{(0)}, 
\mathbf{p}_F^{(0)}, {\mathbf{p}_F^{(0)}}^{\prime}) \, \Phi_0(\phi^{\prime})
\end{equation}
to $\delta \lambda_0$. We decompose \eqref{eq:116} as a sum of contributions
with fixed HO quantum numbers at the intermediate state of the second order
scattering process, $\delta \lambda_0^{(B)} = \sum_{n, n^{\prime}} \delta
\lambda_{n, n^{\prime}}^{( B)}$, where 
\begin{multline}  \label{eq:117}
\delta \lambda_{n, n^{\prime}}^{(B)} = -\nu \int \frac{d \phi}{2 \pi} \int 
\frac{d \phi^{\prime}}{2 \pi} \int \Phi_0(\phi) \, \Phi_0(\phi^{\prime}) \\
\times \frac{V_{n, n^{\prime}} \mskip-.5\thinmuskip (\mathbf{p}_F^{(0)} - 
\mathbf{q}) \, V_{n, n^{\prime}} \mskip-.5\thinmuskip (\mathbf{q} - {\mathbf{%
p}_F^{(0)}}^{\prime})}{2 \, \xi(\mathbf{q}) + \omega_0 (n + n^{\prime}) - i 0%
} \frac{d \mathbf{q}}{(2 \pi)^2}.
\end{multline}
For $n = n^{\prime}= 0$ we find the asymptotic expression 
\begin{multline}  \label{eq:118}
\delta \lambda_{0,0}^{(B)} \overset{\eta \rightarrow 0}{\sim} g^2 (0.35 -
1.56 \sin^2 \! \theta_0 + 1.79 \sin^4 \! \theta_0) \\
+ g^2 \ln(\eta) \left( \tfrac{1}{4} - \tfrac{9}{8} \sin^2 \! \theta_0 + 
\tfrac{41}{32} \sin^4 \! \theta_0 \right) - i \tfrac{\pi}{2} \lambda_0.
\end{multline}
Note that despite the smallness of $\eta = p_F l_0$, terms which are
proportional to $g^2 \ln(\eta)$ represent a small correction to $\lambda_0 =
O(g)$ in the limit where $g/\eta = r_d/l_0 \ll 1$.

We are left with the calculation of the contribution to the second order
Born correction that involves excited states of the transverse trapping
potential, $\delta \lambda_{*}^{(B)} = \delta \lambda_0^{(B)} - \delta
\lambda_{0,0}^{(B)}$. In terms of new summation indexes that may be
interpreted as ``relative'' and ``center of mass'' HO quantum numbers we put
it in the form 
\begin{multline}  \label{eq:119}
\delta \lambda_{*}^{(B)} = \sum_{N = 1}^{\infty} \sum_{n = -N}^{N} \delta
\lambda_{N + n, N - n}^{(B)} \\
+ \sum_{N = 1}^{\infty} \sum_{n = - N + 1}^{N} \delta \lambda_{N + n - 1, N
- n}^{(B)},
\end{multline}
i.\,e., we separate parts in which the sum of HO quantum numbers is even and
odd, respectively.

For the present purpose we rewrite the WKB matrix elements \eqref{eq:146}
and \eqref{eq:147} as 
\begin{multline}  \label{eq:120}
V_{N + n, N - n}(\mathbf{p}) = \sqrt{\frac{1}{2N}} \frac{g}{\pi \eta \nu} \, 
\mathrm{e}^{-n^2/2N} \\
\times \left[ u(\mathbf{p}) \sin^2 \! \theta_0 -2 \, P_2(\cos \theta_0) %
\right] \frac{(p l_0)^2}{(p l_0)^2 + 2 N},
\end{multline}
and 
\begin{multline}  \label{eq:121}
V_{N + n - 1, N - n}(\mathbf{p}) = 2i \frac{g}{\pi \eta \nu}(-1)^{1 +n} \, 
\mathrm{e}^{-(n - 1/2)^2/2 N} \\
\times \sin (2\theta_0) \, v(\mathbf{p}) \frac{p l_0}{(p l_0)^2 + 2 N}.
\end{multline}

Inserting these expressions in Eq.~\eqref{eq:117} and performing the
integrals we find, in the limit $\eta \rightarrow 0$, 
\begin{gather}
\delta \lambda _{N+n,N-n}^{(B)}=-(\tfrac{g}{\pi N})^{2}\,\mathrm{e}%
^{-n^{2}/N}\left( \tfrac{2}{3}-4\sin ^{2}\!\theta _{0}+\tfrac{43}{12}\sin
^{4}\!\theta _{0}\right) ,   \label{eq:122} \\
\delta \lambda _{N+n-1,N-n}^{(B)}=-\tfrac{7}{6}(\tfrac{g}{4\pi N})^{2}\,%
\mathrm{e}^{-(n-1/2)^{2}/N}\sin (2\theta _{0})^{2}.   \label{eq:123}
\end{gather}%
In the second line we are actually restricting ourselves to the asymptotic
behavior of $\delta \lambda _{N+n-1,N-n}^{(B)}$ for $N\rightarrow \infty $,
which, however, gives a sufficiently accurate approximation even for $N=1$.
We insert Eqs.~\eqref{eq:122} and \eqref{eq:123} in \eqref{eq:119} to obtain 
\begin{multline}
\delta \lambda _{\ast }^{(B)}=-\tfrac{1}{16\pi ^{2}}g^{2}  \label{eq:124} \\
\times \left[ S_{1}\left( \tfrac{2}{3}-4\sin ^{2}\!\theta _{0}+\tfrac{43}{12}%
\sin ^{4}\!\theta _{0}\right) +\tfrac{7}{6}S_{2}\sin (2\theta _{0})^{2}%
\right] ,
\end{multline}%
where the sums $S_{\alpha }$ for $\alpha =1,2$ are given by 
\begin{equation}
S_{\alpha }=\sum_{N=1}^{\infty }S_{\alpha }(N)/N^{2},  \label{eq:125}
\end{equation}%
with 
\begin{equation}
S_{1}(N)=\!\sum_{n=-N}^{N}\mathrm{e}^{-n^{2}/N},\quad
S_{2}(N)=\sum_{n=-N+1}^{N}\!\mathrm{e}^{-(n-1/2)^{2}/N}.  \label{eq:126}
\end{equation}%
These sums can be calculated semi-analytically with the result 
\begin{equation}
\delta \lambda _{\ast }^{(B)}=g^{2}\left( -0.02-0.01\sin ^{2}\!\theta
_{0}+0.03\sin ^{4}\!\theta _{0}\right) .  \label{eq:127}
\end{equation}

Adding the contributions of \eqref{eq:109}, \eqref{eq:115}, \eqref{eq:118},
and \eqref{eq:127} we obtain the correction to the eigenvalue $\lambda_0$, 
\begin{multline}  \label{eq:128}
\delta \lambda_0 = g^2 \left( 0.52 - 2.47 \sin^2 \! \theta_0 + 2.83 \sin^4
\! \theta_0 \right) \\
+ g^2 \ln(\eta) \left( 0.25 - 1.13 \sin^2 \! \theta_0 + 1.28 \sin^4 \!
\theta_0 \right).
\end{multline}

In the third term on the RHS of \eqref{eq:104} we express the gap $\Delta
(\xi^{\prime},\phi^{\prime})$ via \eqref{eq:102} and perform a numerical
integration. We find that this term gives a contribution 
\begin{equation}  \label{eq:129}
(g/\lambda_0)^2 \left( -0.14 + 0.63 \sin^2 \! \theta_0 - 0.71 \sin^4 \!
\theta_0 \right)
\end{equation}
to $\ln(\varpi /\varepsilon_F^{(0)})$.

The combination of \eqref{eq:104}, \eqref{eq:128} and \eqref{eq:129} yields
the final expression for the critical temperature in the GM approach ($%
\theta_0 \gtrsim \theta_c$), 
\begin{equation}  \label{eq:130}
T_c \approx \frac{2 \, \mathrm{e}^{\gamma} \varepsilon_{F }^{(0)}}{\pi} \,
f(\theta_0) \, \eta^{g(\theta_0)} \exp \bigl( -\pi/g \! \left| 4/3 -3 \sin^2
\! \theta_0 \right| \bigr),
\end{equation}
where 
\begin{equation}  \label{eq:131}
f(\theta_0) = \exp \biggl[ \frac{0.52 -2.47 \sin^2 \! \theta_0 +2.83 \sin^4
\! \theta_0}{0.18 -0.81 \sin^2 \! \theta_0 +0.91 \sin^4 \! \theta_0} \biggr],
\end{equation}
and 
\begin{equation}  \label{eq:132}
g(\theta_0) = \frac{0.25 -1.13 \sin^2 \! \theta_0 +1.28 \sin^4 \! \theta_0}{%
0.18 -0.81 \sin^2 \! \theta_0 +0.91 \sin^4 \! \theta_0}.
\end{equation}
Fig.~\ref{fig:TcKRb} shows the critical temperature as a function of the
tilting angle for values of $g$ and $\eta$ that correspond to a gas of polar
KRb molecules. The corresponding eigenfunction $\Phi_0(\phi)$ for the order
parameter is given by \eqref{eq:99}. 
\begin{figure}[tbp]
\psfrag{ylabel}[b][t]{\small $T_c/\mathrm{nK}$} 
\psfrag{xlabel}[t][]{\small
$\theta_0$} \psfrag{y1}[r][r]{\small $0$} \psfrag{y2}[r][r]{\small $1$} %
\psfrag{y3}[r][r]{\small $2$} \psfrag{y4}[r][r]{\small $3$} %
\psfrag{y5}[r][r]{\small $4$} \psfrag{x1}[t][t]{\small $0$} %
\psfrag{x2}[t][t]{\small $\pi/4$} \psfrag{x3}[t][t]{\small $\pi/2$} 
\centering
\includegraphics[width=.9\linewidth]{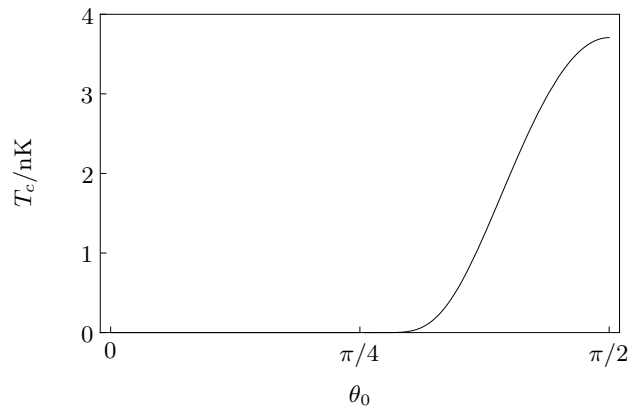}
\caption{The critical temperature as a function of $\protect\theta_0$ for $g
\approx 1.2$ and $\protect\eta \approx 0.2$. These values correspond to a
gas of KRb molecules, with $n_{2D} = 4 \cdot 10^8 \, \mathrm{cm}^{-2}$ and $%
\protect\omega_0 = 2 \protect\pi \cdot 100 \, \mathrm{kHz}$, see discussion
in Sec.~\protect\ref{sec:concluding-remarks}.}
\label{fig:TcKRb}
\end{figure}

\section{Summary and conclusion}

\label{sec:concluding-remarks} 
\begin{figure}[tbp]
  \psfrag{a}[t][b]{$g$} \psfrag{b}[b][t]{$\theta_0$} \psfrag{c}[t][]{\small $0$}
  \psfrag{d}[t][]{\small $0.5$} \psfrag{e}[t][]{\small $1$} %
  \psfrag{f}[t][]{\small $1.5$} \psfrag{g}[t][]{\small $2$} %
  \psfrag{h}[r][]{\small $0$} \psfrag{i}[r][]{\small $\pi/8$} %
  \psfrag{j}[r][]{\small $\pi/4$} \psfrag{k}[r][]{\small $3 \pi/8$} %
  \psfrag{l}[r][]{\small $\pi/2$} \psfrag{m}[][]{NFL} \psfrag{n}[][]{SF} %
  \psfrag{o}[][]{LWI} \psfrag{p}[][]{DW} \centering
\includegraphics[width=\linewidth]{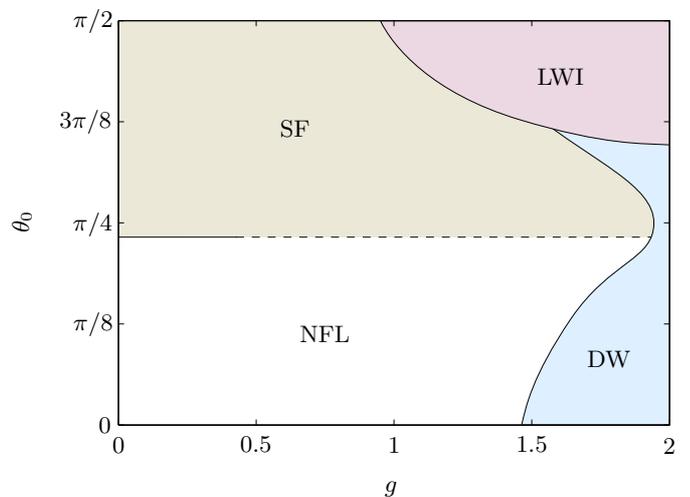}
\caption{(Color online) Phase diagram of the quasi-2D dipolar Fermi gas at $T =
  0$. For $%
  0\leq g \lesssim 1.45$ and small tilting angles the system is a normal Fermi
  liquid (NFL). The transition to the superfluid phase (SF) occurs at the
  critical angle $\protect\theta_c \approx 0.7217$. In our perturbative
  approach, which is -- strictly speaking -- valid only for $g \ll 1$, this
  value is independent of $g$. At moderately strong interactions there appear a
  density wave phase (DW) in the lower part of the phase diagram and a long
  wavelength instability (LWI) in the upper part.}
\label{fig:phase_diagram}
\end{figure}
In this paper, we studied various properties of a quasi-2D dipolar Fermi
gas. We found that the normal phase is characterized by an anisotropic Fermi
surface, which -- in order to minimize the interaction energy -- is
elongated along the projection of the dipoles on the plane of confinement
(here: the $x$-axis). Consequently, the dispersion relation of
single-particle excitations is anisotropic too, with an increased effective
mass in the $x$-direction and a decreased effective mass in the $y$%
-direction (see Fig.~\ref{fig:quasi-particle-dispersion}).

Long wavelength collective excitations (zero sound), corresponding to
deformations of the equilibrium Fermi surface, can propagate through the
medium only when the tilting angle exceeds a value of $\theta_0 \approx 0.5$%
, which depends only weakly on $g$ for (at least) $0.1 \leq g \leq 2$. In
this regime we found two distinct modes, symmetric and antisymmetric about
the propagation direction, respectively.

The experimental observation of zero sound requires $T$ so low that the
attenuation is negligible. Note, however, that for $\theta_0 > \theta_c
\approx 0.7$ (the critical angle for the superfluid transition) we also
require $T > T_c$ for the observation of zero sound, although under certain
conditions, zero sound can still persist below $T_c$ (see discussion in \cite%
{ronen10:_zero_fermi} and references given therein).

At higher values of $g$ and for large tilting angles, we found complex
eigenfrequencies of the long wavelength collective modes indicating the
instability of the system towards collapse in this part of the phase
diagram, see Fig.~\ref{fig:phase_diagram}. Instabilities with finite
momentum that drive the system towards a state with periodic density
modulation, which breaks translational symmetry, appear at small tilting
angles. In the isotropic case with $\theta_0 = 0$ (dipoles are aligned
perpendicular to the plane of confinement) also rotational symmetry is
broken. This phase can be detected experimentally using Bragg spectroscopy.

Finally, let us discuss prospects for observing the superfluid transitions
in the quasi-2D dipolar Fermi gas. In current experiments with KRb at JILA,
electric dipole moments of $d \approx 0.3 \, \mathrm{D}$ are available.
Then, assuming an area density of $n_{2D} \approx 4 \cdot 10^8 \, \mathrm{cm}%
^2$ corresponding to $\varepsilon_{F}^{(0)} \approx 100 \, \mathrm{nK}$, the
dimensionless coupling constant takes the value $g \approx 1.2$. Note that
strictly speaking this value of $g$ lies beyond the weak coupling regime in
which Eq.~\eqref{eq:130} is valid. However, this formula still provides an
estimate for the onset of superfluidity in the intermediate coupling regime.

The strong confinement condition is realized for a transverse trapping frequency
of $\omega_0 = 2 \pi \cdot 100 \, \mathrm{kHz}$, so that $\eta \approx 0.2$. For
the above value of $g$ the system is stable up to a tilting angle of $\theta_0
\approx 1.3$ (see Fig.~\ref{fig:phase_diagram}).  Then we have $T_c \approx 0.04
\, \varepsilon_F^{(0)} \approx 4 \, \mathrm{nK%
}$, which is comparable to the values of $T_c$ that are to be expected in
bilayer systems (see Refs.~\cite%
{pikovski10:_inter_super_bilay_system_fermion_polar_molec,%
  zinner10:_bcs_bec_cross_bilay_cold,lutchyn10:_spont,%
  potter10:_super_dimer_multil_system_fermion_polar_molec,%
  baranov11:_bilay,levinsen11:_topol}).

\section*{Acknowledgments}

The authors would like to thank G.\,V.\,Shlyapnikov, G.\,M.\,Bruun,
E.\,Demler, S.\,G.\,Hofer, A.\,W.\,Glaetzle and P.\,Zoller for insightful
discussions. This work was supported by the Austrian Science Fund FWF (SFB
FOQUS F 4015-N16). \appendix

\section{Calculation of the HF self-energy}

\label{sec:calculation-hf-self} In order to solve Eqs.~\eqref{eq:22} and %
\eqref{eq:24} for $p_F$ it is convenient introduce another unknown function $%
f(\phi)$ according to 
\begin{equation}  \label{eq:133}
f(\phi) \equiv p_F(\phi)^2/p_F^{(0)2} - 1.
\end{equation}
The expansion of $f$ in a Fourier series takes the form 
\begin{equation}  \label{eq:134}
f(\phi) = \sum_{n \in \mathbb{N}} c_n \cos(2 n \phi),
\end{equation}
i.\,e., the constant term is absent, as is immediately apparent by inserting
the definition \eqref{eq:133} in Eq.~\eqref{eq:23}. Moreover, due to the
symmetry of the problem, terms which are proportional to $\sin (n \phi)$ or $%
\cos(n \phi)$ with odd $n$ do not appear.

We solve for the Fourier coefficients $c_n$ by iteration: As initial values
we take the results from first order perturbation theory (see Sec.~\ref%
{sec:single-part-excit}), 
\begin{equation}  \label{eq:135}
c_n^{(1)} = 
\begin{cases}
\tfrac{16}{15 \pi} g \sin^2 \! \theta_0 & \text{for} \quad n = 1, \\ 
0 & \text{for} \quad n = 2,3,\dotsc .%
\end{cases}%
\end{equation}
The iteration scheme consists in repeatedly carrying out the integrals
[which result from the combination of Eqs.~\eqref{eq:22}, \eqref{eq:24}, %
\eqref{eq:133}, and \eqref{eq:134}] 
\begin{multline}  \label{eq:136}
c_n^{(i+1)} = \frac{1}{\pi^2} \int_0^1 dx \int d \phi \, d
\phi^{\prime}\cos(2 n \phi) \, x \left[ 1 + f^{(i)}(\phi^{\prime}) \right] \\
\times \nu \, V_0 \Bigl( p_F^{(0)} \sqrt{1 +f^{(i)}(\phi)} \, \hat{\mathbf{p}%
} - p_F^{(0)} x \sqrt{1 + f^{(i)}(\phi^{\prime})} \, \hat{\mathbf{p}}%
^{\prime}\Bigr),
\end{multline}
for given values of $g$ and $\theta_0$, where we keep coefficients up to $n
= 4$ [i.\,e., terms in the series \eqref{eq:134} up to $\cos(8\phi)$] and $%
f^{(i)}(\phi)$ is obtained by replacing $c_n$ in Eq.~\eqref{eq:134} by $%
c_n^{(i)}$. A measure for the convergence of this procedure is the relative
change in $f$ in one step of the iteration process, 
\begin{equation}  \label{eq:137}
\frac{\left\| f^{(i+1)} - f^{(i)} \right\|}{\left\| f^{(i)} \right\|} = 
\sqrt{\frac{\sum_n \left( c_n^{(i+1)} - c_n^{(i)} \right)^2}{ \sum_n
c_n^{(i)}}},
\end{equation}
and we terminate the iteration when this quantity drops below $10^{-3}$.

From the final result for $f$ we immediately obtain $p_F$ by inverting Eq.~%
\eqref{eq:133}. Then, $\Sigma$ and $\varepsilon$ follow from Eq.~%
\eqref{eq:19}, and by taking the gradient in \eqref{eq:19} we obtain the
Fermi velocity \eqref{eq:26} as 
\begin{equation}  \label{eq:138}
\mathbf{v}_F = \frac{1}{m} \biggl[ \mathbf{p}_F - \int \frac{d \phi^{\prime}%
}{2 \pi} \int_0^{p_F^{\prime}} dp^{\prime}\, p^{\prime}\, \nu \, \nabla V_0(%
\mathbf{p}_F - \mathbf{p}^{\prime}) \biggr].
\end{equation}

\section{Numerical solution to the Bethe-Salpeter equation}

\label{sec:numer-solut-bse} To solve Eq.~\eqref{eq:32} numerically it is
advantageous to rewrite it in a form that is symmetrized with respect to the
transferred momentum $\mathbf{q}$. To this end, we introduce the shifted
function $\tilde{\chi}_{\mathbf{q}}(\mathbf{p})\equiv \chi (\mathbf{p}-%
\mathbf{q}/2;\mathbf{q})$. Then we have 
\begin{multline}
\tilde{\chi}_{\mathbf{q}}(\mathbf{p})=\int \frac{d\mathbf{p^{\prime }}}{%
(2\pi )^{2}}\,\tilde{\Gamma}_{\mathrm{ph}}(\mathbf{p}-\mathbf{p}^{\prime },%
\mathbf{q})  \label{eq:139} \\
\times \frac{n(\mathbf{p}^{\prime }-\mathbf{q}/2)-n(\mathbf{p}^{\prime }+%
\mathbf{q}/2)}{\omega +\varepsilon (\mathbf{p}^{\prime }-\mathbf{q}%
/2)-\varepsilon (\mathbf{p}^{\prime }+\mathbf{q}/2)}\,\tilde{\chi}_{\mathbf{q%
}}(\mathbf{p}^{\prime }).
\end{multline}%
The factor $n(\mathbf{p}^{\prime }-\mathbf{q}/2)-n(\mathbf{p}^{\prime }+%
\mathbf{q}/2)$ in the numerator on the RHS restricts the area of integration
as depicted in Fig.~\ref{fig:phi_plus_minus}. Rewriting the integral in
polar coordinates $\mathbf{p}^{\prime }=p^{\prime }\,(\cos \phi ^{\prime
},\sin \phi ^{\prime })$, and in terms of a new function 
\begin{equation}
\nu _{\mathbf{q}}(\mathbf{p})\equiv \frac{\tilde{\chi}_{\mathbf{q}}(\mathbf{p%
})}{\omega +\varepsilon (\mathbf{p}-\mathbf{q}/2)-\varepsilon (\mathbf{p}+%
\mathbf{q}/2)},  \label{eq:140}
\end{equation}%
Eq.~\eqref{eq:139} becomes (for the definition of the angles $\phi _{\pm }$
that limit the $\phi ^{\prime }$-integration see Fig.~\ref%
{fig:phi_plus_minus}; $p_{\pm }(\phi ^{\prime })$ are the angle-dependent
upper and lower boundaries for the integration over $p^{\prime }$) 
\begin{figure}[tbp]
\psfrag{a}[][]{\small $\mathbf{q}/2$} \psfrag{b}[][]{\small $-\mathbf{q}/2$}
\psfrag{c}[][]{\small $\phi_{+}$} \psfrag{d}[][]{\small $\phi_{-}$} 
\centering
\includegraphics[width=\linewidth]{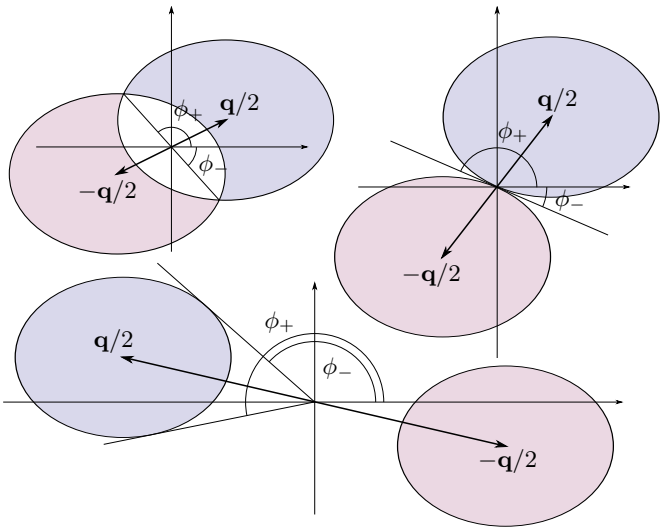}
\caption{(Color online) Area of integration in Eq.~\eqref{eq:139} and definition
  of the limiting angles $\protect\phi _{\pm }$, for three different values of
  $%
  \mathbf{q}$, with $\left| \mathbf{q}\right| <2\,p_{F}(\protect\phi _{\mathbf{%
      q}})$ (top left), $\left| \mathbf{q}\right| =2\,p_{F}(\protect\phi
  _{\mathbf{%
      q}})$ (top right), and $\left| \mathbf{q}\right| >2\,p_{F}(\protect\phi
  _{%
    \mathbf{q}})$ (bottom).}
\label{fig:phi_plus_minus}
\end{figure}
\begin{multline}
\left[ \omega +\varepsilon (\mathbf{p}-\mathbf{q}/2)-\varepsilon (\mathbf{p}+%
\mathbf{q}/2)\right] \nu _{\mathbf{q}}(\mathbf{p})  \label{eq:141} \\
=\frac{1}{2\pi }\int_{\phi _{-}}^{\phi _{+}}\frac{d\phi ^{\prime }}{2\pi }%
\int_{p_{-}(\phi ^{\prime })}^{p_{+}(\phi ^{\prime })}dp^{\prime
}\,p^{\prime }\,\tilde{\Gamma}_{\mathrm{ph}}(\mathbf{p}-\mathbf{p}^{\prime },%
\mathbf{q})\,\nu _{\mathbf{q}}(\mathbf{p}^{\prime }) \\
-\frac{1}{2\pi }\int_{\pi +\phi _{-}}^{\pi +\phi _{+}}\frac{d\phi ^{\prime }%
}{2\pi }\int_{p_{-}(\phi ^{\prime })}^{p_{+}(\phi ^{\prime })}dp^{\prime
}\,p^{\prime }\,\tilde{\Gamma}_{\mathrm{ph}}(\mathbf{p}-\mathbf{p}^{\prime },%
\mathbf{q})\,\nu _{\mathbf{q}}(\mathbf{p}^{\prime }).
\end{multline}%
We introduce a new variable $x\in \lbrack 0,1]$ that parametrizes the
momentum as $p(x,\phi )=p_{-}(\phi )+[p_{+}(\phi )-p_{-}(\phi )]\,x$, and
discretize the resulting integrals over $x^{\prime }$ and $\phi ^{\prime }$
according to the trapezoidal quadrature rule in two dimensions. Here we
choose a number of 20 grid points in the variable $x^{\prime }$ and $140$
grid points in $\phi ^{\prime }$. Then, for the values of the parameters of
Fig.~\ref{fig:cmdr}, an addition of 10 grid points in either variable leads
to an absolute change in the result for $\omega /\varepsilon _{F}^{(0)}$
that is less than $10^{-2}$.

\section{The matrix elements $V_{n, n^{\prime}}$}

\label{sec:matr-elem-vdd} 
\begin{figure}[tbp]
\psfrag{p}[][]{\small $p l_0$} 
\psfrag{Vddeven}[][b]{\small
    $V^{(\mathrm{r})}_{n, n'} (p)$} 
\psfrag{Vddodd}[][b]{\small
    $V^{(\mathrm{r})}_{n, n'} (p)$} \psfrag{y0p0}[r][r]{\small $0$}  %
\psfrag{y0p2}[r][r]{\small $0.2$} \psfrag{y0p4}[r][r]{\small $0.4$}  %
\psfrag{y0p6}[r][r]{\small $0.6$} \psfrag{y0p8}[r][r]{\small $0.8$}  %
\psfrag{y1p0}[r][r]{\small $1.0$} \psfrag{y1p2}[r][r]{\small $1.2$}  %
\psfrag{y0p5}[r][r]{\small $0.5$} \psfrag{y1p5}[r][r]{\small $1.5$}  %
\psfrag{y2p0}[r][r]{\small $2.0$} \psfrag{y2p5}[r][r]{\small $2.5$}  %
\psfrag{x00}[][]{\small $0$} \psfrag{x02}[][]{\small $2$}  %
\psfrag{x04}[][]{\small $4$} \psfrag{x06}[][]{\small $6$}  %
\psfrag{x08}[][]{\small $8$} \psfrag{x10}[][]{\small $10$} \centering
\includegraphics[width=.8
      \linewidth]{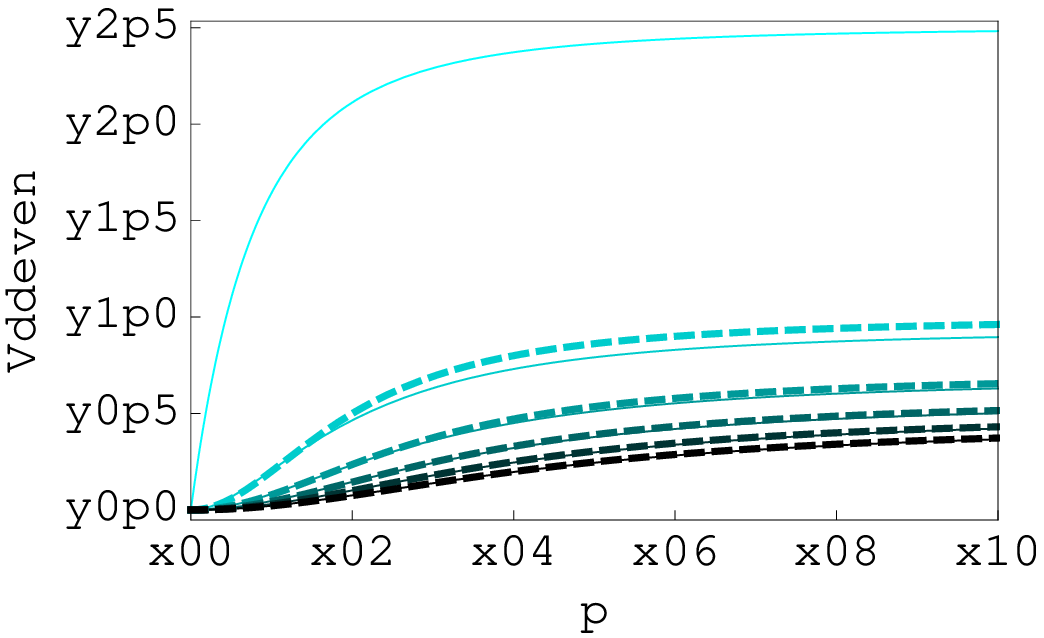} \\[4mm]
\includegraphics[width=.8\linewidth]{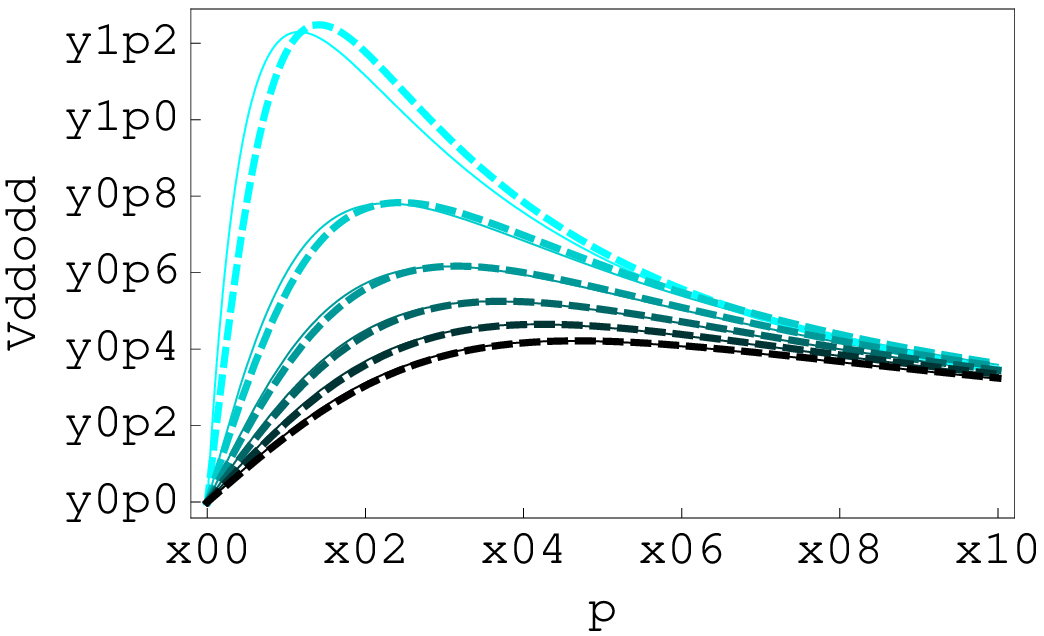}
\caption{(Color online) Top: Solid and dashed lines correspond to
  $V_{n,n}^{(\mathrm{r}%
    )}(p) $ with $n=0,2,\dotsc ,10$, calculated exactly and in WKB
  approximation, respectively. (We do not use the WKB approximation in the case
  $n=0$, hence there is no dashed line corresponding to uppermost solid line.)
  Bottom: Same as above, for $V_{n,n+1}^{(\mathrm{r})}(p)$ with $%
  n=0,2,\dotsc ,10$. In both figures higher values of $n$ correspond to darker
  colors.}
\label{fig:Vdd_matrix_elements}
\end{figure}
In this appendix we present explicit expressions for the matrix elements %
\eqref{eq:5} with $n_{3}=n_{4}=0$, which appear in the second order term %
\eqref{eq:85} in the Born series for the two-body vertex function. Omitting
the momentum independent contribution (see footnote \ref{fn:1}), with the
aid of the integral table \cite{Prudnikov/Brychkov/Marichev:II} we find, for 
$n+n^{\prime }\in 2\mathbb{N}_{0}$, 
\begin{multline}
V_{n,n^{\prime }}(\mathbf{p})=(2^{n+n^{\prime }}n!\,n^{\prime
}!)^{-1/2}\,\Gamma (\tfrac{1+n+n^{\prime }}{2})  \label{eq:142} \\
\times (d^{2}/l_{0})\left[ u(\mathbf{p})\sin ^{2}\!\theta _{0}-2\,P_{2}(\cos
\theta _{0})\right] \\
\times (pl_{0})^{1+n+n^{\prime }}\,\mathrm{e}^{p^{2}l_{0}^{2}/2}\,\Gamma (%
\tfrac{1-n-n^{\prime }}{2},p^{2}l_{0}^{2}/2),
\end{multline}%
where $u(\mathbf{p})\equiv (p_{x}^{2}-p_{y}^{2})/p^{2}$, and $\Gamma (z)$
and $\Gamma (\nu ,z)$ are the complete and incomplete Gamma functions
respectively. For $n+n^{\prime }\in 2\mathbb{N}_{0}+1$, with $v(\mathbf{p}%
)\equiv p_{1}/p$, we have 
\begin{multline}
V_{n,n^{\prime }}(\mathbf{p})=[(-1)^{n-n^{\prime }}/2^{n+n^{\prime
}}n!\,n^{\prime }!]^{1/2}\,\Gamma (1+\tfrac{n+n^{\prime }}{2})
\label{eq:143} \\
\times (2d^{2}/l_{0})\,v(\mathbf{p})\sin (2\theta _{0}) \\
\times (pl_{0})^{1+n+n^{\prime }}\,\mathrm{e}^{p^{2}l_{0}^{2}/2}\,\Gamma (-%
\tfrac{n+n^{\prime }}{2},p^{2}l_{0}^{2}/2).
\end{multline}

The relatively complicated functional dependence of the matrix elements $%
V_{n, n^{\prime}}$ on the momentum $\mathbf{p}$ prevents us from calculating
the integrals in \eqref{eq:117} analytically. We avoid this problem by using
WKB matrix elements \cite{LL:III} instead of the exact ones.

The WKB eigenfunctions of the one-dimensional HO, which give good
approximations to the exact eigenfunctions for $n \geq 1$, are \cite%
{baranov04:_bcs_pairin_in_trapp_dipol_fermi_gas} 
\begin{equation}  \label{eq:144}
\phi_n(z) = (-1)^{n} \sqrt{2m \omega_0/\pi \, p_n(z)} \cos[\Psi_n(z)-\pi/4],
\end{equation}
where the position dependent momentum $p_n(z)$ and the phase $\Psi_n(z)$ are
given by 
\begin{align}
p_n(z) & = \sqrt{2 m [ E_n - V(z) ]}, \\
\Psi_n(z) & = \int_{-z_n}^z d z^{\prime}\, p_n(z^{\prime}),
\end{align}
with the potential $V(z) = m \omega_0^2 z^2/2$ and the HO energy levels $E_n
= \omega_0 (n + 1/2)$. Above expression for $\phi_n(z)$ is valid in the
classically allowed region [i.\,e., for values of $z$ such that $E_n > V(z)$%
] and far away from the two turning points at $\pm z_n = \pm l_0\sqrt{2 n + 1%
}$, at which the kinetic energy is zero, $p_n(\pm z_n) = 0$ , or,
equivalently, $E_n = V(\pm z_n)$.

The HO ground state wave function decays exponentially on a scale that is
set by $l_{0}$. Therefore, the integration in the matrix elements $%
V_{n,n^{\prime }}$ is essentially restricted to the interval $\lvert z\rvert
\lesssim l_{0}$. For these values of $z$ we may use the approximations 
\begin{equation}
p_{n}(z)\approx p_{n}(0)\equiv p_{n}, \quad \Psi _{n}(z)\approx p_{n}z+(\pi
/4)(2n+1).
\end{equation}%
The WKB wave function \eqref{eq:144} then becomes 
\begin{equation}
\phi _{n}(z)\approx (-1)^{n}\sqrt{2m\omega _{0}/\pi p_{n}}\cos (p_{n}z+n\pi
/2).  \label{eq:145}
\end{equation}
Consequently, for the matrix elements $V_{n,n^{\prime }}$ in WKB
approximation we find 
\begin{multline}
V_{n,n^{\prime }}(\mathbf{p})\approx 2(n+n^{\prime })^{-1/2}\,\mathrm{e}%
^{-(n-n^{\prime })^{2}/4(n+n^{\prime })}(d^{2}/l_{0})  \label{eq:146} \\
\times \left[ u(\mathbf{p})\sin ^{2}\!\theta _{0}-2\,P_{2}(\cos \theta _{0})%
\right] \frac{(pl_{0})^{2}}{(pl_{0})^{2}+n+n^{\prime }}
\end{multline}%
for even $n+n^{\prime }$, and 
\begin{multline}
V_{n,n^{\prime }}(\mathbf{p})\approx 4i^{n-n^{\prime }}\,\mathrm{e}%
^{-(n-n^{\prime })^{2}/4(1+n+n^{\prime })}  \label{eq:147} \\
\times \frac{d^{2}}{l_{0}}\sin (2\theta _{0})\,v(\mathbf{p})\frac{pl_{0}}{%
(pl_{0})^{2}+1+n+n^{\prime }}
\end{multline}%
for odd $n+n^{\prime }$.

In order to conveniently compare the exact results for $V_{n,n^{\prime }}$
with those obtained in WKB approximation we introduce reduced matrix
elements which are real, dimensionless and depend only on the magnitude $p$
of the momentum. For even $n+n^{\prime }$ the reduced matrix element $%
V_{n,n^{\prime }}^{(\mathrm{r})}(p)$ is defined via the relation 
\begin{equation}
V_{n,n^{\prime }}(\mathbf{p})=\tfrac{d^{2}}{l_{0}}\left[ u(\mathbf{p})\sin
^{2}\!\theta _{0}-2\,P_{2}(\cos \theta _{0})\right] \,V_{n,n^{\prime }}^{(%
\mathrm{r})}(p),  \label{eq:148}
\end{equation}%
and, for odd $n+n^{\prime }$, we set 
\begin{equation}
V_{n,n^{\prime }}(\mathbf{p})=i\tfrac{d^{2}}{l_{0}}\,v(\mathbf{p}%
)\,V_{n,n^{\prime }}^{(\mathrm{r})}(p).  \label{eq:149}
\end{equation}%
Figure \ref{fig:Vdd_matrix_elements} compares the reduced matrix elements in
WKB approximation with the exact ones. Quantitative agreement improves with
increasing HO quantum numbers (corresponding to darker colors in Fig.~\ref%
{fig:Vdd_matrix_elements}) and is, however, satisfactory even for $%
n,n^{\prime }\geq 1$.

\end{fmfshrink}
\end{fmffile}

\bibliography{bibliography}

\end{document}